\documentclass[appendixfloats, iop]{emulateapj-rtx4}
\usepackage{lscape}
\usepackage{natbib}
\usepackage{hhline}
\usepackage{times}
\usepackage{graphicx}
\usepackage{xspace}
\usepackage{hyperref}
\usepackage{color}

\newcommand{\um}{\,$\mu$m\xspace}
\newcommand{\uma}{$\mu$m\xspace}

\newcommand{\umb}{\,\mu m\xspace}
\newcommand{\mh}{H$_{2}$\xspace}

\newcommand{\chandra}{\emph{Chandra}\xspace}
\newcommand{\xmm}{\emph{XMM-Newton}\xspace}
\newcommand{\spitzer}{\emph{Spitzer}\xspace}
\newcommand{\galex}{\emph{GALEX}\xspace}
\newcommand{\wise}{\emph{WISE}\xspace}
\newcommand{\herschel}{\emph{Herschel}\xspace}

\newcommand{\magphys}{\textsc{magphys}\xspace}

\shorttitle{SF Suppression in MOHEGs}
\shortauthors{Lanz et al.}

\begin{document}

\title{Star Formation Suppression Due to Jet Feedback\\ in Radio Galaxies with Shocked Warm Molecular Gas}

\author{Lauranne Lanz$^{1}$, Patrick M. Ogle$^{1}$, Katherine Alatalo$^{2}$\altaffilmark{$\dagger$}, Philip N. Appleton$^{1,3}$}

\affil{
$^{1}$ Infrared Processing and Analysis Center, California Institute of Technology, MC100-22, Pasadena, CA 91125, USA; llanz@caltech.edu \\
$^{2}$ Observatories of the Carnegie Institution of Washington, 813 Santa Barbara Street, Pasadena, CA 91101, USA \\
$^{3}$ NASA Herschel Science Center, IPAC, California Institute of Technology, MC100-22, Pasadena, CA 91125, USA\\
}
\altaffiltext{$\dagger$}{Hubble fellow}

\begin{abstract}

We present \herschel observations of 22 radio galaxies, selected for the presence of shocked, warm molecular hydrogen emission. We measured and modeled spectral energy distributions in 33 bands from the ultraviolet to the far-infrared to investigate the impact of jet feedback on star formation activity. These galaxies are massive, early-type galaxies with normal gas-to-dust ratios, covering a range of optical and infrared colors. We find that the star formation rate (SFR) is suppressed  by a factor of $\sim3-6$, depending on how molecular gas mass is estimated. We suggest this suppression is due to the shocks driven by the radio jets injecting turbulence into the interstellar medium (ISM), which also powers the luminous warm \mh line emission. Approximately 25\% of the sample shows suppression by more than a factor of 10. However, the degree of SFR suppression does not correlate with indicators of jet feedback including jet power, diffuse X-ray emission, or intensity of warm molecular \mh emission, suggesting that while injected turbulence likely impacts star formation, the process is not purely parameterized by the amount of mechanical energy dissipated into the ISM. Radio galaxies with shocked warm molecular gas cover a wide range in SFR--stellar mass space, indicating that these galaxies are in a variety of evolutionary states, from actively star-forming and gas-rich to quiescent and gas-poor. SFR suppression appears to have the largest impact on the evolution of galaxies that are moderately gas-rich.
\end{abstract}

\keywords{galaxies: active -  galaxies: evolution - galaxies: ISM - galaxies: jets - galaxies: star formation} 

\section{INTRODUCTION}

\subsection{AGN Feedback via Radio Jets}

 Active galactic nucleus (AGN) feedback on the interstellar medium (ISM) is thought to be an important factor in regulating star formation activity in galaxies \citep[e.g.,][]{hopkins06}. In our current paradigm of galaxy evolution, supported by numerical simulations, feedback can clear galaxies of gas and thereby suppress the star formation activity as well as supermassive black hole growth \citep[e.g.][]{silk98, dimatteo05}. However, our understanding of the details involved and the variety of means via which the AGN can impact its host remains incomplete
  
 One type of feedback is the interaction between radio jets and the ISM, which may have either positive or negative effects on the star formation rate \citep[SFR; ][]{wagner11}. Hydrodynamical simulations of such interactions have shown that a radio jet may couple strongly to an inhomogeneous, clumpy ISM, injecting turbulence and depositing energy by creating cocoons of hot X-ray emitting gas \citep{sutherland07}. The expansion of these bubbles can then spread the effects of the radio jet across the host galaxy. The net effect may suppress star formation by driving shocks and turbulence into the ISM, thereby rendering the molecular gas infertile to star formation, or by driving outflows that can remove the raw materials for new stars \citep[e.g., ][]{guillard12}. Neutral, ionized, and molecular outflows have all been found in radio galaxies \citep[e.g.][]{emonts05, feruglio10, mahony13, morganti13, morganti15, garcia14, alatalo15mrk}. 
 
Star formation suppression has been conclusively measured, based on resolved molecular observations and detailed modeling of the spectral energy distributions (SEDs) in only a few galaxies. It was demonstrated in NGC\,1266, whose AGN has a small radio jet \citep{nyland13} and is driving a massive molecular outflow \citep{alatalo11}, where star formation activity is suppressed by a factor of 50--150 \citep{alatalo15SF}. \citet{karouzos13} found hints that radio-loud AGN hosts have a lower SFR than inactive galaxies but could not examine the star formation efficiency. \citet{guillard15} recently discussed how the turbulence being injected into the ISM of 3C\,326N could explain the quenching of its star formation activity.

\subsection{Molecular Hydrogen Emission Galaxies}

To study the impact of jet feedback on the star formation activity in their host galaxies, the ideal laboratories are galaxies where we already have evidence of interaction of the jets with the ISM. One class of such galaxies are molecular hydrogen emission galaxies \citep[MOHEGs; ][]{ogle07, ogle10}. These galaxies are identified by the {\em Spitzer Space Telescope} \citep{werner04} by their high mid-infrared (MIR) \mh emission relative to their star formation-related emission (24\um or polycyclic aromatic hydrocarbons (PAHs)). Specifically, MOHEGs are defined to have $L$(H$_2$ 0--0 S(0)--S(3))/$L$(PAH$_{7.7\mu m})>$ 0.04, a ratio that is too large to be produced solely by photoelectric heating in photodissociation regions \citep{ogle10}. Using this criterion to select galaxies therefore excludes those where the warm molecular emission is predominantly due to heating by star formation activity.

\citet{ogle10} explored potential heating mechanisms for this warm molecular emission in a sample of radio galaxies. They ascertained that X-ray heating by an AGN was insufficient, since these radio galaxies do not contain the high-luminosity, high-ionization AGN necessary. They could not rule out the mechanism of cosmic ray heating but calculated that a very high cosmic ray density would be required to explain the observed \mh emission. Therefore, they determined that the most likely mechanism was shock heating, which has been seen in radio galaxies \citep[e.g.,][]{labiano13, scharwachter13} and is a likely result of the interaction between the radio jet and the ISM. This picture is further supported by the correlation found by \citet{lanz15} between the MIR \mh luminosity and the diffuse X-ray luminosity in radio MOHEGs, as both would be powered by the dissipation of the jet's mechanical energy into the ISM. Therefore, radio MOHEGs provide an excellent sample for investigating the effect of jet feedback on star formation activity.  

We present ultraviolet (UV) to far-infrared (FIR) SEDs of 22 radio MOHEGs, which we use to analyze the properties of the host galaxies. We describe the sample selection and the data analysis, including new \herschel photometry, in \S2. In \S3, we discuss our SED fitting methodology and the caveats involved and test the reliability of our SED-derived parameters. We use these galaxy parameters to examine the colors, ISM properties, and star formation activity of this sample in \S4 and summarize our conclusions in \S5. We comment on individual galaxies and present the details of UV--FIR images and fitted SEDs in the appendix.

\section{OBSERVATIONS AND ANALYSIS}
\subsection{Sample}
Our sample is derived from the surveys of \citeauthor{ogle10} (\citeyear{ogle10}; 15 sources) and \citeauthor{guillard12} (\citeyear{guillard12}; 7 sources) of radio galaxies observed with the \spitzer Infrared Spectograph \citep[IRS;][]{houck04}, containing both core-dominated \citep[i.e., FR\,I;][]{fanaroff74} and lobe-dominated (FR\,II) sources. The Ogle et al. galaxies were selected from the 3CRR catalog with a redshift ($z<$0.13 for FR\,I and $z<$0.22 for FR\,II) and flux cuts ($S_{\nu}{\rm(178\,MHz)}>$15\,Jy for FR\,I and $>$16.4\,Jy for FR\,II). The redshift cut insured that the purely rotational quadrupole transitions \mh line series (0--0 S(0) to 0--0 S(7)) was observable with IRS.  The Guillard et al. galaxies were selected to have neutral outflows and have some sources in common with the Ogle et al. sample. 

We specifically focus on those galaxies identified as MOHEGs.\footnote{Although 3C\,31 falls just outside the MOHEG criterion on the H$_{2}$/PAH ratio, \citet{ogle10} argued that it should also be called a radio MOHEG, since it has a larger ratio than the SINGS galaxies, but has a lower ratio than most MOHEGs due to strong PAH7.7\um emission. PKS1549-79 has a similar ratio and likewise has strong PAH7.7\um emission. } Table \ref{sample} presents the sample, including their morphologies, environments, and distances. The Guillard galaxies extend the range of star formation activity, but are not systematically different from the Ogle galaxies. Throughout this paper, we assume a cosmology with Hubble constant $H_{0} = 70 $\,km\,s$^{-1}$\,Mpc$^{-1}$, matter density
parameter $\Omega _M = 0.3$, and dark energy density $\Omega_{\Lambda} = 0.7$ \citep{spergel07}. For the two galaxies with  $z<$ 0.01, we use redshift-independent distances calculated by \citet{tonry01}.

\begin{turnpage}
\begin{deluxetable*}{llcrllrrlc}
\tabletypesize{\scriptsize}
\tablecaption{Sample\label{sample}}
\tablecolumns{9} 
\tablewidth{0pt}
\centering
\tablehead{
\colhead{} & \colhead{Other} &  \colhead{} & \colhead{$D_{L}$\tablenotemark{c}} & \colhead{} & \colhead{} &  \multicolumn{3}{c}{Aperture} & \colhead{Available} \\
\cline{7-9}
\colhead{Name\tablenotemark{a}} & \colhead{Names} &\colhead{$z$\tablenotemark{b}} & \colhead{(Mpc)} &  \colhead{Morphology\tablenotemark{d}} & 
\colhead{Environment\tablenotemark{e}}  & \colhead{R.A.(J2000)} &	\colhead{Dec.(J2000)} & \colhead{Size\tablenotemark{f}} &	 \colhead{Data\tablenotemark{g}}
}
\startdata
3C\,31   		& N\,383			&  0.0170	& 73.8~~	& Ep; I	& Close pair		& 01:07:24.959		 &$+$32:24:45.21	& $~\,98\farcs8\times~\,73\farcs5$(40$^{\circ}$)\tablenotemark{h}&1,2,3,4,5,6,7,8	\\ 
3C\,84   		& N\,1275, Per. A	&  0.0176	&  76.4~~	& cD+D; I	& Perseus CC		& 03:19:48.160		 &$+$41:30:42.11	& $~\,84\farcs2\times~\,77\farcs9$(27$^{\circ}$)\tablenotemark{h}	&	1,2,3,4,5,6,7,8	\\ 
3C\,218 		& Hyd. A			&  0.0549 	&  245~~~~ & cD; I	& Abell780 CC		& 09:18:05.688		 &$-$12:05:43.39	& $~\,27\farcs5\times~\,23\farcs9$(51$^{\circ}$)\tablenotemark{h}	& ~~~~1,~~\,3,4,5,6,7,8,\,10 \\
3C\,236$^{*}$ 	& ...				&  0.1005	& 463~~~~ & E; II	& Single			& 10:06:01.735		 &$+$34:54:10.43	& $~\,23\farcs9\times~\,19\farcs6$($322^{\circ}$)	&	~~~~~~~~~\,2,3,4,5,6,7,8,~~~\,11 \\
3C\,270 		& N\,4261 		&  0.0074	& 32.0$^{\dagger}$ & Ep; I& Virgo Member& 12:19:23.245	 &$+$05:49:29.63	& $106\farcs6\times~\,90\farcs4$($66^{\circ}$)\tablenotemark{h}	& ~~1,2,3,4,5,6,~~~\,~9 \\
3C\,272.1 		& N\,4374, M\,84 	& 0.0034	& 18.0$^{\dagger}$ & Ep; I& Virgo Member& 12:25:03.707	 &$+$12:53:12.92	& $117\farcs8\times112\farcs3$($30^{\circ}$)\tablenotemark{h}	& ~~1,2,3,4,5,6,~~~\,~9 \\
4C\,12.50$^{*}$ & PKS\,1549$+$12	&  0.1217	& 568~~~~ & S0; II	& Merger			& 13:47:33.360		 &$+$12:17:24.04	& $~~33\farcs0\times~\,33\farcs0$($0^{\circ}$)	&     1,2,3,4,5,6,7,8 \\
3C\,293		& U\,8782			&  0.0450	& 199~~~~ & S0; I	& Pair			& 13:52:17.821	 	 &$+$31:26:46.50	& $~~47\farcs5\times~\,22\farcs5$(333$^{\circ}$)\tablenotemark{h}&~~1,2,3,4,5,6,~~~\,~9\\
MRK\,668$^{*}$ & OQ\,208		&  0.0766	& 347~~~~  & Sa; CSO & Pair			& 14:07:00.400		 &$+$28:27:14.70	& $~\,29\farcs4\times~\,26\farcs7$(76$^{\circ}$)\tablenotemark{h}	&	1,2,3,4,5,6,7,8	\\
3C\,305$^{*}$ 	& IC\,1065		&  0.0416	& 184~~~~ & Ep; I	& Single			& 14:49:21.625		 &$+$63:16:14.43	& $~\,27\farcs4\times~\,23\farcs1$(9.3$^{\circ}$)	&	1,2,3,4,5,6,7,8	\\
3C\,310 		& VV\,204b		&  0.0538	& 240~~~~ & Ep+D; II& Poor cluster		& 15:04:57.179	 	 &$+$26:00:58.33	& $~\,42\farcs4\times~\,40\farcs5$(313$^{\circ}$)\tablenotemark{i}&1,2,3,4,5,6,7,8\\
3C\,315 		&  ...				&  0.1083 	& 501~~~~ & S0; II	& Close pair		& 15:13:40.055		 &$+$26:07:30.44	& $~~~~9\farcs5\times~~~~6\farcs8$(308$^{\circ}$)\tablenotemark{h}& ~\,~2,3,~~\,5,6,7,8\\
3C\,317 		&U\,9799   		&  0.0345	& 152~~~~ & cD+D; I& A2052 CC		& 15:16:44.498		 &$+$07:01:17.62	& $145\farcs8\times~\,72\farcs3$(40$^{\circ}$)\tablenotemark{h}&	1,2,3,4,5,6,7,8	\\
3C\,326N 		& ...				&  0.0895	& 409~~~~ & Ep+D; II& Pair			& 15:52:09.140		 &$+$20:05:47.24	& $~\,30\farcs9\times~\,22\farcs0$(20$^{\circ}$)\tablenotemark{h}	&	1,2,3,4,5,6,7,8	\\
PKS\,1549-79$^{*}$ 	& ...			&  0.1522	& 725~~~~ & E; CSO&Single			& 15:56:58.900		 &$-$79:14:04.30	& $~\,87\farcs7\times~\,75\farcs5$(327$^{\circ}$)\tablenotemark{h, j}& ~~~~~~1,~~\,3,~~5,~~\,7,8,9,10	\\
3C\,338 		& N\,6166			&  0.0304	& 133~~~~ & cD; Ip	& A2199 CC		& 16:28:38.202		 &$+$39:33:04.70	& $139\farcs4\times~\,93\farcs6$(302$^{\circ}$)\tablenotemark{i}&	1,2,3,4,5,6,7,8	 \\
3C\,386 		& ...				&  0.0169	& 73.3~~ 	& E; I	& Single			& 18:38:26.251		 &$+$17:11:49.94	& $~\,67\farcs3\times~\,54\farcs8$(337$^{\circ}$)\tablenotemark{h} &	1,~~\,3,4,5,6,7,8	\\ 
3C\,424 	    	& ...				&  0.1270	& 595~~~~ & Ep; I	& Group			& 20:48:12.099	 	 &$+$07:01:17.05	& $~\,26\farcs0\times~\,21\farcs9$(37$^{\circ}$)\tablenotemark{h}	&	\,~~~~1,~~\,3,~\,~5,~\,~7,8,\,10\\
IC\,5063$^{*}$ 	 & ...				&  0.0113	& 48.8~~ 	& Ep; II	& Pair			& 20:52:02.402		 &$-$57:04:07.58	& $~\,89\farcs4\times~\,84\farcs8$(313$^{\circ}$)	&	\,~~~~1,~~\,3,4,5,6,7,8,\,10	\\
3C\,433 	    	& ...				&  0.1016	& 468~~~~ & S0; II	& Group			& 21:23:44.565		 &$+$25:04:27.56	& $~\,44\farcs3\times~\,42\farcs3$(295$^{\circ}$)\tablenotemark{h}	&	\,~~~~1,~~\,3,~\,~5,~\,~7,8,\,10	\\ 
3C\,436 	    	& ...				&  0.2145	& 1060~~~~ & Ep; II	& Single			& 21:44:11.700		 &$+$28:10:19.00	& $~\,10\farcs9\times~~~~9\farcs1$(276$^{\circ}$)	&   	1,2,3,~~\,5,~\,~7,8	\\ 
3C\,459$^{*}$ 	& ...				&  0.2201	& 1090~~~~ & Ep; II	& Single			& 23:16:35.230		 &$+$04:05:18.50	& $~~~~6\farcs7\times~~~~6\farcs6$(78$^{\circ}$)\tablenotemark{k} 	&   	1,2,3,~~\,5,6,7,8	
\enddata
 \tablenotetext{a}{The $^{*}$ indicates that the source is from \citet{guillard12} rather than \citet{ogle10}. }
 \tablenotetext{b}{Redshifts were taken from NED.}
 \tablenotetext{c}{{Luminosity} distance calculated assuming $H_{0}=70 {\rm km~s}^{-1}~{\rm Mpc}^{-1}$, $\Omega_{\rm M}=0.3$, and $\Omega_{\Lambda }$=0.7 \citep{wright06,spergel07}, except for galaxies at $z<$0.01  ($^{\dagger}$), where distances were taken from \citet{tonry01}.}
 \tablenotetext{d}{Host morphology from \citet{ogle10}: E$-$elliptical, Ep$-$peculiar elliptical, cD$-$cluster dominant, S0$-$lenticular. The notation ``+D'' indicates a significant exponential disk component. I or II indicate  the \citet{fanaroff74} type of radio jet (p$-$peculiar), while CSO indicates a compact symmetric object.}
 \tablenotetext{e}{Cluster, group, or pair membership from \citet{ogle10}. CC$-$cool X-ray core cluster. }
\tablenotetext{f}{Apertures are given as semi-major axis $\times$ semi-minor axis (position angle given counter-clockwise from north; see also Section 2.2.6).}
\tablenotetext{g}{The numbers listed correspond to the following observations being available: 1: \galex, 2: SDSS, 3: 2MASS, 4: \spitzer IRAC, 5: {\em WISE}, 6: \spitzer MIPS, 7: \herschel PACS, and 8: \herschel SPIRE. Ancillary literature photometry is also marked if used: 9: {\em IRAS}, 10: optical, and 11: UV. }
\tablenotetext{h}{Exclusion regions were placed on the companion(s) or nearby background/foregound sources.}
\tablenotetext{i}{These galaxies lie at the center of clusters in a nest with several close companions, which we have excluded; however, as a result we may be excluding some of the source flux or some contamination may remain.  }
\tablenotetext{j}{The PACS and SPIRE images have significant background/foreground structure in this aperture, and the galaxy is a point source at \herschel wavelength, so we use the point source aperture.}
\tablenotetext{k}{This aperture is smaller than the point source apertures in MIR--FIR, but there is a nearby optically bright point source. Therefore, this aperture is used in UV--NIR bands with larger point source apertures at longer wavelengths. }
\end{deluxetable*}
\end{turnpage}

\subsection{Observations and Data Reduction}
In order to examine the properties of the host galaxies of these radio MOHEGs, we created UV--FIR SEDs, based on observations from {\em Galaxy Evolution Explorer} \citep[\galex;][]{martin05}, Sloan Digital Sky Survey \citep[SDSS;][]{york00}, the 2 Micron All Sky Survey \citep[2MASS;][]{skrutskie06}, {\em Spitzer} \citep{werner04}, {\em Wide-field Infrared Survey Explorer} \citep[\wise;][]{wright10}, and \herschel \citep{pilbratt10}. In the next sections, we describe the reduction we performed on these data. For ease of reading, we confine the details of the observations (Table \ref{obs_desc}) and measured photometry (Table \ref{phot}) to the appendix, where we also comment on peculiarities of the individual galaxies and present UV--FIR images and fitted SEDs. 

\subsubsection{Ultraviolet (GALEX) Photometry}

All but two of our galaxies were observed by \galex. Mosaics of the longest observations were retrieved from the Mikulski Archive for Space Telescopes using GalexView version 1.4.10. In the case of 3C\,433, only an near-UV (NUV) observation is available. We used the conversions from count rate to fluxes provided by \citet{goddard04}\footnote{\url{http://galexgi.gsfc.nasa.gov/docs/galex/FAQ/counts_background.html}} and corrected for foreground extinction due to the Milky Way dust using the extinction laws given by \citet{wyder05} and the $N_{\rm H}$ of \citet{kalberla05}\footnote{Obtained from \url{http://heasarc.gsfc.nasa.gov/cgi-bin/Tools/w3nh/w3nh.pl}}. Backgrounds estimates were measured near each galaxy in source-free regions. Photometric uncertainties consist of the Poisson uncertainty added in quadrature with a 10\% calibration uncertainty \citep{goddard04}.

\subsubsection{Optical (SDSS) Photometry}
Sixteen of our galaxies have SDSS images available in DR12 \citep{alam15}. We retrieved mosaics of each galaxy in all five {\em ugriz} filters from the DR12 Science Archive Server\footnote{\url{http://dr12.sdss3.org./fields}}. All were taken in Drift mode with 53.9\,s exposure times. We corrected for foreground extinction, using the extinction corrections of \citet{stoughton02} with the same $N_{\rm H}$ as for the \galex corrections. Background estimates were measured near each galaxy in source-free regions. Photometric uncertainties consist of the Poisson uncertainty added in quadrature with a 3\% ({\em gri}) or 5\% ({\em uz}) calibration uncertainty \citep{stoughton02}. For galaxies lacking SDSS observations (six galaxies), {\em UBVR} photometry or limits were obtained from the literature (see \S 2.2.7).

\subsubsection{Near-infrared (2MASS) Photometry}
Our sample has complete near-infrared (NIR) coverage from 2MASS. We retrieved mosaics from the NASA/IPAC Infrared Science Archive (IRSA), preferably from the Large Galaxy Atlas \citep{jarrett03}. The counts measured in the images were converted to magnitudes using the magnitude zeropoints given in the header of each image and then to Janskys using the flux conversions of \citet{cohen03_2mass}. Backgrounds estimates were measured near each galaxy in source-free regions and photometric uncertainties are the sum in quadrature of the uncertainty due the flux conversion factor uncertainty, a calibration uncertainty of 3\%, and Poisson uncertainty  \citep{2massExSup}.

\subsubsection{Mid-infrared Photometry}

\paragraph{Spitzer {\em IRAC}}Sixteen of our galaxies were observed with the Infrared Array Camera \citep[IRAC;][]{fazio04} as part of nine different programs. Pipeline-created mosaics (S18.25.0) were retrieved from the Spitzer Heritage Archive. The \spitzer fluxes required aperture corrections. We determined the effective radius of each elliptical aperture\footnote{r$_{{\rm eff}}=\sqrt{a \times b}$ for semimajor axis \emph{a} and semiminor axis \emph{b}} and used the extended source flux corrections given in the IRAC Instrument Handbook\footnote{\url{http://irsa.ipac.caltech.edu/data/SPITZER/docs/irac/iracinstrumenthandbook/30/}}. Background estimates were measured in the same field in regions selected to mimic the content of background and foreground objects in the apertures in the outskirts of the galaxies. Photometric uncertainties consist of the sum in quadrature of the 3\% calibration uncertainty, which typically dominates, and the error measured from the uncertainty images \citep{cohen03_irac}.
\vspace{-4.1mm}
\paragraph{WISE}Our sample has complete coverage by \wise. We retrieved mosaics from IRSA.  Counts were converted to fluxes via magnitudes using the zeropoints given in the All-Sky Explanatory Supplement \citep{allwiseExSup}. \wise fluxes require both an aperture and a color correction, depending on the shape of the SED. We first determined which power-law or blackbody model best fits the photometry and then applied those color corrections \citep{wright10}. Aperture corrections are given for the default point source aperture ($8.25''$ for 3.4\um-12\um and $16.5''$ for 22\um)\footnote{\url{http://wise2.ipac.caltech.edu/docs/release/allsky/expsup/sec4_4c.html}}  in the All-Sky Explanatory Supplement \citep{allwiseExSup}, as are point spread function (PSF) images. We derive aperture corrections by measuring the ratio of the flux contained in the default point source aperture in the PSF images to the flux contained in our desired aperture, and multiplying that ratio by the standard correction. Photometric uncertainties consist of the sum in quadrature of the uncertainty in the flux conversion factor, the Poisson uncertainty, and calibration uncertainties of 2.4\%, 2.8\%, 4.5\%, and 5.7\%, respectively, in order of increasing wavelength \citep{allwiseExSup}.

\subsubsection{Far-infrared Photometry}

\paragraph{Spitzer {\em MIPS}}Eighteen of our galaxies were observed with {\em Spitzer}'s Multiband Imaging Photometer \citep[MIPS;][]{rieke04}, as part of nine different programs. Pipeline-created mosaics (S18.12.0 or S18.13.0) were retrieved from the Spitzer Heritage Archive. At 70 and 160\um, we use the filtered mosaics, which are better corrected for artifacts and are recommended for point sources, as none of our galaxies are resolved by MIPS at these longer wavelengths.  MIPS fluxes require aperture corrections. Aperture corrections at several radii are given in the MIPS instrument Handbook\footnote{\url{http://irsa.ipac.caltech.edu/data/SPITZER/docs/mips/mipsinstrumenthandbook/50/}}. We estimate the radius as the effective radius of the aperture (see footnote 9) and interpolate between the available aperture corrections. Photometric uncertainties consist of the sum in quadrature of the calibration uncertainty (4\% at 24\um and 15\% at 70 and 160\um) and the error measured from the uncertainty images \citep{engelbracht07}. MIPS 70\um photometry is only used in the absence of Photoconductor Array Camera and Spectrometer (PACS) 70\um photometry; MIPS 160\um photometry is only used in the absence of PACS160\um photometry.
\vspace{-4.1mm}
\paragraph{Herschel {\em PACS}}
Nineteen of our galaxies were observed with the PACS \citep{poglitsch10} instrument on \herschel, which observed at 160\um in conjunction with either 70\um or 100\um. All 19 were observed at 100\um, but only nine were observed at 70\um. About 75\% were taken as part of a Cycle 1 Open Time (OT1) program on radio jet feedback (P.I. Ogle), but we also use PACS observations taken as part of five other programs. Level 0 data were retrieved from the Herschel Science Archive and processed to  Level 1 using the calibration trees of version 12.1.0 of the Herschel Interactive Processing Environment \citep[HIPE;][]{ott10} to prepare the products necessary to create mosaics using the 2013 July 31 version of Scanamorphos \citep{roussel13}. 
PACS photometry requires both color and aperture corrections. Color corrections are available for a range of blackbody models with temperatures between 5 and 1000\,K \citep{pacscolor}; we used the color correction for the blackbody whose temperature best fit our photometry. HIPE contains aperture corrections for 140 different radii. We derive aperture corrections for our elliptical apertures at their effective radii (see footnote 9) by interpolating between the HIPE values. Photometric uncertainties for PACS bands consists the sum in quadrature of a statistical uncertainty based on the background fluctuations (following the method of \citealt{dale12}) and a 10\% calibration uncertainty \citep{pacsunc}. We find good agreement between the MIPS and PACS photometry at 70 and 160\um for those galaxies observed with both, but preferentially use the PACS photometry where available.
\vspace{-4.1mm}

\begin{turnpage}
\begin{deluxetable*}{lccccccccccccccc}
\tabletypesize{\scriptsize}
\tablecaption{Literature Properties\label{litprop}}
\tablewidth{0pt}
\tablehead{
\colhead{} &  \multicolumn{3}{c}{Warm H$_2$\tablenotemark{a}}&  \multicolumn{3}{c}{Cold H$_{2}$} & \colhead{Stellar} & \multicolumn{2}{c}{X-rays\tablenotemark{f}} & \multicolumn{3}{c}{Jet Power}\\
\cline{2-4}  \cline{5-7} \cline{9-10} \cline{11-13}
\colhead{} &\colhead{Log($L$)} &\colhead{Log($M$)}&\colhead{H$_{2}$/} &
 \colhead{Log($M$)\tablenotemark{b}} & \colhead{Size\tablenotemark{c}} & \colhead{References\tablenotemark{d}} &\colhead{Size\tablenotemark{e}} &
\colhead{Log($L_{\rm X,\,diff.}$)}	& \colhead{Log($L_{\rm X,\,AGN}$)} &
\colhead{S$_{\rm 178\,MHz}$\tablenotemark{g}} &\colhead{$P_{\rm Jet}$\tablenotemark{h}} & \colhead{References\tablenotemark{i}}  \\ 
\colhead{Galaxy} &\colhead{(erg\,s$^{-1}$)} & \colhead{[$M_{\odot}$]} 		   &\colhead{PAH$_{{\rm 7.7\mu m}}$} &
 \colhead{[$M_{\odot}$]} 		   & \colhead{(kpc)} 	     & \colhead{} 	&\colhead{(kpc)} &
\colhead{(erg\,s$^{-1}$)}	& \colhead{(erg\,s$^{-1}$)}	& 
\colhead{(Jy)} 				&\colhead{(10$^{43}$\,erg\,s$^{-1}$)} & \colhead{} 
}
\startdata
3C\,31   		&40.32			& ~~\,8.37	& 0.030	 	& ~\,8.95	& ~\,1.0 (1)		& (1)		& $14\times10$ (1)	&41.07 	& 40.67	 	& ~\,18.3~\,				& ~~~0.44~\,	&  (1)	\\
3C\,84   		&41.81			&$<$8.91	& 0.560	 	& 10.47	&$14.3\times7.2$ (1)& (2)		& $13\times11$ (1)	&44.10 	& 42.91	 	& ~\,68.2~\,				& ~~~1.8~~~	&  (1)	\\
3C\,218 		&41.10			& ~~\,9.30	& 0.124	 	& ~\,9.26	& ~\,8.5 (2)		& (3)		& $14\times11$ (3)	&43.67 	& 41.69	 	& ~228~~~\,\tablenotemark{j}	& ~\,54~~~~~	&  (2)	\\	
3C\,236 		& ~\,41.73$^{*}$	& ~~\,9.26	& $>$0.469~~\, & ~\,9.31	& ~\,1.3 (1) 		& (4)		& $15\times11$ (1)	&42.11 	& 43.02	 	& ~\,20.5~\,				& ~\,17~~~~~	&  (2) 	\\
3C\,270 		&39.30			&$<$7.48	& 0.096	 	& $<$6.82	& ~\, ... (4) 		& (5)		& $11\times~\,9$ (1)	&40.96 	& 41.08	 	& ~\,53.3~\,				& ~~~0.30~\,	&  (2) 	\\	
3C\,272.1 		&39.01			& ~~\,6.90	& 0.126	 	& ~\,6.77	& ~\, ... (4)		& (6)		& ~\,8 (1)			&41.46 	& 39.34	 	& ~\,21.3~\,				& ~~~0.037	&  (1) 	\\	
4C\,12.50 		& ~\,42.50$^{*}$	& ~10.61	& 0.213 	 	& 10.73	& ~\,4.2 (1)		& (7)		&  16 (1)			&42.29 	& 43.34	 	& ~~~4.60					& ~~~5.7~~~	&  (3) 	\\	
3C\,293		&41.76			& ~~\,9.57	& 0.242	 	& 10.32	& 10.6 (1)			& (8)		& $19\times11$ (1)	&41.39 	& 42.78	 	& ~\,13.8~\,				& ~~~2.2~~~	&  (1) 	\\	
MRK\,668 	& ~\,41.89$^{*}$ 	&$<$9.20 	& 0.104 		& 10.19	& ~\, ... (4)		& (6)		& $10\times~\,7$( 1)	&41.7~\,	& 42.5~\,	 	& ~~~~~~0.12 \tablenotemark{k} & ~~~0.058	&  (3) 	\\	
3C\,305 		& ~\,41.59$^{*}$ 	&$<$8.03 	& 0.153 		& ~\,9.31& $20.6\times8.2$ (2)	& (6)		& $21\times15$ (1)	&41.29 	& 41.23	 	& ~\,17.1~\,				& ~~~2.3~~~	&  (1)	\\
3C\,310 		&40.86			& ~~\,8.23	& $>$0.734~~\,& ...		& ~\, ... (4)		& ...  		& 10 (1)			&41.47 	& 40.11	 	& ~\,61.0~\,				& ~\,14~~~~~	& (1)\\
3C\,315 		&41.83			&$<$8.52	& 0.625		& ...		& ~\, ... (4)		&  ... 		& $10\times~\,6$ (1)	&41.47  	& 41.68	 	& ~\,20.6~\,				& ~\,20.~~~~\,	&  (1) 	\\
3C\,317 		&40.59			&$<$8.31	&  $>$0.707~~\,& ~\,7.93	& ~\,2.6 (1)		& (9)		& $17\times10$ (1)	&43.30  	& 41.30	 	& ~\,49.0~\, 				& ~~~4.6~~~	&  (4)	\\
3C\,326N 		&41.73			& ~~\,9.34	&$>$4.43~\,~~\, & ~\,9.14	& ~\, ... (4)		& (10) 	& $10\times~\,7$ (1)	&41.37  	& 40.63	 	& ~\,22.2~\,				& ~\,14~~~~~ 	&  (5)	\\
PKS\,1549-79	& ~\,42.61$^{*}$ 	&$<$9.93 	& 0.035 	 	& ...		& ~\, ... (4)		&  ... 		& 11 (5)			&43.1~\, 	& 44.7~\, 	 	& ~~\,22.0~\,\tablenotemark{l}	& ~\,44~~~~~	&  (2) 	\\
3C\,338 		&40.59			&$<$8.30	&  $>$0.613~~\, & ~\,7.89	& ~\, ... (4)		& (11)	& $12\times~\,9$ (1)	&43.45  	& 40.30	 	& ~\,51.1~\,				& ~~~3.7~~~	&  (1) 	\\
3C\,386 		&39.90			&$<$7.60	&  $>$0.613~~\, & ~\,8.21	& ~\,2.7 (2)		& (6)		& ~\,$7\times~\,5$ (2)&40.24  	& 39.75	 	& ~\,26.1~\, 				& ~~~0.62~\,	&  (1) 	\\
3C\,424 	    	&41.97			& ~~\,9.52	&  $>$3.68~\,~~\, & $<$9.79	& ~\, ... (4)	& (12)	& ~\,9 (5)			&42.12  	& 42.44	 	& ~\,15.9~\,				& ~\,21~~~~~	&  (4)	\\
IC\,5063 		& ~\,40.87$^{*}$  	& ~~\,8.94& 0.137 		& ~\,8.69& $4.6\times2.3$ (3)	&(13)	& ~\,$7\times~\,6$ (1)&41.34 	& 42.97	 	& ~~~~~7.47 \tablenotemark{m}& ~~~0.086	&  (6, 7)	\\
3C\,433 	    	&42.13			& ~10.36	& 0.648		& $<$9.90	& ~\, ... (4)		& (14)	& ~\,9 (4)			&42.62  	& 43.90	 	& ~\,61.3~\, 				& ~\,51~~~~~	&  (1) 	\\
3C\,436 	    	&42.31			& ~10.21	& 0.478 		& ...		& ~\, ... (4)		& ...  		& $10\times~\,9$ (1)	&42.30  	& 43.53	 	& ~\,19.4~\, 				& ~\,82~~~~~	&  (1) 	\\	
3C\,459 		& ~\,42.38$^{*}$  	&$<$9.96 & 0.075 		& ...		& ~\, ... (4)		&  ... 		& 11 (1)			&42.80  	& 43.24	 	& ~\,30.8~\, 				& 140~~~~~	&  (2) 	
\enddata
\tablenotetext{a}{H$_{2}$ luminosities (0-0 S(0)-S(3)) marked with $^{*}$ are from \citet{guillard12}. All others are from \citet{ogle10}, corrected as described in those papers for undetected lines. The warm masses are corrected for differences in distance assumptions and come from thermal modeling of the IRS SLEDs.  Ratios of the \mh luminosities to the 7.7\um PAH luminosities are from the same papers.} 
\tablenotetext{b}{Mass is calculated assuming a typical $\alpha_{\rm CO}=4.3$\,M$_{\odot}$\,(K\,km\,s$^{-1}$\,pc$^{2}$)$^{-1}$, corresponding to $X_{\rm CO}=2\times10^{20}$\,cm$^{-2}\,{\rm(K\,km\,s^{-1})^{-1}}$ \citep{bolatto13}. } 
\tablenotetext{c}{If a single number is given, it is the radius of the disk. Otherwise, we give semimajor $\times$ semiminor axis. (1) extent from resolved CO;  (2) extent from IRAC 8\um; (3) extent from UV; (4) no information on extent, so we assume 1 kpc in radius (see also Section 2.2.8).}
\tablenotetext{d}{References for H$_2$ masses and extent from CO observations: (1) \citet{okuda05}; (2) \citet{salome06}; (3) \citet{salome03}; (4) \citet{labiano13}; (5) \citet{okuda13}; (6) \citet{ocana10}; (7) \citet{dasyra14}; (8) \citet{labiano14}; (9) \citet{braine94}; (10) \citet{nesvadba10}; (11) \citet{smolcic11}; (12) \citet{saripalli07}; (13) \citet{morganti13}; and (14) \citet{evans05}.}
\tablenotetext{e}{If a single number is given, it is the radius of the disk. Otherwise, we give semimajor $\times$ semiminor axis. Extent from (in order of preference): (1) SDSS g;  (2) IRAC 3.6\um; (3) IRAC 4.5\um; (4) 2MASS K; or (5) unresolved so we assume 4$''$ (the 2MASS resolution), corresponding to the sizes given.}
\tablenotetext{f}{Diffuse 0.5--8\,keV luminosities as calculated in \citet{lanz15}, except for Mrk\,668 and PKS\,1549-79 (see Appendix B). AGN luminosities (2--10\,keV) are likewise from \citet{lanz15} and references therein.}
\tablenotetext{g}{The 178 MHz flux density on \citet{baars77} scale.}
\tablenotetext{h}{Jet power calculated using the formula of \citet{punsly05}: P$_{\rm jet}=6.7\times10^{44}\,(1+z)^2\,Z^2\,S_{\rm 178 MHz}$\,erg\,s$^{-1}$, where $Z=3.31-3.65\times([(1+z)^4 -0.203(1+z)^3+0.749(1+z)^2+0.444(1+z)+0.205]^{-0.125})$. \citet{guillard12} used the version of equation 2 (defining $Z$) of \citet{punsly05} containing a typographical error, which is corrected in arXiv:astro-ph/0503267 (Punsly 2015, private communication). The use of the incorrect formula results in a median factor of $\sim$30 difference in the jet power, increasing with proximity.}
\tablenotetext{i}{References for the 178\,MHz flux: (1) \citet{laing80}, (2) \citet{kuhr81}, (3) \citet{stanghellini98}, (4) \citet{kellermann69}, (5) \citet{laing83}, (6) \citet{large81}, and (7) \citet{mauch03}.}
\tablenotetext{j}{Extrapolated assuming a power-law from measurements at 160 and 468\,MHz.}
\tablenotetext{k}{Extrapolated assuming a power-law from measurements at 327, 365, and 610\,MHz.}
\tablenotetext{l}{Extrapolated assuming a power-law from measurements at 468 and 960\,MHz.}
\tablenotetext{m}{Extrapolated assuming a power-law from measurements at 408 and 843\,MHz.}
\end{deluxetable*}
\end{turnpage}

\paragraph{Herschel {\em SPIRE}}The same nineteen galaxies observed with PACS were also observed with the Spectral and Photometric Imaging Receiver \citep[SPIRE;][]{griffin10}, with about 75\% taken as part of the OT1 program of P. Ogle. Additional data from four other programs were also used. The data were retrieved from the Herschel Science Archive and processed through HIPE using the default pipeline scripts to create Small Map mode mosaics (calibration trees v11.0). SPIRE photometry requires both color and aperture corrections. Color corrections are available for a range of power-law models with indices between --4 and 5 and include color-dependent beam shape corrections, since image units are in Jy per beam \citep{spirehandbook}. In contrast to PACS, SPIRE documentation only had aperture corrections for its default point source apertures (22$''$, 30$''$, and 42$''$ at 250\um, 350\um, and 500\um, respectively). Therefore, we obtained PSF images\footnote{\url{https://nhscsci.ipac.caltech.edu/sc/index.php/Spire/PhotBeamProfileDataAndAnalysis}} and derived aperture corrections in the same manner as for the \wise photometry. For seven of the 19 galaxies with SPIRE data, the aperture determined at optical/MIR wavelengths (see \S2.2.6) is smaller than the point source aperture at one or more SPIRE bands. In these cases, we measured the SPIRE photometry in the point source aperture instead. Photometric uncertainties consist of the sum in quadrature of a statistical uncertainty calculated in the same manner as for the PACS bands and a 10\% systematic uncertainty\footnote{SPIRE literature indicates the calibration uncertainty is 4--5\%. However, \citet{pearson14}, amongst others, argues that aperture photometry is a less reliable method than some of the tools found in HIPE, such as Timeline Fitter, which are not easily applicable to our study, which seeks to use consistent extraction apertures. Therefore, we use a higher systematic uncertainty.}  \citep{pearson14}. The \herschel photometry of three of our galaxies has previously been published (3C\,84 by \citealt{mittal12}, 3C\,326N by \citealt{guillard15}, and IC\,5063 by \citealt{melendez14}), with which we typically have good agreement.

\subsubsection{Aperture Determination}

For consistency, we sought to use matched apertures across our SEDs. In order to determine the aperture necessary to fully capture both the optical and infrared (IR) emission, we used the SExtractor algorithm \citep{bertin96} to determine Kron apertures in the SDSS and \wise images. In the absence of SDSS images, we used the IRAC images. We measured the {\em ugriz} and \wise photometry in the largest SDSS- and \wise-derived apertures, determined whether there was a significant difference in the photometry, and examined the extent of the apertures relative to other sources in the field. Using this information, we selected the aperture that captured all the flux at both optical and MIR wavelengths, while minimizing contamination due to foreground or background sources. Some of our galaxies have companions or exist in clusters. We used the results of SExtractor to help define exclusion regions to minimize the contamination to the flux of these other sources. In Table \ref{sample}, we provide the size and orientation of the apertures used and note which also have exclusion regions applied. 3C\,310 and 3C\,338 lie in the centers of clusters in a nest of galaxies, making it particularly difficult to exclude all of the flux from neighboring galaxies without removing flux from the host galaxy. Therefore, the photometry of these galaxies should be treated with caution. Similarly, 3C\,459 lies near an optically bright foreground star. Therefore, the best aperture at short wavelengths is quite small in order to minimize contamination. At longer wavelengths, we use the (larger) recommended point source apertures. The measured photometry and upper limits (3$\sigma$) are given in Table \ref{phot}.

\begin{deluxetable*}{lccccccccc}
\tabletypesize{\scriptsize}
\tablecaption{Galaxy Parameters\label{sedpar}}
\tablewidth{\linewidth}
\tablehead{
\colhead{} & \multicolumn{6}{c}{{\sc magphys}\tablenotemark{a}}	& & 	\multicolumn{2}{c}{AGN} \\  
\cline{2-7} \cline{9-10}
\colhead{} & \colhead{$M_{*}$} & \colhead{SFR}& \colhead{$L_{\rm Dust}$}& \colhead{$M_{\rm Dust}$}&\colhead{$T_{\rm Warm}$} &\colhead{$T_{\rm Cold}$} & & \colhead{$\alpha$\tablenotemark{d}} 	& \colhead{Log($\nu L_{\nu}$}\\
\colhead{Galaxy}& \colhead{(10$^{11}\,M_{\odot}$)}	& \colhead{($M_{\odot}$\,yr$^{-1}$)}	& \colhead{(10$^{10}\,L_{\odot}$)}	& \colhead{(10$^{7}\,M_{\odot}$)}	& \colhead{(K)\tablenotemark{b}} &	 \colhead{(K)\tablenotemark{c}}	 &
& \colhead{}	& \colhead{(6\uma; $L_{\odot}$))}		}
\startdata
\vspace{0.6mm}
3C\,31		&$2.95^{+0.07}_{-0.07}$	&$~~~0.162~\,^{+~\,0.004~\,}_{-~\,0.001~\,}$	& $~~~0.871~\,^{+~~~0.006~\,}_{-~~~0.026~\,}$		& $~~~1.32~~~^{+~\,0.27~~~}_{-~\,0.14~~~}$	& $39.8^{+~\,9.6}_{-~\,6.3}$	&$20.8^{+0.5}_{-0.8}$ 	&	&	2.0	&	42.3	 \\ \vspace{0.6 mm}
3C\,84		&$2.40^{+0.06}_{-0.40}$	&$~~~7.76~~~^{+~\,0.31~~~}_{-~\,4.11~~~}$	& $~\,10.2~~~~\,^{+~~~0.5~~~~\,}_{-~~~5.6~~~~\,}$  	& $~\,11.0~~~~\,^{+~\,6.9~~~~\,}_{-~\,6.5~~~~\,}$	& $59.6^{+~\,0.1}_{-~\,0.1}$	&$19.4^{+0.4}_{-0.1}$ 	&	&	3.0	&	43.6	 \\ \vspace{0.6 mm}
3C\,218		&$1.45^{+0.25}_{-0.10}$	&$~~~3.63~~~^{+~\,0.08~~~}_{-~\,0.68~~~}$	& $~~~2.19~~~^{+~~~0.05~~~}_{-~~~0.15~~~}$		& $~~~7.76~~~^{+~\,0.89~~~}_{-~\,0.68~~~}$	& $39.2^{+~\,8.3}_{-~\,1.4}$	&$15.4^{+0.1}_{-0.2}$ 	&	&	...	&  	... 	\\ \vspace{0.6 mm}
3C\,236		&$1.00^{+1.51}_{-0.26}$	&$~~~0.251~\,^{+~\,0.876~\,}_{-~\,0.217~\,}$	& $~~~2.88~~~^{+~~~1.90~~~}_{-~~~0.31~~~}$		& $~\,10.7~~~~\,^{+~\,1.8~~~~\,}_{-~\,4.4~~~~\,}$	& $55.2^{+~\,3.6}_{-~\,6.1}$	&$18.3^{+0.7}_{-0.5}$ 	&	&	2.3	&	43.5	 \\ \vspace{0.6 mm}
3C\,270\tablenotemark{e}	&$0.91^{+0.53}_{-0.58}$ &$~~~0.0794^{+~\,0.0206}_{-~\,0.0519}$	& $~~~0.0813^{+~~~0.0187}_{-~~~0.0254}$		& $~~~0.0724^{+~\,0.1181}_{-~\,0.0537}$		& $46.6^{+~\,9.0}_{-10.6}$ 	&$21.4^{+2.6}_{-3.9}$	&	&	2.0	&	41.7	 \\ \vspace{0.6 mm}
3C\,272.1\tablenotemark{e}&$0.79^{+0.44}_{-0.19}$	&$~~~0.0437^{+~\,0.0288}_{-~\,0.0327}$	& $~~~0.0794^{+~~~0.0097}_{-~~~0.0308}$		& $~~~0.0661^{+~\,0.1201}_{-~\,0.0510}$		& $44.0^{+~\,9.3}_{-~\,9.5}$ 	&$21.3^{+2.7}_{-4.0}$	&	&	2.0	&	41.3	 \\ \vspace{0.6 mm}
4C\,12.50	  	&$2.45^{+0.36}_{-0.06}$ &$~\,24.5~~~~\,^{+~\,0.6~~~~\,}_{-18.9~~~~\,}$	& $182~~~~~~\,^{+~~~9~~~~~~\,}_{-153~~~~~~\,}$	& $100.~~~~~~^{+~\,1~~~~~~\,}_{-49~~~~~~\,}$	& $59.7^{+~\,0.1}_{-~\,1.2}$	&$15.8^{+0.1}_{-0.6}$ 	&	&	3.0	&	44.5	\\ \vspace{0.6 mm}
3C\,293		&$0.65^{+0.29}_{-0.25}$	&$~~~0.871~\,^{+~\,0.453~\,}_{-~\,0.195~\,}$	& $~~~3.31~~~^{+~~~0.40~~~}_{-~~~0.49~~~}$		& $~~~2.00~~~^{+~\,0.58~~~}_{-~\,0.92~~~}$	& $58.4^{+~\,0.8}_{-~\,4.7}$	&$24.1^{+0.3}_{-1.6}$ 	&	&	1.6	&	43.1	\\ \vspace{0.6 mm}
MRK\,668		&$0.91^{+0.04}_{-0.02}$	&$~~~1.23~~~^{+~\,0.08~~~}_{-~\,0.01~~~}$	& $~\,17.4~~~~\,^{+~~~0.8~~~~\,}_{-~~~0.1~~~~\,}$	& $~~~6.46~~~^{+~\,0.18~~~}_{-~\,0.29~~~}$	& $54.9^{+~\,1.9}_{-~\,11.3}$	&$24.7^{+0.2}_{-0.1}$ 	&	&	2.0	&	44.6	\\ \vspace{0.6 mm}
3C\,305		&$0.93^{+0.27}_{-0.40}$	&$~~~0.295~\,^{+~\,1.157~\,}_{-~\,0.007~\,}$	& $~~~2.75~~~^{+~~~0.88~~~}_{-~~~0.57~~~}$		& $~~~2.24~~~^{+~\,0.33~~~}_{-~\,0.20~~~}$	& $45.8^{+~\,6.4}_{-~\,8.8}$	&$23.8^{+0.4}_{-1.0}$ 	&	&	2.5	&	43.1	\\ \vspace{0.6 mm}
3C\,310\tablenotemark{e}		&$2.24^{+0.05}_{-0.69}$	&$~~~0.0398^{+~\,0.0002}_{-~\,0.0207}$		& $~~~0.0741^{+~~~0.0003}_{-~~~0.0078}$			& $~~~0.0776^{+~\,0.0604}_{-~\,0.0239}$		& $42.7^{+11.0}_{-~\,9.0}$ 	&$22.7^{+1.6}_{-2.0}$	&	&	...	&  	... 	\\ \vspace{0.6 mm}
3C\,315		&$0.25^{+0.01}_{-0.06}$	&$~~~2.00~~~^{+~\,0.05~~~}_{-~\,0.65~~~}$	& $~~~1.82~~~^{+~~~0.04~~~}_{-~~~0.12~~~}$		& $~~~1.12~~~^{+~\,0.83~~~}_{-~\,0.45~~~}$	& $55.5^{+~\,3.2}_{-~\,6.5}$	&$20.4^{+2.2}_{-2.3}$ 	&	&	...	&  	... 	 \\ \vspace{0.6 mm}
3C\,317		&$3.39^{+0.08}_{-1.44}$	&$~~~0.513~\,^{+~\,0.002~\,}_{-~\,0.169~\,}$	& $~~~0.891~\,^{+~~~0.006~\,}_{-~~~0.155~\,}$		& $~~~0.309~\,^{+~\,0.001~\,}_{-~\,0.068~\,}$	& $58.7^{+~\,0.1}_{-~\,0.1}$	&$23.3^{+0.1}_{-0.1}$ 	&	&	...	&  	... 	\\ \vspace{0.6 mm}
3C\,326N		&$1.55^{+0.04}_{-0.04}$	&$~~~0.087~\,^{+~\,0.106~\,}_{-~\,0.046~\,}$	& $~~~0.454~\,^{+~~~0.249~\,}_{-~~~0.157~\,}$		& $~~~0.605~\,^{+~\,0.376~\,}_{-~\,0.219~\,}$	& $56.6^{+~\,2.4}_{-~\,6.2}$	&$20.8^{+1.4}_{-1.8}$ 	&	&	...	&  	... 	 \\ \vspace{0.6 mm}
PKS\,1549-79 	&$0.23^{+0.62}_{-0.13}$&$~\,38.0~~~~\,^{+23.6~~~~\,}_{-15.6~~~~\,}$& $112~~~~~~\,^{+~\,62~~~~~~\,}_{-~\,14~~~~~~\,}$& $~\,12.0~~~~\,^{+~\,3.7~~~~\,}_{-~\,2.3~~~~\,}$	& $53.5^{+~\,4.7}_{-~\,5.8}$	&$23.7^{+1.0}_{-2.4}$ 	&	&	2.0	&	45.2	 \\ \vspace{0.6 mm}
3C\,338\tablenotemark{f}&$2.00^{+0.05}_{-0.09}$&$~~~0.603~\,^{+~\,0.004~\,}_{-~\,0.024~\,}$& $~~~0.339~\,^{+~~~0.002~\,}_{-~~~0.014~\,}$	& $~~~0.871~\,^{+~\,0.006~\,}_{-~\,0.589~\,}$	& ...						& ...				 	&	&	...	&  	... 	 \\ \vspace{0.6 mm}
3C\,386		&$0.20^{+0.02}_{-0.04}$	&$~~~0.0794^{+~\,0.0865}_{-~\,0.0716}$		& $~~~0.132~\,^{+~~~0.009~\,}_{-~~~0.020~\,}$		& $~~~0.191~\,^{+~\,0.072~\,}_{-~\,0.043~\,}$	& $57.4^{+~\,1.9}_{-~\,3.0}$	&$19.5^{+0.9}_{-1.0}$ 	&	&	...	&  	... 	 \\ \vspace{0.6 mm}
3C\,424\tablenotemark{e}		&$0.26^{+0.20}_{-0.11}$	&$~~~0.0501^{+~\,0.0879}_{-~\,0.0363}$		& $~~~0.324~\,^{+~~~0.105~\,}_{-~~~0.054~\,}$		& $~~~3.55~~~^{+~\,0.72~~~}_{-~\,1.26~~~}$	& $48.2^{+~\,8.2}_{-11.0}$	&$15.4^{+0.7}_{-0.3}$ 	&	&	2.0	&	42.9	 \\ \vspace{0.6 mm}
IC\,5063		&$0.40^{+0.02}_{-0.08}$	&$~~~0.759~\,^{+~\,0.356~\,}_{-~\,0.004~\,}$	& $~~~3.16~~~^{+~~~1.30~~~}_{-~~~0.01~~~}$		& $~~~1.91~~~^{+~\,1.06~~~}_{-~\,0.01~~~}$	& $60.0^{+~\,0.1}_{-~\,1.0}$	&$19.4^{+0.1}_{-0.1}$ 	&	&	3.0	&	43.4	\\ \vspace{0.6 mm}
3C\,433		&$1.12^{+0.03}_{-0.38}$	&$~~~3.63~~~^{+~\,0.65~~~}_{-~\,0.15~~~}$	& $~~~9.55~~~^{+~~~0.00~~~}_{-~~~0.43~~~}$		& $~~~2.57~~~^{+~\,0.04~~~}_{-~\,0.10~~~}$	& $59.9^{+~\,0.1}_{-~\,0.1}$	&$24.7^{+0.1}_{-0.1}$ 	&	&	2.4	&	44.3	\\ \vspace{0.6 mm}
3C\,436		&$1.55^{+0.49}_{-0.14}$ 	&$~~~0.427~\,^{+~\,0.031~\,}_{-~\,0.322~\,}$	& $~~~4.57~~~^{+~~~0.22~~~}_{-~~~0.31~~~}$		& $~~~3.31~~~^{+~\,0.58~~~}_{-~\,0.29~~~}$	& $46.4^{+~\,7.8}_{-~\,9.1}$	&$24.4^{+0.4}_{-0.8}$ 	&	&	2.0	&	43.4	 \\ \vspace{0.6 mm}
3C\,459	  	&$0.36^{+0.01}_{-0.02}$&$195~~~~~~\,^{+~\,5~~~~~~\,}_{-48~~~~~~\,}$	& $182~~~~~~\,^{+~~~4~~~~~~\,}_{-~\,23~~~~~~\,}$& $~\,28.8~~~~\,^{+~\,5.2~~~~\,}_{-~\,0.5~~~~\,}$	& $59.8^{+~\,0.1}_{-~\,0.1}$	&$24.6^{+0.1}_{-0.1}$ 	&	&	3.0	&	44.1	 
\enddata	
\tablenotetext{a}{Uncertainties take into account both the uncertainty in the best fit as well as the variation during the iterative fitting.}
\tablenotetext{b}{Warm component restricted to 30--60\,K and assumes $\beta=1.5$.}
\tablenotetext{c}{Cold component restricted to 15--25\,K and assumes $\beta=2$.}
\tablenotetext{d}{Power-law index of the AGN model restricted to be between 1 (approximating a face-on torus) and 3 (effectively an edge-on torus).}
\tablenotetext{e}{These SEDs are poorly sampled in the IR and are not very well fit in the FIR, so the parameters should be used with caution. }
\tablenotetext{f}{3C\,338 only has upper limits at $\lambda>$60\um, so the derived dust mass and luminosity and SFR should be considered upper limits. Similarly, we do not have concrete information on its dust temperatures. }
\end{deluxetable*}

\begin{figure*}[t]
\centerline{\includegraphics[trim=0.7cm 0.1cm 0.cm 0.cm, clip, width=\linewidth]{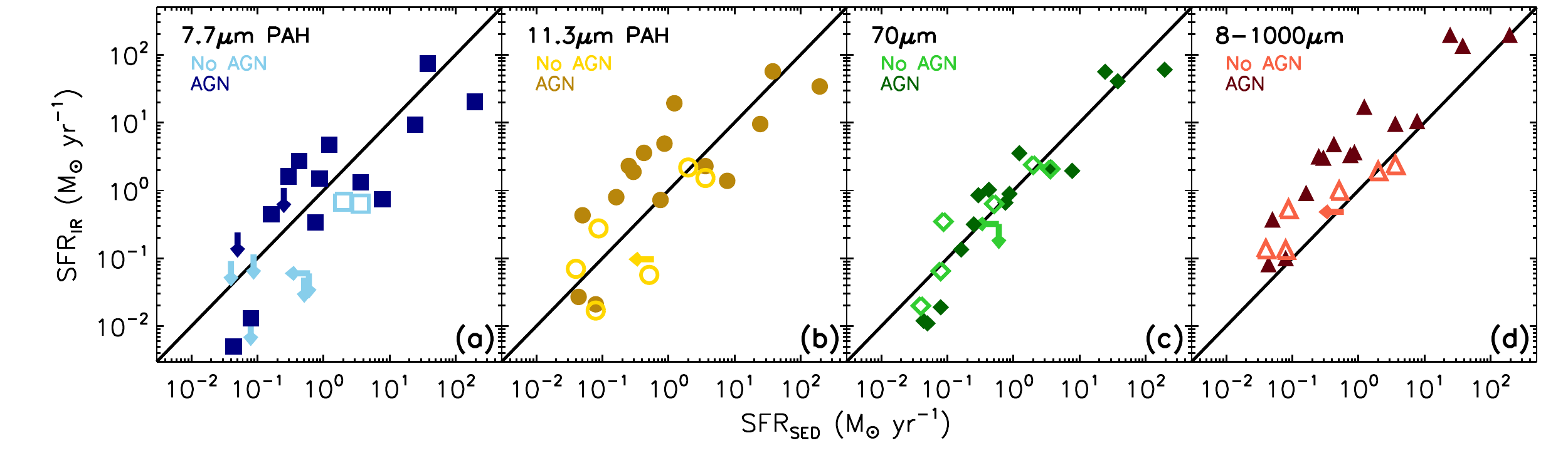}} 
\caption{Comparison of the SED-derived SFR with SFR calculated from several IR relations, with solid, darker symbols indicating that the SED fit included an AGN component and unfilled, lighter-colored symbols indicating an AGN component did not improve the SED fit.  SFRs calculated based on the luminosity of the 7.7\um PAH {\textbf{(a)}} and 11.3\um PAH {\textbf{(b)}}  show some dispersion but correlate well with with the SED-derived SFR. {\textbf{(c)}} The SFR calculated from the 70\um relation of \citet{calzetti10} agrees well with the SED-derived SFR, as the SED at this wavelength is generally dominated by the host galaxy. In contrast, the SFR calculated from the 8--1000\um luminosity \citep[{\textbf{(d)}}][]{kennicutt98} is often too large, particularly when the SED is better fit with a MIR AGN component.}
\label{sfr_comp}
\end{figure*}

\subsubsection{Literature Photometry and Properties}

Table \ref{phot} also gives the photometry we gathered from the literature. We only use {\em Infrared Astronomical Satellite} ({\em IRAS}) photometry when PACS data are lacking, particularly at 70\um. For those galaxies without optical images from SDSS, we first searched for {\em UBV} photometry in the Third Reference Catalog \citep[RC3; ][]{devaucouleurs91}. Thereafter, we used the NASA Extragalactic Database (NED)\footnote{\url{https://ned.ipac.caltech.edu/}} to find {\em UBVR} photometry available in the literature. Since these photometry were not measured in the same aperture we used, we only use these data if they agree with the shape of the SED traced by measured UV and NIR photometry and with larger uncertainties than reported to reduce the weight of these points in our fits. Only 3C\,236's lack of UV photometry from \galex can be remedied by existing literature photometry, in this case from the {\em Hubble Space Telescope} \citep{tremblay10}. For galaxies without IRAC or MIPS observations, we obtained photometry at 8 and 24\um estimated from the IRS spectra as part of the IRS enhanced products in the Spitzer Heritage Archive. These galaxies are typically at sufficiently high redshift that the whole galaxy is contained within the slit.

We also collected CO-derived molecular masses and extents from the literature. We corrected these masses for discrepancies in the distances assumed in these papers and here. Much uncertainty still remains on the precise conversion, and its dependence on galactic parameters such as metallicity \citep[e.g.,][]{narayanan12, bolatto13} or radiation intensity \citep[e.g. in (ultra-)luminous IR galaxies; ][and references therein]{bolatto13}, between CO luminosity (or integrated line intensity, $I_{\rm CO}$) and the associated mass (or column) of molecular hydrogen, which is typically accumulated in the $\alpha_{\rm CO}$ (or $X_{\rm CO}$) parameter. The literature-derived CO masses were calculated with a variety of  $\alpha_{\rm CO}$ or $X_{\rm CO}$, so we also adjusted these masses to assume a common  $\alpha_{\rm CO}=4.3$\,M$_{\odot}\,{\rm(K\,km\,s^{-1}\,pc^{2})^{-1}}$ equivalent to $X_{\rm CO}=2\times10^{20}$\,cm$^{-2}\,{\rm(K\,km\,s^{-1})^{-1}}$ \citep{bolatto13}. The resulting masses are given in Table \ref{litprop}. We also give warm molecular masses and luminosities, as well as the ratio in luminosity between \mh and PAHs.

Table \ref{litprop} also summarizes the X-ray and radio properties of these galaxies. \citet{lanz15} measured the diffuse X-ray emission, excluding the AGN, for the 20 galaxies from this sample that had \chandra observations. Since that paper, an observation of Mrk\,668 (Obs. ID 16071) has become public and was analyzed in an identical manner. PKS\,1549-79 has only been observed with \xmm, which has much poorer spatial resolution. For this galaxy, we fit the spectrum with a thermal component (effectively our diffuse emission) and an absorbed power-law (effectively the AGN). Appendix B provides additional details of this reduction. In the last columns of Table \ref{litprop}, we also give the jet power calculated with the formula of \citet{punsly05} from the 178\,MHz flux density.

\subsubsection{Extent of the Star-forming Region}

The most difficult aspect of localizing galaxies on the Kennicutt-Schmidt \citep[K--S;][]{kennicutt98} plot of the surface density of star formation versus the surface density of molecular gas is measuring the surface area. Although 17 of our 22 galaxies have CO line intensities or limits, only six of these were observed with instruments capable of spatially resolving the molecular emission (typically an interferometer). For these galaxies, we assume, as is commonly done, that the extent of the star-forming region is the same as that of the molecular disk. For an additional four galaxies, we can estimate the extent based on the size of the PAH (IRAC 8\um) or UV emission as a proxy. For the rest, no information on the extent of the star-forming disk exists within current observations. Since these galaxies are typically early-type galaxies (ETGs), we use the typical radius of 1\,kpc measured in ATLAS$^{3D}$ galaxies \citep{davis14}. Further, \citet{davis13a} found that ATLAS$^{3D}$ typically had a CO radius $\sim20$\% of the stellar extent radius, which for our galaxies correspond to 1--2\,kpc. The position of these galaxies on the K--S diagram should be taken as preliminary, pending resolved molecular observations. The sizes we use, and from where they were determined,  are given in Table \ref{litprop}. \\\\

\vspace{-7mm}
\section{SED FITTING}

\subsection{Method}
To estimate SFR, stellar and dust masses, and dust temperatures, we used the SED fitting code \magphys \citep{dacunha08}. \magphys fits SEDs with a combination of UV$-$NIR stellar spectral libraries from \citet{bruzual03} and a simple, physically motivated model for IR emission from dust developed in \citet{dacunha08}. It models the ISM as a mix of diffuse dust interspersed with denser, warmer stellar birth clouds. The IR dust libraries have five components: a fixed PAH spectrum shape derived from the M17 SW star-forming region \citep{madden06}, a NIR continuum associated with the PAH emission modeled by a modified blackbody ($\beta=1$) at 850 K, a hot MIR continuum modeled by the sum of two modified blackbodies ($\beta=1$) at 130 K and 250 K, a warm (30--60 K) dust component modeled as a modified blackbody ($\beta=1.5$), and a cold (15--25 K) dust component modeled as a modified blackbody ($\beta=2$). The warm dust component is assumed to exist both in the diffuse ISM and in denser birth clouds, while the cold dust exists only in the diffuse ISM. The model shape of the different dust components (i.e. $\beta$) is determined based on the likely size of the grains emitting primarily at those temperatures (see \citealt{dacunha08} for further details).

In order to determine the physical parameters associated with each SED, {\sc magphys} combines UV--NIR and IR spectral libraries, each with 50,000 models calculated for a range across each parameter, such that the energy absorbed in the UV/visible regime is re-emitted in the IR. These models are convolved with the response functions of each filter for which the user has provided photometry.  {\sc magphys} does not so much fit for its derived parameters as determine how well the photometry of each of its models (and their associated parameters) match observations. As a result, in addition to determining which model matches best, {\sc magphys} creates probability distribution functions (PDFs) for each of its parameters, indicating the likelihood of its value. The parameters we use in this work are the median of these distributions with the range of parameter values with 16\%--84\% likelihood providing the bulk of the uncertainty. 

\magphys does not currently include an AGN component, and many of our galaxies have a significant MIR contribution likely due to an AGN. \citet{sajina12} described an empirical SED model for UV--FIR SEDs which includes a component associated with AGN tori, modeled as a broken, tapered, power-law:
\begin{equation}
F_{AGN} = \frac{\nu}{ (\frac{\nu}{\nu_{0}})^{\alpha}~e^{0.5\nu} + (\frac{\nu}{\nu_{0}})^{-0.5} + (\frac{\nu}{0.3 \nu_{0}})^{-3.0}}
\end{equation}
The effect of dust sublimation is captured by the exponential tapering, and the $\nu^{-3}$ component acts as the Rayleigh-Jeans tail of a dust component (with $\beta=1$), softened by the flatter component ($\nu^{-0.5}$).

Including another component such as this one into \magphys is an endeavor beyond the scope of this paper. Instead, we used an iterative method to fit both the Sajina AGN component and a host galaxy (via \magphys) in order to better model our SEDs. First, we fit the SED purely with \magphys.Then we subtracted the photometry associated with the best MAGPHYS fit and fit the MIR residuals (5--24\um) with the Sajina AGN. We fix $\nu_{0}$ in order to have a peak in the MIR, within the range used by \citet{sajina12}. We tried both fits with $\alpha=2$ and with a free $\alpha=[1,3]$.\footnote{$\alpha = 1$ corresponds approximately to a face-on torus, whereas $\alpha=3$ better models edge-on tori.} The photometry associated with the better AGN fit is subtracted from the observed fluxes, and the resulting photometry is fit again with \magphys.\footnote{In some cases, the model photometry in a band is larger than the observed photometry. In these cases, we treat the observed flux as an upper limit in the subsequent fit, which adds substantially to the $\chi^2$ value if the model of that fit is larger than this data point.}  We iterate several times between the \magphys fits and the MIR AGN fits until the \magphys fit no longer improves. If adding the AGN component has not significantly improved the fit, then we use the original (pre-AGN) fit; otherwise we include the AGN component. Fifteen of the 22 SED fits improve significantly with the inclusion of an AGN component. We add the parameter variation seen over the iterations to the uncertainties from the {\sc magphys} 16\%-84\% parameter likelihood range, as a means of estimating the impact of different AGN models on the parameters. Table \ref{sedpar} summarizes the derived parameters. 

We tested whether beginning with the AGN component, rather than the host component, affects the fit. When the AGN component is weak (e.g., 3C\,31) or moderate (e.g, 3C\,236), there is very little difference. For strong AGNs (e.g., IC\,5063), fit iterations starting with the AGN component converge to a model with a slightly stronger AGN contribution and an SFR lower by a factor of $\sim2$, within the reliability of our SFR (see \S 3.2.1). Additionally, if the SFR are indeed lower than those derived with our methodology in the 3--8 of our galaxies with the strongest MIR AGN, then this will only strengthen the results we discuss in \S4.2.

We tested the reliability of the AGN contribution by fitting a subset of our sample with another IR-only SED fitting program, DecompIR \citep{mullaney11}, which pairs one of five host galaxy templates with a piece-wise AGN model consisting of two power laws and a modified blackbody. Our DecompIR fits of galaxies requiring significant AGN contributions with our fitting method (e.g., 3C\,236 and IC\,5063) yield AGN fractional contributions to the 8--35\um of $\sim75$\%, similar to the MIR AGN fractions in our fits. Further, DecompIR fits of galaxies where our method does not require an AGN (e.g., 3C\,218 and 3C\,326N) yield MIR AGN fractions $<20$\%.


\begin{figure}
\centerline{\includegraphics[width=\linewidth]{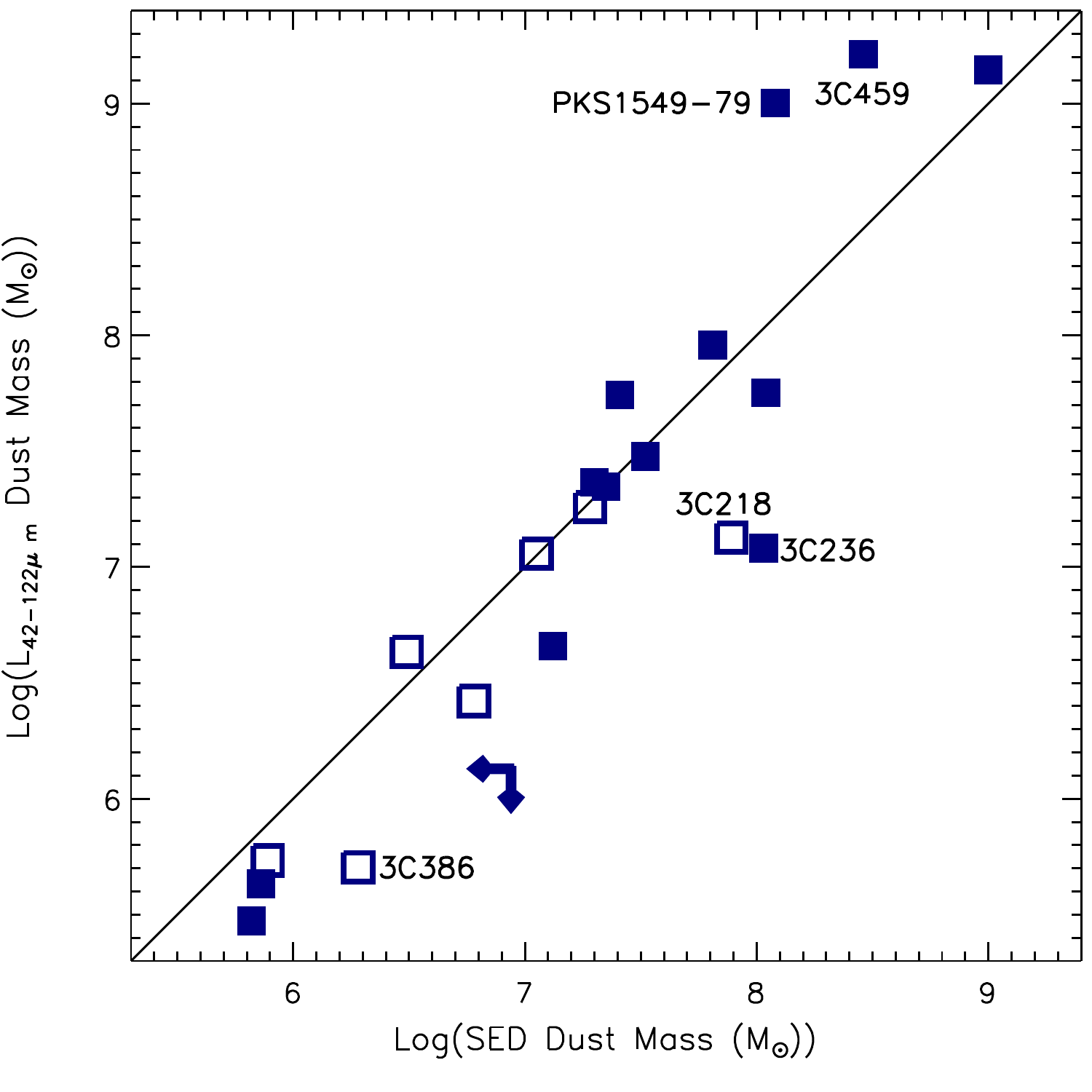}} 
\caption{Comparison of the SED-derived dust mass with a estimate of dust mass based on the 42--122\um luminosity. Solid symbols include an AGN in the fit. The five labeled galaxies have differences greater than a factor of 3. 3C\,459 and PKS\,1549-79 are both ULIRGs and peak at shorter wavelengths, corresponding to warmer temperatures. The difference in derived mass for 3C\,218, 3C\,236, and 3C\,386 is more likely due to SEDs dominated by a colder component than the 25\,K assumed in the IR luminosity-based estimate of dust mass (see \S3.2.2 for further details).}
\label{dm_comp}
\end{figure}

\subsection{Parameter Comparison}
In our examination of galaxy properties, we will use the parameters derived from the SED fitting. Since the SFR, dust mass, and stellar mass will be key properties, we first examine how these SED-derived values compare to those from simpler methods. This comparison will also provide some guidance in future studies with more limited observational data sets. 

\subsubsection{SFR}

Figure \ref{sfr_comp} shows the comparison of the SFR obtained from the \magphys fit\footnote{\textbf{{\sc magphys} calculates SFR from the average SFR of the last 100\,Myr of the star formation history.}} with four other measures of SFR. We examined how the \magphys SFR  compares to SFR calculated from the PAH fluxes from \citet{ogle10} and \citet{guillard12}.\footnote{We use the same formulae as \citet{ogle10} but with our assumed distances: SFR(PAH$_{7.7\mu m})=2.4\times10^{-9}\,L({\rm PAH}_{7.7\mu m})/L_{\odot}$ and SFR(PAH$_{11.3\mu m})=9.2\times10^{-9}\,L({\rm PAH}_{11.3\mu m})/L_{\odot}$.} We find that these SFRs correlate well with the SED-derived values but with dispersions of $\sim0.7$ dex (Fig. \ref{sfr_comp}ab). Some of this dispersion may also be due to the fact that the IRS slit did not fully cover the host for all of our galaxies.\footnote{This issue was previously noted for a subset of these radio galaxies in the appendix of \citet{alatalo15SF}.}

One of the most common estimates of SFR is based on the IR luminosity, such as the 8--1000\um relation of \citet{kennicutt98} (Fig. \ref{sfr_comp}d). However, many of these galaxies have sizable MIR contributions likely due to warm dust heated by an AGN. Therefore, it is not surprising that the total IR SFR is biased high compared to the SED-derived value for our galaxies. In contrast, if we only examine FIR emission where the host galaxy dominates (e.g. 70\um; Fig. \ref{sfr_comp}c), the SFRs correlate much better with a dispersion of 0.4 dex.

\begin{figure}
\centerline{\includegraphics[width=\linewidth]{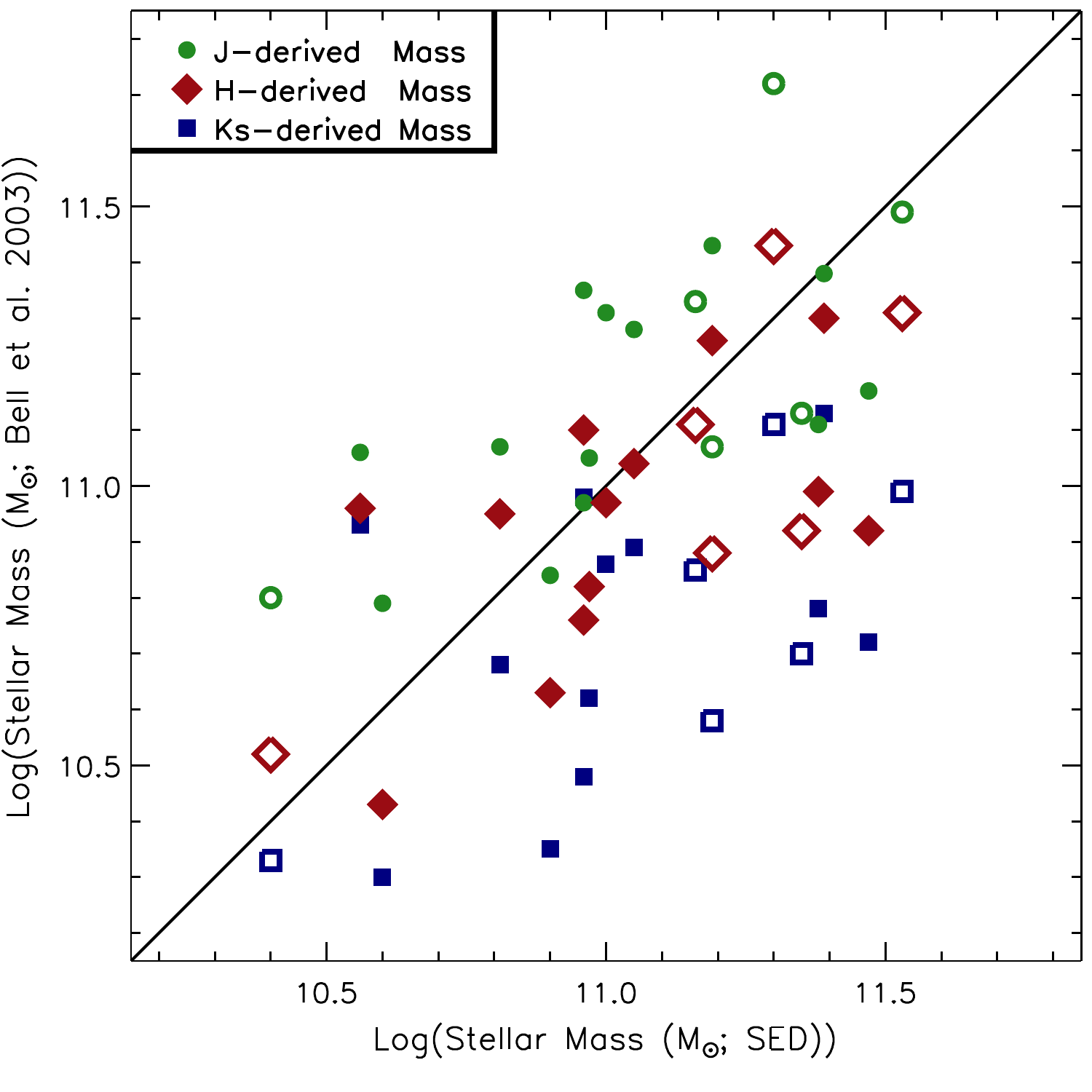}} 
\caption{Comparison of the SED-derived stellar mass with the mass calculated from a color-dependent mass-to-light relation for the three 2MASS bands. Solid symbols include an AGN in the fit. The SED-derived masses agree best with the {\em H}-band derived masses within a factor of 2.}
\label{sm_comp}
\end{figure}

\subsubsection{Dust Mass}

Figure \ref{dm_comp} compares the \magphys-derived dust mass with an estimate of the dust mass based on the 42-122\um luminosity. We assume a simple model of a single modified blackbody with a dust emissivity power-law index of $\beta=1.8$ and a typical temperature of 25\,K (as suggested for example by \citealt{scoville14}). We find that there is good agreement (dispersion of 0.46 dex), despite the simple assumptions of the second estimate.

The galaxies that deviate the most provide interesting insights on the estimation of dust mass. In Figure \ref{dm_comp}, we identified the five galaxies whose estimates of dust mass differ by more than a factor of three. The three with lower dust masses in the simple model (3C\,218, 3C\,236, and 3C\,386) all have cold dust temperatures from \magphys below 20\,K. Since dust mass varies with dust temperature as $M_D\propto T^{-(4+\beta)}$, a decrease in temperature from 25 to 20\,K increases the derived dust mass by a factor of 3.6 (assuming $\beta=1.8$). PKS\,1549-79 and 3C\,459 are two of our three ultra-luminous IR galaxies (ULIRGs), which tend to have typically hotter dust \citep[e.g., ][]{clements10} and therefore peak at shorter wavelengths than less luminous galaxies. As a result, the dust mass derived from their 42-122\um luminosity is overestimated. The third ULIRG, 4C\,12.50, is well fit by a 15.8\,K cold dust temperature, so the two effects cancel out. Similarly, 3C\,84 has a temperature just under 20\,K, but it has a relative high luminosity as a luminous IR galaxy (LIRG). We conclude that the SED-derived values provide a good estimate of the dust masses due to the ability of this method to fit multiple thermal components. \\

\begin{figure}
\centerline{\includegraphics[width=\linewidth]{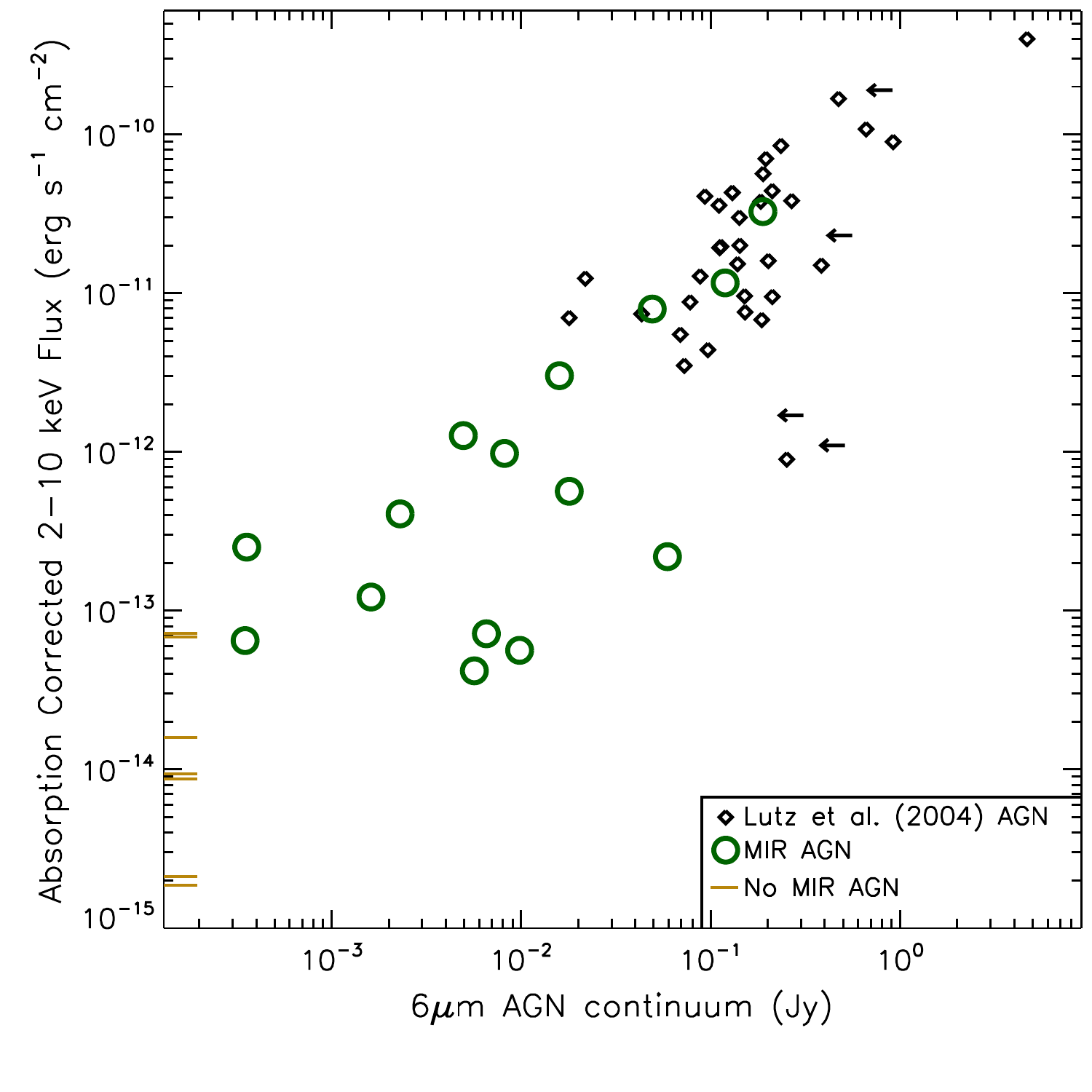}} 
\caption{Comparison of the SED-derived 6\um AGN continuum flux with the 2--10\,keV AGN flux, showing that our galaxies fall along the correlation seen for more luminous AGN by \citet{lutz04}. }
\label{agn_comp}
\end{figure}

\subsubsection{Stellar Mass}

We also compare the \magphys-derived stellar mass determined based on the star formation history associated with the best fit model with masses calculated from a color-dependent mass-to-light relation \citep[Fig. 3;][]{bell03}. We use {\em g--r} colors if we have SDSS observations, or {\em B--V} colors if available, along with the luminosity in the three 2MASS bands.\footnote{The \citet{bell03} relations assume a different initial mass function than {\magphys}, but the effect on the mass calculated is small ($<1\% $ difference).} We find good agreement with {\em H}-band derived masses, which are intermediate between the {\em J}-band and {\em Ks}-band masses, with a dispersion of 0.26 dex. The dispersion across the set of 2MASS-derived masses is 0.32 dex. We do not see significant differences between the best fits of those galaxies better modeled with and without a MIR AGN.

\begin{figure}
\centerline{\includegraphics[trim=0.1cm 0.1cm 0.1cm 0.1cm, clip, width=\linewidth]{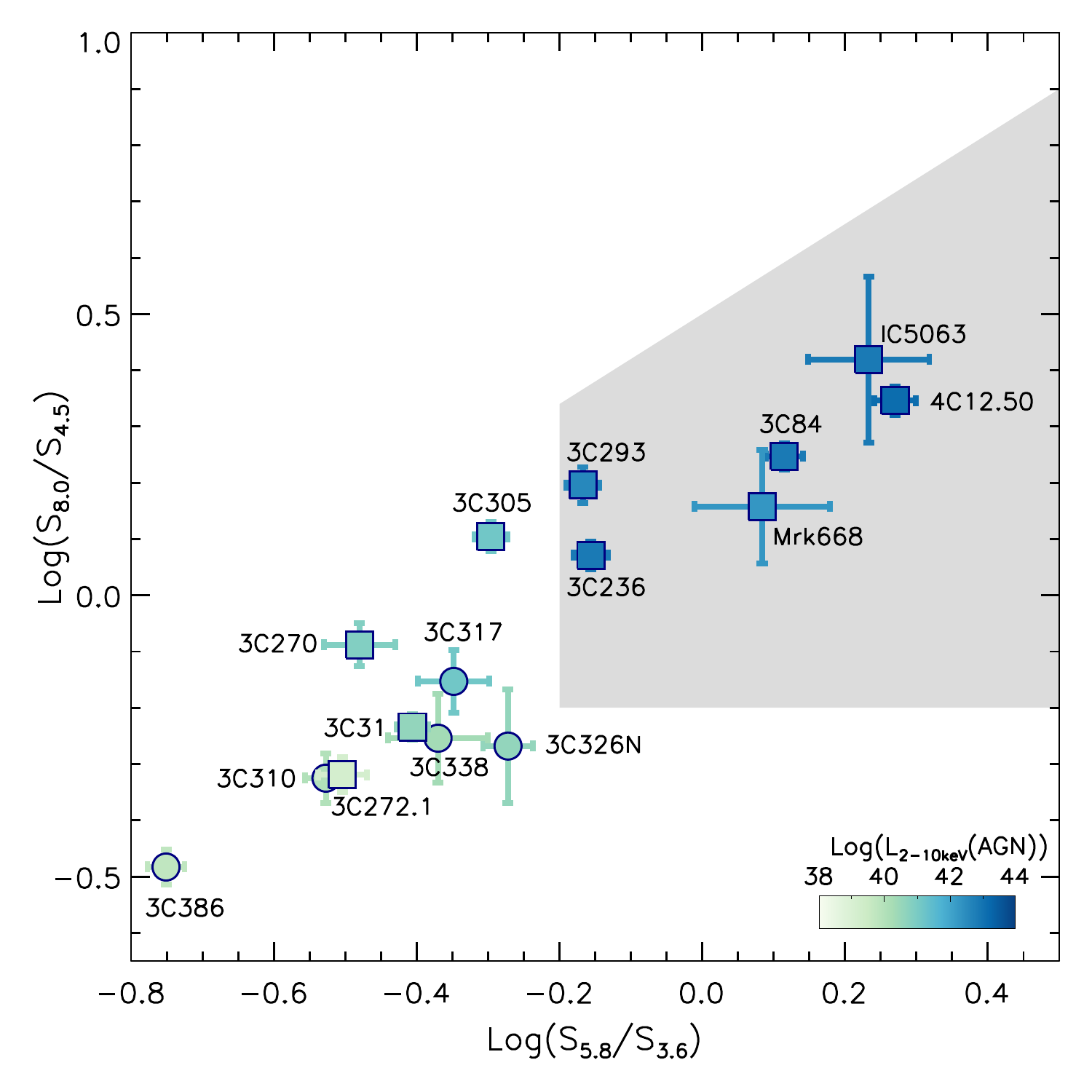}} 
\caption{The $S_{5.8\umb}/S_{3.6\umb}$ versus $S_{8.0\umb}/S_{4.5\umb}$ IRAC colors with the AGN-dominated region defined by \citet{lacy04} shown in light gray. AGNs with X-ray (2--10\,keV) luminosities (which defines the color of the points) all fall within the ``Lacy'' wedge. However, not all  MIR AGNs (squares) do.  }
\label{irac}
\end{figure}

\subsubsection{AGN}

Approximately two-thirds (15/22) of our sample was better fit with the inclusion of a MIR AGN component.  \citet{lutz04} found a correlation between 6\um and 2--10\,keV fluxes and luminosities \citep[see also][]{goulding11}. In Figure \ref{agn_comp}, we find that our galaxies broadly fall along this correlation, despite our AGN being typically weaker than those looked at by \citet{lutz04}. Those galaxies whose SEDs do not require a MIR AGN have the lowest hard X-ray flux. This test bolsters the dependability of our SED decomposition. 

We also examine the IRAC colors of our galaxies in Figure \ref{irac}. Only fifteen galaxies were observed with all four bands of IRAC, but we find that those galaxies with the most X-ray luminous AGN all fall within the region identified by \citet{lacy04} as AGN-dominated. Four galaxies whose SEDs are best fit with a MIR AGN fall outside this wedge, but these are amongst the weakest in our sample in both their X-ray and IR emission associated with an AGN.

\begin{figure}
\centerline{\includegraphics[trim=0.4cm 0.2cm 0.1cm 0.1cm,clip, width=\linewidth]{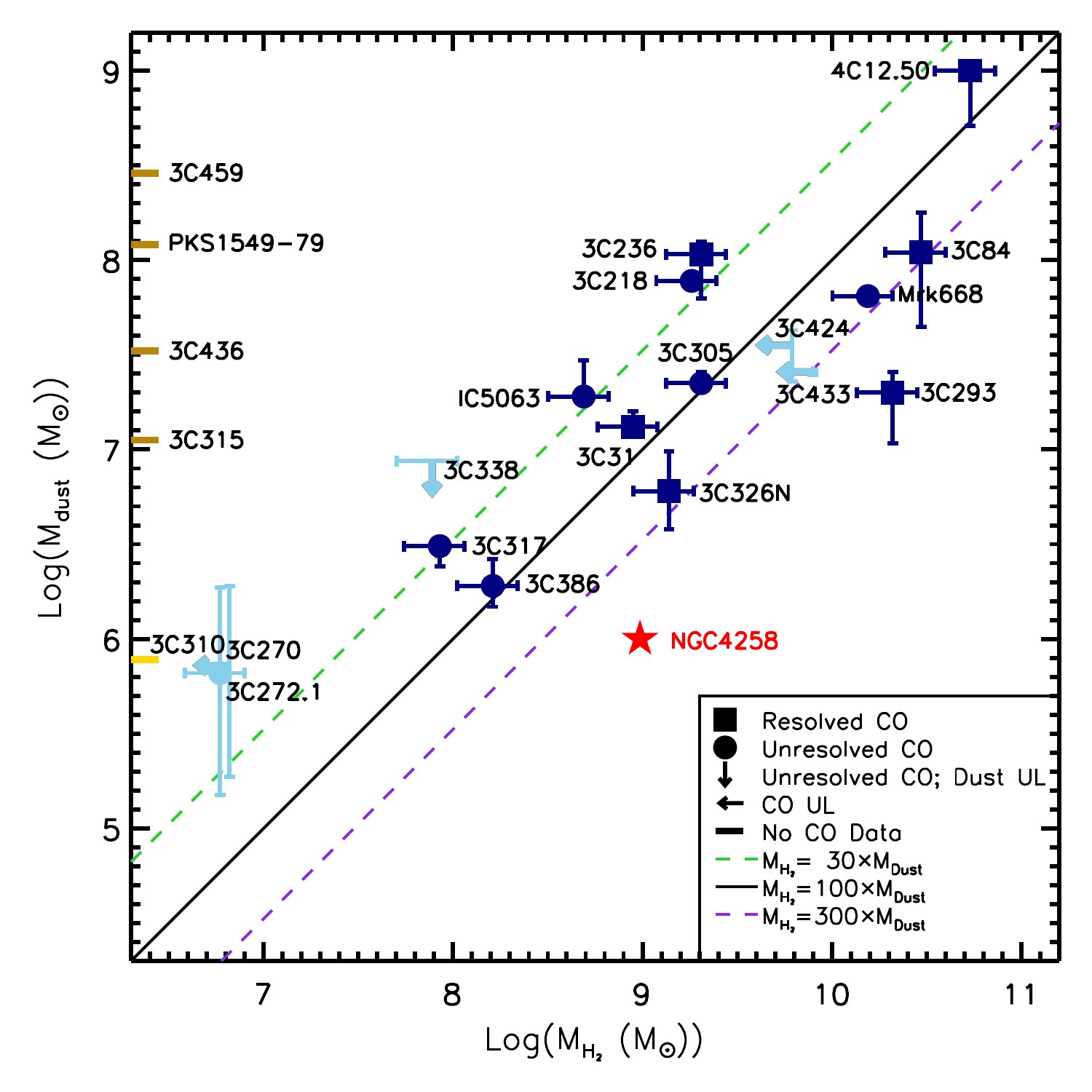}} 
\caption{Molecular gas mass (calculated with a common X$_{\rm CO}=2\times10^{20}$\,cm$^{-2}\,{\rm(K\,km\,s^{-1})^{-1}}$) compared to the dust mass derived from the SED fits. Most of our galaxies have normal GDRs, although 3C\,293 has a particularly high ratio (\citealt{lanz15}; \citealt{papadopoulos08} has suggested its CO emission is enhanced by shocks). Darker symbols have more reliable dust masses. 3C\,270 and 3C\,272.1, in particular, should be used with caution. For comparison, we show the MOHEG NGC\,4258 (red star) discussed by \citet{ogle14}.}
\label{gdr}
\end{figure}

\subsection{Caveats}
We discuss individual peculiarities of each galaxy in Appendix A, but there several common caveats, which we discuss below.

\subsubsection{MIR Spectral Variations}

\magphys includes a PAH component via the use of a template based on the spectrum of the star-forming region M17 SW \citep{madden06, dacunha08}. Therefore, it assumes a particular ratio between the different PAH lines, which is kept fixed. For example, this template has a 7.7\um/11.3\um ratio of $\sim3.9$.\footnote{Based on a {\sc pahfit} \citep{smith07pah} fit of the template.} In our sample, that ratio ranges from 0.73--5.0 with a median value of 2.3 \citep{ogle10}. This difference in ratios may explain fit discrepancies in IRAC 8\um and \wise 12\um bands (e.g., the fits of 3C\,218 and 3C\,386 show an excess of modeled PAH emission in the long-wavelength IRAC bands.) However, the PAH component contributes very little to the dust mass (\citealt{dacunha08} estimates it contributes at most a few percent), and the SFR is primarily influenced by the UV and FIR emission. Therefore, the uncertainty introduced by this fixed PAH ratio has little impact on the final estimated SFRs and dust masses. Similarly, the spectra of our galaxies \citep{ogle10} show a range of silicate emission and absorption at 10\um, which is not taken into account in our fitting. However, like the PAH features, this spectral component will not greatly affect the SFRs and dust masses we derive.

\begin{figure}
\centerline{\includegraphics[trim=1.2cm 1.7cm 1cm 1.cm,clip, width=\linewidth]{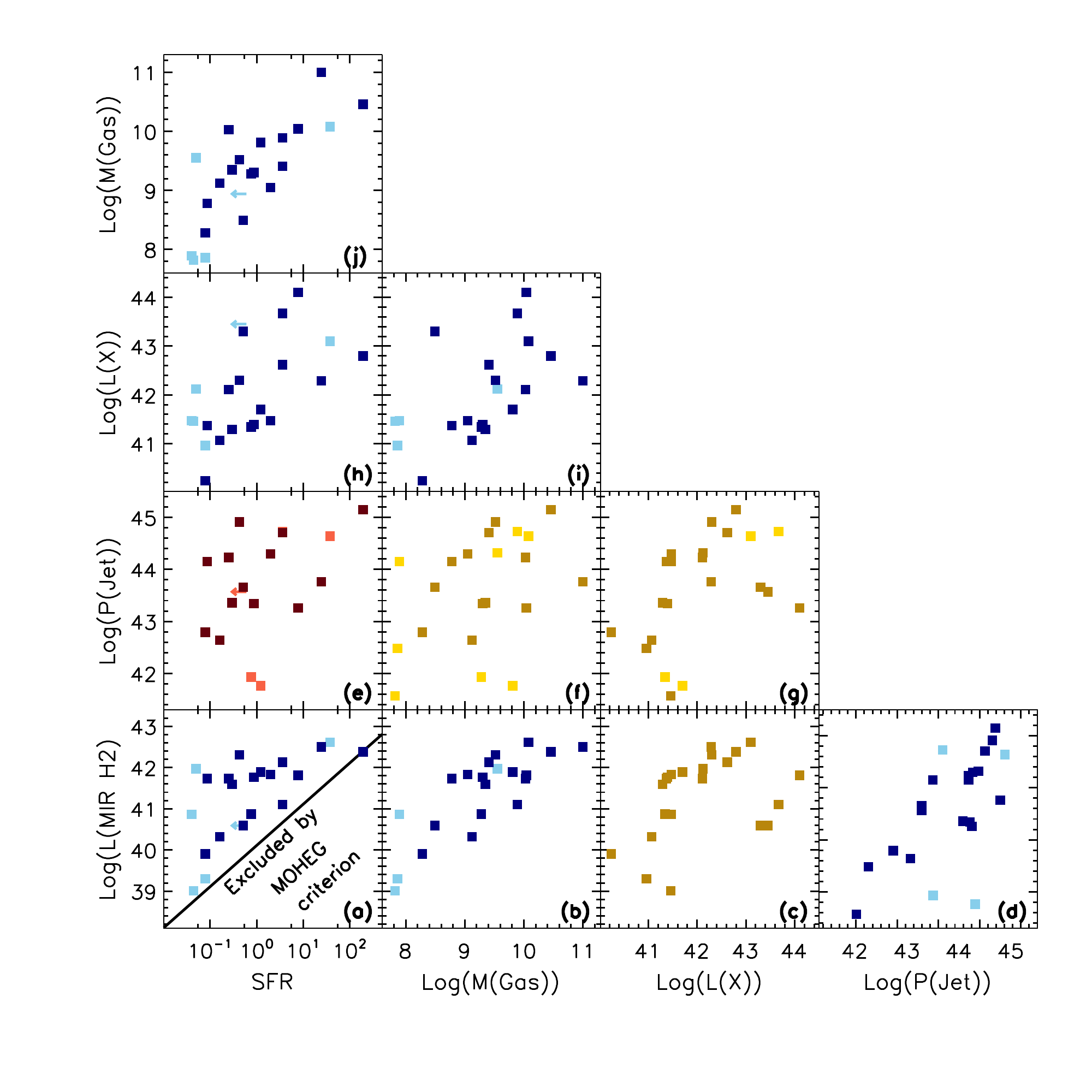}} 
\caption{Comparison of galaxy and feedback properties: \mh luminosity from IRS (${\rm erg\,s^{-1}}$) in the S(0)--S(3) lines; SFR from {\sc magphys} (${M_{\odot}\,yr^{-1}}$); cold molecular mass calculated from the dust mass ($M_{\odot}$); diffuse X-ray luminosity (${\rm erg\,s^{-1}}$) in the 0.5--8\,keV band; and jet power (${\rm erg\,s^{-1}}$). Blue symbols have significant correlations ($p<0.01$) and yellow symbols indicate suggestive correlations ($p<0.05$), calculated based on darker symbols that have more secure parameter values.}
\label{parcomp}
\end{figure}

\subsubsection{FIR Excess}

Several of our galaxies show excess emission over the Rayleigh-Jeans tail of the {\sc magphys} dust continuum, likely due to one of two causes. First, since our sample consists of radio galaxies, we would not be surprised to detect synchrotron emission, possibly even at wavelengths observed with \herschel. However, analysis of one of these SEDs (3C\,84) by \citet{leipski09} concluded that there was little synchrotron contribution to the FIR emission. Second, these galaxies could contain additional dust cooler than the 15\,K limit of {\sc magphys}. {\sc magphys} does not currently contain either a synchrotron component or another cooler dust component, and adding a new component would require generating many new models. The good agreement between our SED-derived SFRs and the 70\um SFR (where this excess is not present), as well as the good match of our fits to the FIR peaks, suggests that the impact of this unmodeled component is unlikely to have a significant impact on our SFR. There may be a more important impact on the dust mass, as a cold component could contribute significant additional mass. However, an increase in dust mass would only strengthen the results we discuss in \S4.2.

\subsubsection{Limited SED Coverage}

A minority of our galaxies have limited FIR coverage. Three were not observed with \herschel. In the case of 3C\,293, the combination of MIPS and {\em IRAS} photometry yields a good fit, but for 3C\,270 only two of these bands are detected. Its SED, as well as that of 3C\,272.1, are not well fit in the FIR, so their parameters are used with caution. 3C\,338 and 3C\,310 were both observed with \herschel, but were either not detected or their FIR emission could not be disentangled from those of close companions that appear to dominate at these wavelengths. Since 3C\,310 has one FIR data point, we use its derived parameters with caution, but given the complete lack of photometric detections of 3C\,338 at $\lambda>$30\um, we consider its derived SFR, dust mass, and dust luminosities to be upper limits.                               

In the UV regime, only one galaxy, 3C\,315, has no data. However, five others only have upper limits (although these are used to restrict the fits). PKS\,1549-79 has a well defined SED in the IR, but only has one (literature-derived) UV--optical data point. 3C\,424 likewise has a poorly sampled SED overall, so its parameters should be treated with caution and have large uncertainties associated with them.

The parameters associated with these fits are more uncertain than those from better sampled SEDs. Therefore, in the figures that follow, we indicate them with lighter symbols to guide the reader in determining which data points are more reliable.

\section{DISCUSSION}

\subsection{ISM Properties}

Given the uncertainty in the conversion factor from CO luminosity to molecular gas mass, we assume a common X$_{\rm CO}$ \citep[$2\times10^{20}$\,cm$^{-2}\,{\rm(K\,km\,s^{-1})^{-1}}$;][]{bolatto13} to calculate the gas mass in all the galaxies for which literature CO observations exist (Table \ref{litprop}). Figure \ref{gdr} compares this gas mass to the dust mass we derive from the SED fits. Most of our galaxies have gas-to-dust ratios (GDRs) within a factor of a few of the typical ratio found in the Milky Way ($\sim100$). The most extreme ratio is found in 3C\,293  \citep[$\sim10^{3}$;][]{lanz15}. Higher excitation CO lines have been measured in that galaxy \citep{papadopoulos08}, whose spectral line energy distribution (SLED) is consistent with shock-excitation. As a result, the assumed $X_{\rm CO}$ could be too high, as this galaxy could emit more CO per amount of molecular mass than a galaxy whose CO is not shocked. \citet{papadopoulos10} further argued that the effect of shock-induced turbulent heating would have a much larger effect on the gas phase than on the dust phase.  Similarly a high GDR was seen in the inner regions of NGC\,4258, shown for comparison in Figure \ref{gdr}, where warm \mh was mapped along the axis of the radio jet \citep{ogle14}.

Based on Figure \ref{gdr}, however, such an apparently high GDR does not appear to be a common property of all radio MOHEGs. 3C\,293 is not clearly peculiar in the parameters we have examined. It lies in the middle of the range of both galaxy parameters (e.g. SFR, stellar mass, and dust mass) and feedback-associated parameters (e.g. jet power, \mh luminosity, and diffuse X-ray luminosity), and like many of our sample, it has a companion. It is therefore difficult to identify the cause of the particularly large shock-excitation in 3C\,293, although the relative geometry of the jet and molecular gas distribution may play a role. Further exploration of the CO SLED of the full sample would also provide greater insight into whether any others also show indications of shock-excitation in the higher CO lines. 

\begin{figure*}
\centerline{\includegraphics[trim=2.5cm 0.4cm 0.1cm 0.1cm,clip, width=\linewidth]{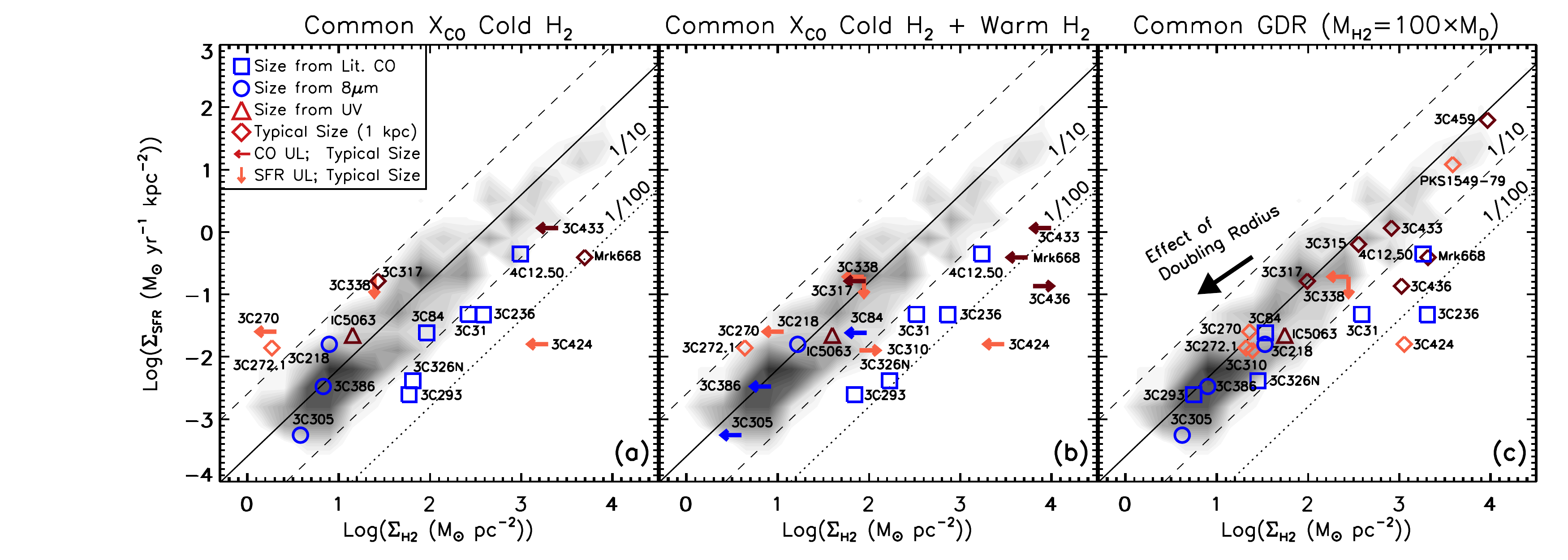}} 
\caption{Surface density of star formation compared to surface density of molecular gas (K--S diagrams; \citealt{kennicutt98}; the solid line is the K-S relation)  calculated \textbf{(a)} from CO luminosity assuming a common $X_{\rm CO}$ ($2\times10^{20}$\,cm$^{-2}\,{\rm(K\,km\,s^{-1})^{-1}}$), \textbf{(b)} for the total (cold as in \textbf{a} + warm) molecular gas, or \textbf{(c)} a common GDR. The lower limits on total gas surface density are for those galaxies that have not been observed in CO. For comparison, the underlying grayscale contours show the typical extent of normal galaxies (spirals from \citealt{kennicutt98} and \citealt{fisher13}, CO-detected ETGs from \citealt{davis14}, and the \citealt{shi11} galaxies). We find that MOHEGs tend to fall on the suppressed side of the typical relation, but only three (3C\,31, 3C\,236, and Mrk\,668) have suppressions greater than a factor of ten in all three K-S plots (3C\,436 unfortunately does not have CO data).  Darker symbols have more reliable SFRs and dust masses.}
\label{KS}
\end{figure*}

\begin{figure*}
\centerline{\includegraphics[width=\linewidth]{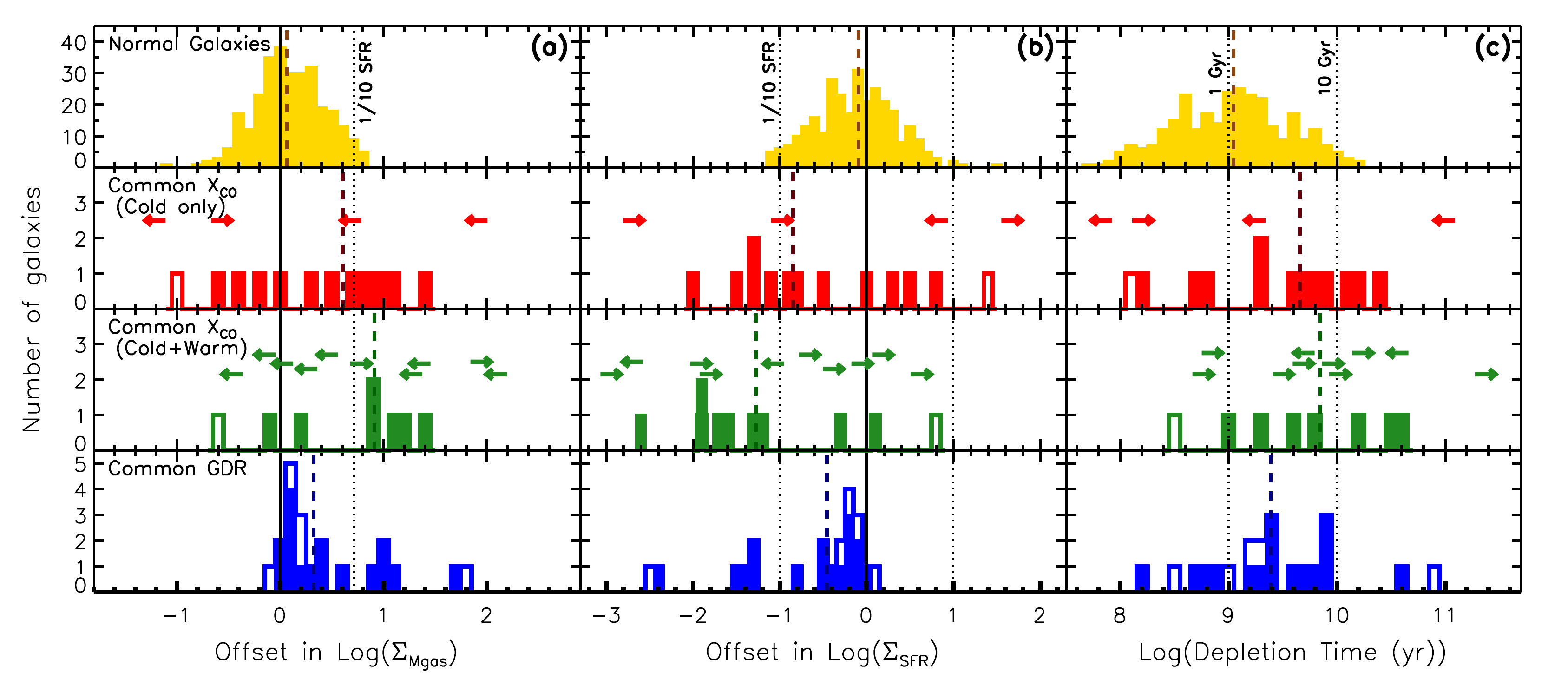}} 
\caption{Histograms comparing the distributions of offsets from the K--S relation ({\textbf{a, b}}) in Figure \ref{KS} and the depletion times of the molecular reservoir ({\textbf{c}}). The dashed line shows the median of each sample. In the lower three rows, the filled histograms correspond to the darker symbols of Figure \ref{KS}, whose SFR and dust mass are more reliable. These histograms show that the median offset (or depletion time) is typically larger (by about a factor of two) when molecular mass are calculated from CO luminosity than from dust mass; adding the warm gas mass further increases the offset. The dust-mass based values have a smaller dispersion and more clearly show an overall shift to lower SFR.}
\label{histKS}
\end{figure*}

We also examine whether galaxy properties, including ISM properties such as GDR and gas fraction, correlate with properties potentially related to jet feedback, such as the diffuse X-ray luminosity (which may be powered by dissipation of mechanical energy of the jet into the ISM) and the jet power. We show in Figure \ref{parcomp} the subset that shows significant or suggestive correlations, as calculated with the Spearman's rank correlation statistic\footnote{We used the r\_correlate routine in IDL. A $p$-value less than 0.05 is suggestive; when $p$-values are less than 0.01 the correlation is more significant.} \citep{numrec}. We do not find significant correlations for either the GDR or the gas fraction, except with \mh luminosity, which is likely due to the strong correlations of all three of these parameters with the gas mass (Fig. \ref{parcomp}b). We also do not find that the fitted cold dust temperature correlates with any other galaxy property. 

This sample was selected in part based on the presence of \mh emission in the purely rotational 0--0 lines in the MIR, so our galaxies all contain significant warm molecular emission. Table \ref{litprop} gives the luminosities of this component, measured in the S(0)28.2\um--S(3)9.66\um lines, which ranges from $7\times10^{38}-8\times10^{42}$\,erg\,s$^{-1}$. For most of our galaxies, the bulk of this gas is at $\sim100$K and comprised between 5\% and 80\% of the molecular reservoir  \citep[see also][]{ogle10}. We do not find correlations between the fraction of gas in the warm component and proxies of jet feedback. Fig. \ref{parcomp}a shows that the $L$(\mh)--SFR parameter space not excluded by the MOHEG criterion is  approximately uniformly covered by our galaxies, suggesting that the correlation is a selection effect.  In contrast, the correlations of L(\mh) with jet power (Fig. \ref{parcomp}d, $p\sim0.008$) and diffuse X-ray luminosity (Fig. \ref{parcomp}c, $p\sim0.027$; see also \citealt{lanz15}) supports the interpretation that these two luminosities are powered by the dissipation of the jet's mechanical energy into the ISM. We do not find correlation between the \mh/PAH ratio and either galaxy or feedback properties, suggesting that while this ratio is indicative of shocked gas, it may not be a good proxy for the strength of jet feedback.

\subsection{Star Formation Suppression in Radio MOHEGs}

To account for the uncertainty in $X_{\rm CO}$ factor and the incomplete availability of molecular observations of our sample, we calculate molecular gas masses and surface densities as well as the associated depletion times in three different ways: (1) from the CO luminosity assuming a common $X_{\rm CO}$ as described earlier, (2) from the sum of the CO-derived cold molecular mass (i.e. method 1) and the warm molecular mass calculated by \citet{ogle10} or \citet{guillard12}, and (3) from the dust mass assuming a common GDR of 100 (which Fig. \ref{gdr} shows to be a reasonable estimate). In Figure \ref{KS}, we plot all three derived molecular surface densities against the surface density of star formation on the K--S \citep{kennicutt98} diagram, compared to the relation found by those authors of $\Sigma_{\rm SFR}\propto\Sigma_{\rm gas}^{1.4}$. For comparison, we also show contours of normal galaxies from \citet{kennicutt98}, \citet{fisher13}, and \citet{shi11} and CO-detected ETGs from \citet{davis14}, whose gas masses and SFRs have been corrected to assume the same $X_{\rm CO}$ and initial mass function as our measurements. These galaxies largely have SFRs within a factor of ten of the predicted value from the K--S relation.

We find that the K--S diagrams show a tendency for radio MOHEGs to lie below the K--S relation. When assuming a common $X_{\rm CO}$, the observed star formation suppression appears larger, with a clustering around a suppression of SFR by a factor of ten. However, the scatter is fairly large, and particularly at low molecular surface densities (e.g. $\lesssim30\,{\rm M_{\odot}\,pc^{-2}}$), what remains of our sample would appear to be in agreement with normal galaxies. Since the cooling timescale of warm \mh is short \citep[$10^{4}$\,years;][]{guillard09} and this gas mass may therefore be quickly available to form stars, we also placed our galaxies on the K--S diagram using the total gas mass (both cold from CO luminosity and warm from IRS observations; Fig. \ref{KS}b). The primary effect is to increase the gas mass and the derived star formation suppression. 

Since our sample was selected to have indications of jet-driven turbulence in the ISM, we might imagine that some (or all) of these galaxies could likewise have CO further excited by the warm, turbulent medium, perhaps resulting in more CO emission on average per mass of molecular gas. Therefore, we also place galaxies on the K--S diagram using a gas mass calculated from the dust mass assuming a common GDR (Fig. \ref{KS}c). We find a smaller scatter with the bulk of our sample falling between the K--S relation and a suppression of SFR by a factor of ten (within the wings of the comparison sample of normal galaxies). However, given that most of our galaxies cluster around a normal GDR \citep[e.g., ][]{sandstrom13} when assuming a typical relation between CO luminosity and molecular mass, the offset we see for many of these galaxies is unlikely to be purely explained by the use of an inaccurate $X_{\rm CO}$. To test this effect, however, we examine the significance of offsets both for the complete sample and without including those galaxies with GDR larger or smaller than typical by a factor of five. 

\begin{figure}
\centerline{\includegraphics[trim=0.1cm 0.4cm 0.1cm 0.1cm,clip, width=\linewidth]{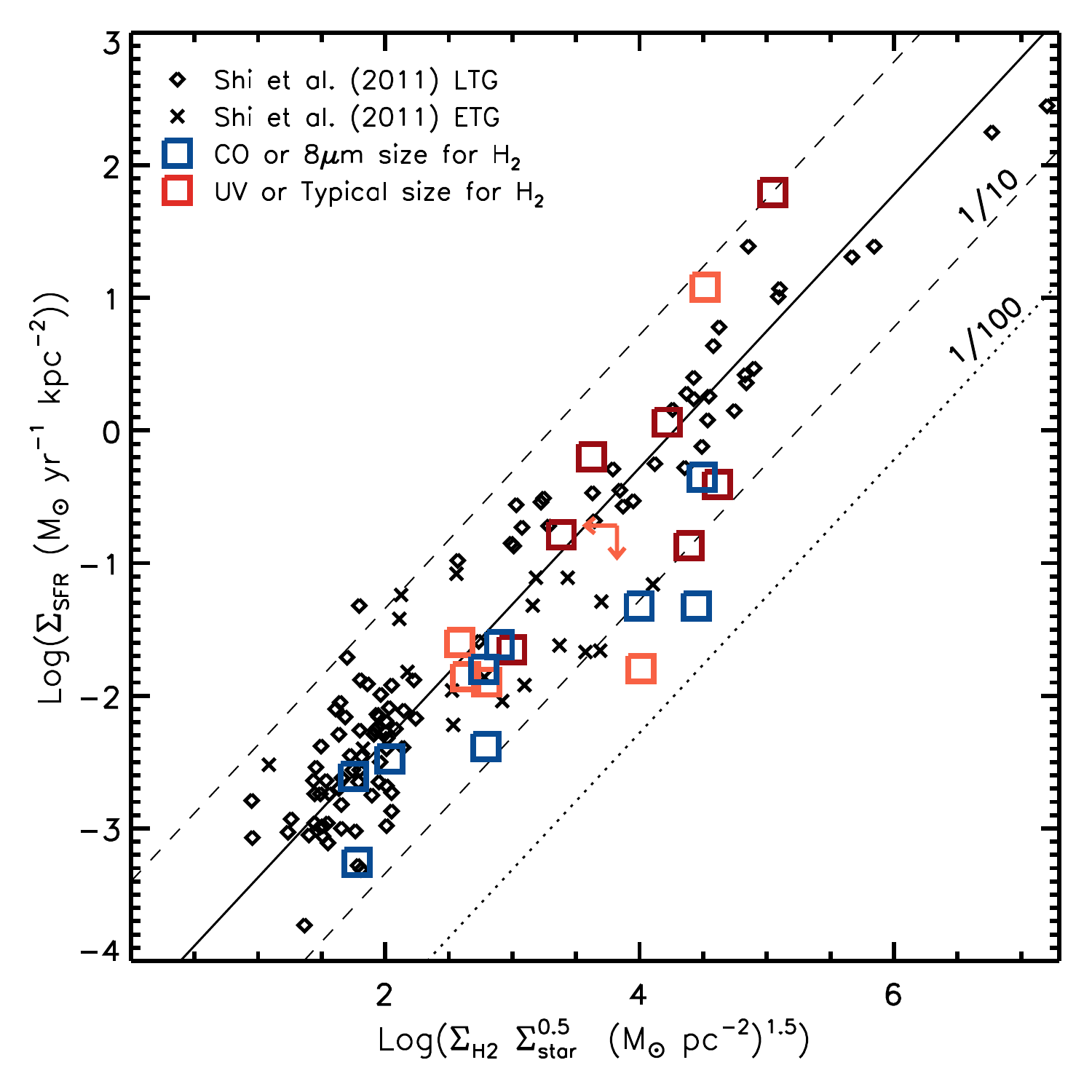}} 
\caption{Surface density of star formation compared to surface density of molecular gas and stellar mass \citep[extended Schmidt law; ][]{shi11}. For comparison, the small black symbols show the \citet{shi11} galaxies. Radio MOHEGs lie below the Shi relation, which may also be due to their hosts being ETGs, although the lenticular ETGs looked at by \citet[][crosses]{shi11} have higher sSFRs (Fig. \ref{histShi}). Darker symbols have more reliable SFRs and dust masses.
\label{shi}}
\end{figure}

\begin{figure*}
\centerline{\includegraphics[width=\linewidth]{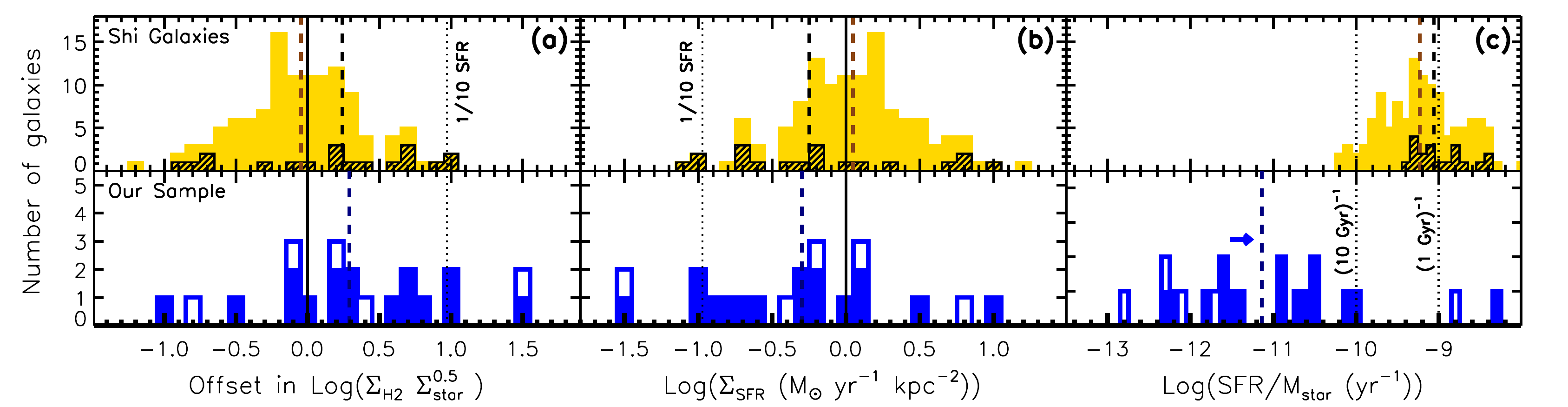}} 
\caption{Histograms comparing the distributions of offsets ({\textbf{a, b}}) in Figure \ref{shi}, as well as the sSFR ({\textbf{c}}). The dashed line shows the median of each sample. For the \citet{shi11} galaxies, we also highlight the ETGs with the black, hashed histograms. In the lower row, the filled histograms correspond to the darker symbols of Figure \ref{shi}, whose SFR, dust mass, and stellar mass are more reliable. Our galaxies have a similar distribution to the Shi ETGs, but their sSFR are noticeably smaller than those of the Shi sample.
\label{histShi}}
\end{figure*}

Figure \ref{histKS} shows histograms of the offsets from the K--S relation in both axes for each sample, quantifying the tendencies we have described above. When assuming $X_{\rm CO}$ (cold gas only), we find offsets of factors of $\sim$3 and $\sim$6 in surface density of molecular gas and SFR, respectively, with scatters of 0.85 and 1.2 dex. Adding the warm component to the molecular reservoir, drives the median offsets to large factors of $\sim$8 and $\sim$18, although these medians do not taken into account the numerous limits. If we only look at the subset of galaxies with the normal GDR, the median offsets change little; if we require the GDR to be even closer to normal GDR, the median offsets become further different from the comparison population. Finally, the median offsets are smaller with common GDR (factors of $\sim$2 and $\sim$3), but the scatter is smaller (0.75 and 1.0 dex). The GDR histograms show that our sample has a large fraction with only slight suppression but with a long tail.

We calculate the Wilcoxon--Mann--Whitney (WMW)\footnote{IDL routine RS\_TEST} and Kolmogorov--Smirnov (KS)\footnote{IDL routine kstwo} statistics, comparing our samples to the normal galaxies. Comparing the cold $X_{\rm CO}$ set with the normal galaxies, both statistics find a suggestive indication that the samples do not come from a common parent distribution (WMW: $p=0.044$; KS: $p=0.0090$). There are too few galaxies with accurate total (cold+warm) molecular masses for the WMW statistics to provide a meaningful assessment, but the KS statistics shows, as expected, a greater departure from the distribution of the normal galaxies ($p=0.0022$). As with the median comparisons, if we exclude those galaxies with GDRs that deviate from typical, we still find statistical differences compared to the normal galaxies (WMW: $p=0.03$; KS: $p=0.004-0.02$). For the larger GDR set, the WMW shows a significant difference from the normal galaxies ($p=0.00054$), while the KS statistic is only suggestive ($p=0.0064$). If we increase the assumed radius where we do not know the molecular extent to 2\,kpc, the significance decreases but is still suggestive (WMW: $p=0.016$; KS: $p=0.06$).

\begin{figure*}
\centerline{\includegraphics[trim=0.1cm 0.1cm 0.1cm 0.1cm, clip, width=\linewidth]{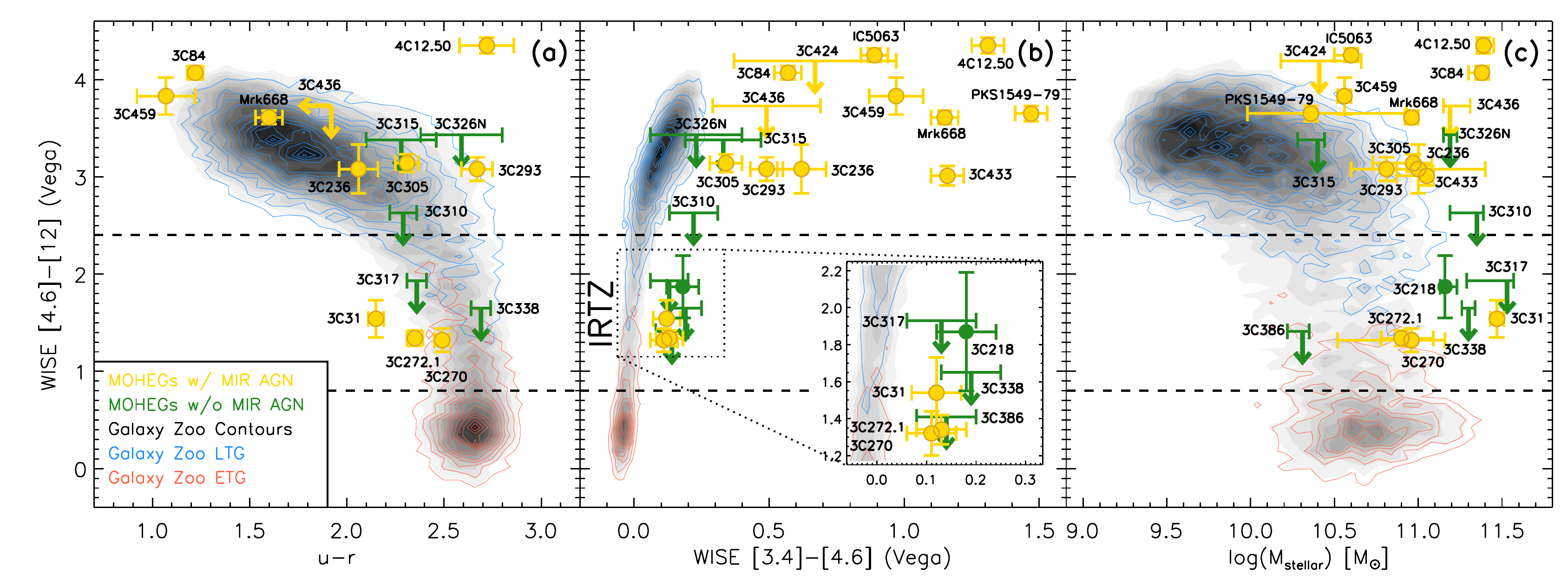}} 
\caption{Radio MOHEGs with (yellow points) and without (green points) MIR AGNs (based on the SED fit) are compared in to ERGs (red contours) and LTGs (blue contours) galaxies from the Galaxy Zoo \citep{lintott08, alatalo14irtf, schawinski14}, in the \wise [4.6]--[12] color vs. {\textbf{(a)}} {\em u-r} color, {\textbf{(b)}} \wise [3.4]--[4.6] color, and {\textbf{(c)}} stellar mass. The presence of a MIR AGN tends to push galaxies to a higher [4.6]--[12] color. Our galaxies show a wide range of optical colors, likely the result of a variety of dust content.  In \wise color-color space, radio MOHEGs have little overlap with Galaxy Zoo galaxies, instead falling into a region primarily occupied by AGNs \citep{stern12}.  The color--mass diagram shows that our sample galaxies are typically more massive than the Galaxy Zoo galaxies. }
\label{colors}
\end{figure*}

Three galaxies show consistent suppression by more than a factor of ten in all three K--S plots. While 3C\,31 is fairly consistent in its position, 3C\,236 shows a suppression (a factor of $\sim20$) based on its CO-derived cold molecular mass, but it has a low GDR, so when a common GDR is assumed, it is pushed past a factor of 200. The extent of the molecular disk in the third galaxies, Mrk\,668, is currently unknown. Assuming a typical radius of 1\,kpc, its SFR is suppressed by a factor of 30--95. As the arrow on Figure \ref{KS}c shows, a change in the assumed radius does not quite move galaxies purely along the K--S relation. Instead, an increase in radius also acts to reduce any suppression observed. For Mrk\,668, we calculate it would need a molecular disk of radius $\geq3.6$\,kpc to bring it back in the range of normal galaxies (i.e. within a factor of 10 of the K--S relation) in the most conservative mass estimate, on the edge of the range of radii (1--4\,kpc) found by \citet{davis13a} for CO-emitting ETGs, suggesting this galaxy is likely to have a large degree of suppression.

3C\,436, which has not been observed in CO, has a surface density of star formation and molecular gas that suggest a star formation suppression by a factor of $\sim30$, requiring a disk with a radius of 4.2\,kpc to bring 3C\,436 within a factor of 10 of K--S (i.e. the range of normal galaxies). Its sizable warm molecular content also positions 3C\,436 significantly away from the K--S relation (Fig. \ref{KS}b), further supporting the presence of significant star formation suppression in this system. In contrast, 3C\,310 (likewise unobserved in CO) lies almost on the K--S relation when using the dust mass to estimate the molecular reservoir; however, its warm molecular gas mass suggests it could be suppressed in star formation by over a factor of 10.

While the galaxies described above often shift in position along the horizontal axis between the different K--S plots, they still generally suggest a consistent shift toward suppressed star formation that cannot be fully explained by the peculiarities of the individual systems. The only galaxy in our sample where the apparently significant star formation suppression (by a factor of $\sim30$ in Fig. \ref{KS}a) can be attributed to an inaccurate $X_{\rm CO}$ factor is 3C\,293, which exhibits a very large GDR of $\sim1000$. In contrast to many in this sample, its CO SLED has been well-mapped to high-J CO lines, and it has a particular shape indicating a strong shock-excited component that may be driving its high CO luminosity \citep[see][]{papadopoulos08}. Indeed, when we assume a normal GDR, it lies on the K--S relation. As noted in the previous section, however, 3C\,293 does not show peculiarities in the other properties we have examined in this study, so the cause of its particular ISM state remains uncertain.

In Figure \ref{histKS}c, we compare the depletion times of the molecular reservoir ($\tau={\rm M_{H2}/SFR}$) of our reference galaxies and for our sample based on our three estimates of the molecular content. Normal galaxies have a median depletion time of $\sim$1\,Gyr \citep[consistent with the findings of][]{leroy08} extending out to 0.1 and 10\,Gyr, with only 2.4\% of the sample with depletion times greater than 10\,Gyr. In contrast, the median depletion times of radio MOHEGs are 2\,Gyr (common GDR), 4\,Gyr (common $X_{\rm CO}$ -- cold only), and  6\,Gyr (common $X_{\rm CO}$ -- cold + warm), with 10-30\% of the sample with depletion times greater than 10\,Gyr. WMW and KS statistics are suggestive that the $X_{\rm CO}$ (cold only: $p=0.015$; 0.023), $X_{\rm CO}$ (cold+warm: KS $p=0.064$), and GDR ($p=0.0047$; 0.028) samples have a different distribution of depletion times from the normal galaxies.

\begin{figure*}
\centerline{\includegraphics[trim=0.5cm 0.1cm 0.1cm 0.1cm,clip, width=\linewidth]{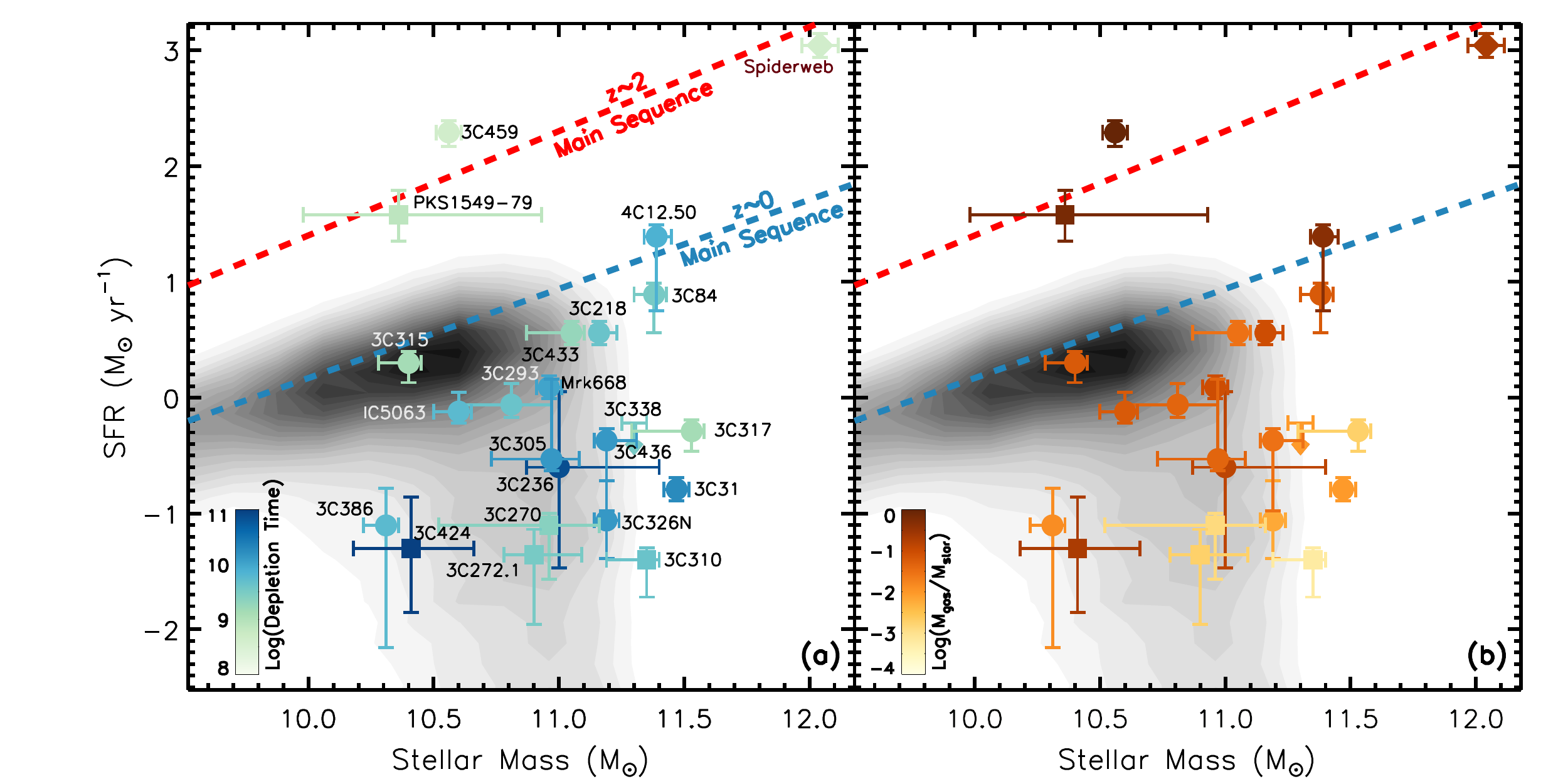}} 
\caption{The galaxy mass--SFR relation of our galaxies compared with a large sample of SDSS galaxies \citep[gray contours;][]{chang15}, whose SFRs and stellar masses were likewise determined with {\sc magphys}, showing the ``star formation main sequence'' with a tail at the high mass and low SFR of ETGs. The blue dashed line show the main sequence at $z\sim0$ \citep{elbaz07}. Our points are colored by their common GDR depletion times (\textbf{a}) and gas-richness (\textbf{b}) with squares having more reliable SFRs. Radio MOHEGs cover a large fraction of the parameter space, although many fall in the ETG tail. Our ULIRGs are found above the main sequence. Depletion time does not clearly correlate with position relative to the main sequence, but gas-poor galaxies tend to fall further off the main sequence. For comparison, we show high redshift radio MOHEG, the Spiderweb Galaxy, along with its associated main sequence at $z\sim2$ \citep{daddi07}.}
\label{sfms}
\end{figure*}

Reassuringly, both statistics indicate that our three samples originate from the same distribution. Further, the KS statistic indicates that those galaxies with a normal GDR could come from the same distribution as those whose GDRs deviate. Since the three methods of calculating molecular mass yield qualitatively similar and quantitatively consistent results, we will use molecular gas masses derived from dust masses for the remainder of the paper, since this method also provides the masses for all our radio galaxies. \\

 \citet{shi11} suggested that stellar mass might also play a role in the regulation of the star formation efficiency and found a tight relation between star formation efficiency and stellar mass surface density that extended the K--S relation to galaxies that the K--S law had previously been unable to explain, including  low surface brightness galaxies. Specifically, while the K--S relation can be re-cast as free fall in a gas-dominated gravitational potential, they argue that the stellar contribution to the gravitational potential can also be important, although they also considered that stellar mass (and its surface density) may be a proxy for more complex physics such as the impact of the kinetic and radiative energy dumped into the ISM by stars. 
 
 In Figure \ref{shi}, we investigate where our sample falls on their ``extended Schmidt law.'' As in the K--S plots, we find that our sample tends to fall below the Shi relation, indicating that the suppression we observe in the K--S plots cannot simply be attributed to galaxy types where the K--S might fail. The offsets between our sample and the Shi galaxies are shown in Figure \ref{histShi}ab, where we also indicate the distribution of the ETGs over the whole Shi sample. The ETGs in the Shi sample (crosses) broadly fall on their relation, although they too tend to be a bit below the relation (with a median SFR surface density a factor of $\sim$2 lower than total sample) and have a suggestive statistical difference with late-type galaxies (LTGs) of the Shi sample (WMW: $p=0.017$; KS: $p=0.0096$). Our galaxies are statistically different from the whole Shi et al. sample (WMW: $p=0.0030$; KS: $p=0.0092$), but cannot be distinguished from the Shi ETGs.

The Shi et al. ETGs, however, tend to be primarily lenticulars, while our sample is dominated by ellipticals.  The turbulence injected into the gas reservoir of radio MOHEGs is likely to further disperse their ISM from a disk into a larger volume. As a result, for these galaxies, the volume density of gas may be more important for determining the star formation efficiency in the case when the gas is not distributed in a disk. For example, some of our galaxies (e.g. 3C\,84) have substantial gas contents in filamentary distributions that are poorly modeled as a disk.

Figure \ref{histShi}c  compares the specific SFRs (sSFRs). The \citet{shi11} galaxies typically have sSFR of $10^{-9.23}$\,yr$^{-1}$, with an even higher median for the ETGs of $10^{-9.06}$\,yr$^{-1}$. In contrast, over 90\% of our sample have sSFR below $10^{-10}$\,yr$^{-1}$ with a median sSFR of $10^{-11.41}$\,yr$^{-1}$. WMW and KS statistics find very significant differences ($p<<0.001$) between the distributions of the sSFRs of our sample compared to the Shi sample, even if we restrict our comparison to the Shi ETGs.  \\

 In short, we find statistical differences between the star formation efficiency and sSFR in our galaxies compared to normal galaxies in the several comparisons based on different estimates of the molecular reservoir.  Our galaxies typically form stars less efficiently than normal galaxies with a suppression that cannot be explained by their stellar mass distribution or excitation of the CO gas resulting in a general over-estimation of the molecular gas mass.  
 
Since we find an overall suppression in star formation, we investigated whether the degree of suppression correlates with galaxy or feedback properties. We parameterize the degree of suppression in two ways: the molecular gas depletion time and the ratio of the expected SFR (if the molecular gas was forming stars as efficiently as predicted by the K--S relation) to the observed SFR. We do not find significant correlations between these values and gas mass, gas fraction, warm \mh luminosity, AGN luminosity, or jet power, suggesting that the relationship between feedback and the degree of suppression is complex and neither process is primarily dependent on or reflected in a single galaxy property.

\subsection{Radio MOHEGs in Galaxy Evolution}

In Figure \ref{colors}, we plot our galaxies on color--color and color--mass diagrams presented in \citet{alatalo14irtf}. Radio MOHEGs span a large range of optical and IR colors and are typically more massive than the underlying Galaxy Zoo distributions shown for comparison  \citep{lintott08,schawinski14}. MOHEGs also tend to be dustier and more gas-rich than less active ETGs, yielding bluer optical colors and larger [4.6]--[12] colors. Our sample also has larger [3.4]--[4.6] colors compared to the Galaxy Zoo galaxies, likely driven by AGN contributions, since MOHEGs with MIR AGNs tend to have particularly larger \wise colors since the AGN contributes more strongly to the longer wavelength of each color. 

Few of our galaxies fall into the optical ``green valley'' or the IR transition zone (IRTZ), typically crossed as galaxies transition from blue actively star-forming galaxies to red-and-dead systems. The five that fall into the IRTZ have minimal AGN contributions and are also the most gas-poor of our sample. We do not find correlations between any of the three colors and measures of star formation suppression.

Figure \ref{sfms} compares the SFRs and stellar masses of our galaxies (colored according to their depletion times and gas-richness) to those of a large sample of SDSS galaxies whose properties were calculated from \magphys \citep{chang15}. The contours show the star formation main sequence \citep{elbaz07, wuyts11} with a tail toward large masses and low SFRs where red-and-dead ETGs are typically found. Radio MOHEGs are found primarily below the main sequence, but they range across more than three orders of magnitude of SFR, indicating that they exist in hosts at a variety of evolutionary stages. 

\citet{alatalo15HCG} found a trend between depletion time and distance from the  star formation main sequence for Hickson Compact Group (HCG) galaxies showing star formation suppression. In contrast, Radio MOHEGs are much more diverse. While those with longer depletion times lie well below the main sequence, those with the shorter depletion times are found both above and below the main sequence, suggesting that the subset with short depletion times is a mix of two populations: gas-rich galaxies with high SFRs  (e.g. 3C\,459, PKS\,1549-79) and gas-poor galaxies with low SFRs (e.g. 3C\,317, 3C\,270). These two groups could also be a single population caught at different evolutionary stages. Radio MOHEGs are therefore likely to be be found in galaxies with a greater variety of gas-richness than the relative homogeneity of HCGs. We find a correlation between gas-richness and distance from the star formation main sequence (Fig. \ref{sfms}b).

For comparison, we also shown in Figure \ref{sfms} the Spiderweb galaxy \citep[PKS1138-26;][]{ogle12}, a much higher redshift ($z = 2.16$) MOHEG, which is a strong radio source in an unvirialized proto-cluster. It has a particularly large reservoir of warm molecular gas (eight times more massive than lower redshift radio galaxies), but with a depletion time similar to the ULIRGs 3C\,459 and PKS1549-79. However, it has a lower gas-richness than these ULIRGs and instead lies close to the main sequence at its redshift, making it a more massive, earlier analog to the radio galaxies just below the main sequence.

\subsection{When is Star Formation Suppression Important?}

\begin{figure}
\centerline{\includegraphics[trim=0.3cm 0.3cm 0.1cm 0.1cm,clip, width=\linewidth]{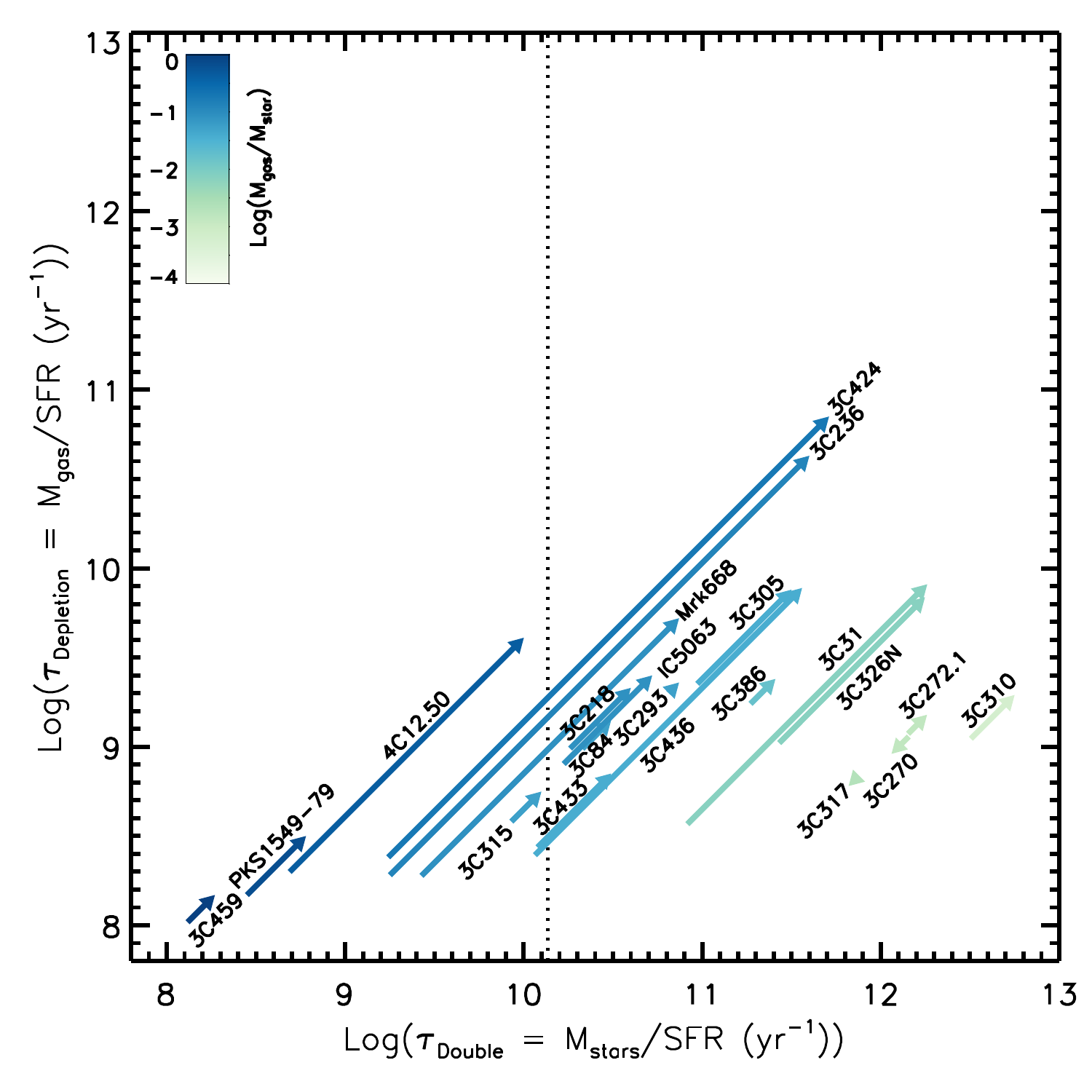}} 
\caption{Change in the time it would take to double the stellar mass and deplete the molecular reservoir \textbf{at the current SFR} from the times assuming star formation at the efficiency predicted by the Kennicutt-Schmidt relation to the times based on the observed SFR (arrow head), colored by gas richness. Longer arrows are more suppressed galaxies. Gas-poor galaxies show little change and have doubling times much longer than a Hubble time (dotted line). Very gas-rich galaxies also change little, with depletion and doubling times much smaller than a Hubble time. Galaxies with intermediate gas-richness tend to show the largest increases in times and are most likely to feel significant impacts on their evolution due to suppression of star formation activity.
\label{sfimp}}
\end{figure}

Given the mass of our galaxies and their low sSFRs (Fig. \ref{histShi}c), it is pertinent to ask whether the suppression of star formation in these systems will have a significant effect on the evolution of these galaxies. To that end, we compare the time it would take to double the stellar mass of the galaxy  and deplete the molecular reservoir to the time it would take if the galaxies were forming stars at the efficiency predicted by the K--S relation based on their molecular content (Fig. \ref{sfimp}). In calculating these timescales, we assume that the SFRs  remain at the observed and calculated (based on the current molecular content) rates, although the SFRs would likely decrease as raw materials are depleted, thereby increasing both timescales.

The most gas-poor galaxies tend to have the longest doubling times because they have so little raw material out of which to form new stars and therefore the evolution of their stellar mass is little changed by a decreased SFR due to suppression. These galaxies are well on their way to being red-and-dead and are the galaxies we find in (or close to) the IRTZ and green valley (Fig. \ref{colors}).

On the other side of the gas-richness spectrum, our most gas-rich galaxies have such large molecular reservoirs that injecting turbulence into their ISM still leaves a sizable portion of their molecular gas in a state where star formation can occur at (or close to) normal efficiency. Therefore, the star formation suppression will have a smaller effect on the evolution of these galaxies.

About a third of our galaxies show significant suppression that has the greatest potential to affect their future stellar mass. They tend to have intermediate gas-richness and, at the SFR predicted by K--S, five of them would have doubled their stellar mass in less than a Hubble time. However, the turbulence injected into their ISM due to jet feedback has likely rendered their molecular gas infertile and thereby sizably increased the time to grow their stellar mass significantly. These galaxies would however expend their molecular reservoirs prior to doubling their stellar mass unless additional gas were to be accreted. Understanding the gas budget would require a better census of the available reservoirs of gas in the environment that could be accreted either through mergers or via gas flows. Further study of the molecular and neutral content of the inter-galactic medium as well as of the companion galaxies is needed to better understand the additional gas that may become available to these galaxies.

Outflows can also reduce star forming activity by stripping galaxies of the necessary raw materials. Many of our galaxies are known to have jet-driven outflows, in multiple gas phases including molecular \citep[e.g., IC\,5063][]{morganti15}, ionized \citep[e.g., 3C\,293][]{emonts05}, and neutral \citep[see][for the eight galaxies from our sample with H{\sc{i}} outflows]{guillard12}. While these outflows can have mass fluxes as high as 10-100\,s\,${M_{\odot}\,yr^{-1}}$,  only a small fraction of this mass is likely to fully escape the gravitational pull of the galaxy \citep[e.g.,][]{alatalo15SF}, leaving the rest to potentially rain back onto the galaxy and reignite star formation or AGN activity. The relative importance of these outflows and the star formation suppression due to the injected turbulence on the evolution of galaxies is still poorly understood. A complete census of the different phases of the ISM in these galaxies and their kinematics would provide key clues with regards to loss of star-forming potential due to outflows compared to injected turbulence.


\section{CONCLUSION}

We modeled the UV--FIR SEDs of 22 radio galaxies with shocked warm molecular emission identified with \spitzer IRS and derived properties of the host galaxies to examine the impact of jet feedback. Figures \ref{img1}-\ref{seds8} show the UV--FIR images and fitted SEDs. The parameters derived from fitting these SEDs are consistent with parameters estimated  via simpler relations relying on only one or a few bands. Based on the properties of these galaxies, our conclusions are as follows. 

\begin{enumerate}
\item We find statistical evidence that star formation activity in radio MOHEGs has been suppressed by a factor of 3--6 compared to normal, star-forming galaxies, depending on whether we calculate molecular content based on dust mass or CO luminosity. Adding the warm molecular gas to our calculation of the molecular reservoir increases the suppression we measure. We do not, however, find a clear correlation between degree of star formation suppression and indicators of jet feedback, including jet power and shocked warm molecular luminosity.

\item Radio MOHEG hosts are typically massive, but in a variety of evolutionary states, covering a large range of optical and MIR colors, indicating a variety of dust content and MIR AGN contributions. These galaxies have normal cold molecular GDRs when assuming a typical X$_{\rm CO}$. 

\item While this sample of radio MOHEGs primarily has early-type morphologies, it covers almost four orders of magnitude of SFR, including several LIRGs and ULIRGs. As a result, radio MOHEGs cover a large range of the SFR--$M_{\rm star}$ space, but are primarily found in the high-mass, low-SFR tail of the star formation main sequence. Gas-rich galaxies tend to be above or near the main sequence, while gas-poor galaxies fall the farthest off of it.

\item Galaxies with an intermediate gas-richness have the greatest potential for large impacts on their future evolution through suppression of star formation by jet-driven turbulence. At least 25\% of our sample will have smaller stellar masses when their star formation ends than if they had continued forming stars at the efficiency of normal star-forming galaxies. Further study is necessary to understand the impact of both gas accretion and gas loss due to outflows. 

\end{enumerate}

\acknowledgements
We thank Mark Lacy for his advice and useful discussion with regards to the Sajina model, as well as George Helou for suggesting the extended Schmidt law as another point of analysis, and Yong Shi, Michael Brown, and Aditya Togi, whose comments improved the revised version of this paper. This work made use of the remote access computing accounts of the NASA Herschel Science Center, which were very helpful in reducing the \herschel data. L.L.  and P.M.O. acknowledges support for this work provided by NASA through an award issued by JPL/Caltech. Support for K.A. is provided by NASA through Hubble Fellowship grant \hbox{\#HST-HF2-51352.001} awarded by the Space Telescope Science Institute, which is operated by the Association of Universities for Research in Astronomy, Inc., for NASA, under contract NAS5-26555.

This work is based in part on observations made with {\em Herschel}, a European Space Agency Cornerstone Mission with significant participation by NASA. This publication used observations made with the {\em Spitzer Space Telescope}, which is operated by the Jet Propulsion Laboratory (JPL)/California Institute of Technology (Caltech) under a contract with NASA. Observations from the {\em Wide-field Infrared Survey Explorer}, which is a joint project of the University of California, Los Angeles, and JPL/Caltech, funded by the NASA, were also used.  This publication makes use of data products from the Two Micron All Sky Survey, which is a joint project of the University of Massachusetts and the Infrared Processing and Analysis Center (IPAC)/Caltech, funded by the NASA and the NSF, as well as from the Sloan Digital Sky Survey (SDSS-III), whose funding has been provided by the Alfred P. Sloan Foundation, the Participating Institutions, the NSF, and the U.S. DOE Office of Science, and which is managed by the Astrophysical Research Consortium for the Participating Institutions of the SDSS-III Collaboration. Finally, this publication makes use of data from the {\em Galaxy Evolution Explorer}, retrieved from the Mikulski Archive for Space Telescopes (MAST), part of the Space Telescope Science Institute, which is operated by the Association of Universities for Research in Astronomy, Inc., under NASA contract NAS5-26555. Support for MAST for non-HST data is provided by the NASA Office of Space Science via grant NNX09AF08G and by other grants and contracts. 

This research has made use of the NASA/IPAC Extragalactic Database (NED) and the Infrared Science Archive (IRSA) which are operated by the JPL/Caltech, under contract with NASA; and NASA's Astrophysics Data System (ADS). We also used software provided by the High Energy Astrophysics Science Archive Research Center (HEASARC), which is a service of the Astrophysics Science Division at NASA/GSFC and the High Energy Astrophysics Division of the Smithsonian Astrophysical Observatory.

{\it Facilities:} \facility{\herschel, \spitzer, \wise, 2MASS, Sloan, \galex}

\bibliography{MOHEG_SED}

\begin{thebibliography}{132}
\expandafter\ifx\csname natexlab\endcsname\relax\def\natexlab#1{#1}\fi

\bibitem[{{Alam} {et~al.}(2015){Alam}, {Albareti}, {Allende Prieto},
  {et~al.}}]{alam15}
{Alam}, S., {Albareti}, F.~D., {Allende Prieto}, C., {et~al.} 2015, \apjs, 219,
  12

\bibitem[{{Alatalo}(2015)}]{alatalo15mrk}
{Alatalo}, K. 2015, \apjl, 801, L17

\bibitem[{{Alatalo} {et~al.}(2015{\natexlab{a}}){Alatalo}, {Appleton},
  {Lisenfeld}, {et~al.}}]{alatalo15HCG}
{Alatalo}, K., {Appleton}, P.~N., {Lisenfeld}, U., {et~al.} 2015{\natexlab{a}},
  \apj, 812, 117

\bibitem[{{Alatalo} {et~al.}(2011){Alatalo}, {Blitz}, {Young},
  {et~al.}}]{alatalo11}
{Alatalo}, K., {Blitz}, L., {Young}, L.~M., {et~al.} 2011, \apj, 735, 88

\bibitem[{{Alatalo} {et~al.}(2014){Alatalo}, {Cales}, {Appleton},
  {et~al.}}]{alatalo14irtf}
{Alatalo}, K., {Cales}, S.~L., {Appleton}, P.~N., {et~al.} 2014, \apjl, 794,
  L13

\bibitem[{{Alatalo} {et~al.}(2015{\natexlab{b}}){Alatalo}, {Lacy}, {Lanz},
  {et~al.}}]{alatalo15SF}
{Alatalo}, K., {Lacy}, M., {Lanz}, L., {et~al.} 2015{\natexlab{b}}, \apj, 798,
  31

\bibitem[{{Baars} {et~al.}(1977){Baars}, {Genzel}, {Pauliny-Toth}, \&
  {Witzel}}]{baars77}
{Baars}, J.~W.~M., {Genzel}, R., {Pauliny-Toth}, I.~I.~K., \& {Witzel}, A.
  1977, \aap, 61, 99

\bibitem[{{Bell} {et~al.}(2003){Bell}, {McIntosh}, {Katz}, \&
  {Weinberg}}]{bell03}
{Bell}, E.~F., {McIntosh}, D.~H., {Katz}, N., \& {Weinberg}, M.~D. 2003, \apjs,
  149, 289

\bibitem[{{Bertin} \& {Arnouts}(1996)}]{bertin96}
{Bertin}, E., \& {Arnouts}, S. 1996, \aaps, 117, 393

\bibitem[{{Bolatto} {et~al.}(2013){Bolatto}, {Wolfire}, \& {Leroy}}]{bolatto13}
{Bolatto}, A.~D., {Wolfire}, M., \& {Leroy}, A.~K. 2013, \araa, 51, 207

\bibitem[{{Braine} \& {Dupraz}(1994)}]{braine94}
{Braine}, J., \& {Dupraz}, C. 1994, \aap, 283, 407

\bibitem[{{Bruzual} \& {Charlot}(2003)}]{bruzual03}
{Bruzual}, G., \& {Charlot}, S. 2003, \mnras, 344, 1000

\bibitem[{{Calzetti} {et~al.}(2010){Calzetti}, {Wu}, {Hong},
  {et~al.}}]{calzetti10}
{Calzetti}, D., {Wu}, S.-Y., {Hong}, S., {et~al.} 2010, \apj, 714, 1256

\bibitem[{{Chang} {et~al.}(2015){Chang}, {van der Wel}, {da Cunha}, \&
  {Rix}}]{chang15}
{Chang}, Y.-Y., {van der Wel}, A., {da Cunha}, E., \& {Rix}, H.-W. 2015, \apjs,
  219, 8

\bibitem[{{Clements} {et~al.}(2010){Clements}, {Dunne}, \&
  {Eales}}]{clements10}
{Clements}, D.~L., {Dunne}, L., \& {Eales}, S. 2010, \mnras, 403, 274

\bibitem[{{Cohen} {et~al.}(2003{\natexlab{a}}){Cohen}, {Megeath}, {Hammersley},
  {Mart{\'{\i}}n-Luis}, \& {Stauffer}}]{cohen03_irac}
{Cohen}, M., {Megeath}, S.~T., {Hammersley}, P.~L., {Mart{\'{\i}}n-Luis}, F.,
  \& {Stauffer}, J. 2003{\natexlab{a}}, \aj, 125, 2645

\bibitem[{{Cohen} {et~al.}(2003{\natexlab{b}}){Cohen}, {Wheaton}, \&
  {Megeath}}]{cohen03_2mass}
{Cohen}, M., {Wheaton}, W.~A., \& {Megeath}, S.~T. 2003{\natexlab{b}}, \aj,
  126, 1090

\bibitem[{{Cutri} {et~al.}(2006){Cutri}, {Skrutskie}, {Van Dyk},
  {et~al.}}]{2massExSup}
{Cutri}, R., {Skrutskie}, M., {Van Dyk}, S., {et~al.} 2006, Explanatory
  Supplement to the 2MASS All Sky Data Release and Extended Mission Products,
  IPAC, Pasadena, CA,
  {\url{http://www.ipac.caltech.edu/2mass/releases/allsky/doc/explsup.html}}

\bibitem[{{Cutri} {et~al.}(2015){Cutri}, {Wright}, {Conrow},
  {et~al.}}]{allwiseExSup}
{Cutri}, R., {Wright}, E., {Conrow}, T., {et~al.} 2015, Explanatory Supplement
  to the WISE All-Sky Data Release Products, IPAC, Pasadena, CA,
  {\url{http://wise2.ipac.caltech.edu/docs/release/allsky/expsup/}}

\bibitem[{{da Cunha} {et~al.}(2008){da Cunha}, {Charlot}, \&
  {Elbaz}}]{dacunha08}
{da Cunha}, E., {Charlot}, S., \& {Elbaz}, D. 2008, \mnras, 388, 1595

\bibitem[{{Daddi} {et~al.}(2007){Daddi}, {Dickinson}, {Morrison},
  {et~al.}}]{daddi07}
{Daddi}, E., {Dickinson}, M., {Morrison}, G., {et~al.} 2007, \apj, 670, 156

\bibitem[{{Dale} {et~al.}(2012){Dale}, {Aniano}, {Engelbracht},
  {et~al.}}]{dale12}
{Dale}, D.~A., {Aniano}, G., {Engelbracht}, C.~W., {et~al.} 2012, \apj, 745, 95

\bibitem[{{Dasyra} {et~al.}(2014){Dasyra}, {Combes}, {Novak},
  {et~al.}}]{dasyra14}
{Dasyra}, K.~M., {Combes}, F., {Novak}, G.~S., {et~al.} 2014, \aap, 565, A46

\bibitem[{{Davis} {et~al.}(2013){Davis}, {Alatalo}, {Bureau},
  {et~al.}}]{davis13a}
{Davis}, T.~A., {Alatalo}, K., {Bureau}, M., {et~al.} 2013, \mnras, 429, 534

\bibitem[{{Davis} {et~al.}(2014){Davis}, {Young}, {Crocker},
  {et~al.}}]{davis14}
{Davis}, T.~A., {Young}, L.~M., {Crocker}, A.~F., {et~al.} 2014, \mnras, 444,
  3427

\bibitem[{{de Vaucouleurs} {et~al.}(1991){de Vaucouleurs}, {de Vaucouleurs},
  {Corwin}, {et~al.}}]{devaucouleurs91}
{de Vaucouleurs}, G., {de Vaucouleurs}, A., {Corwin}, Jr., H.~G., {et~al.}
  1991, {Third Reference Catalogue of Bright Galaxies} ({New York: Springer})

\bibitem[{{Di Matteo} {et~al.}(2005){Di Matteo}, {Springel}, \&
  {Hernquist}}]{dimatteo05}
{Di Matteo}, T., {Springel}, V., \& {Hernquist}, L. 2005, \nat, 433, 604

\bibitem[{{Drake} {et~al.}(2004){Drake}, {McGregor}, \& {Dopita}}]{drake04}
{Drake}, C.~L., {McGregor}, P.~J., \& {Dopita}, M.~A. 2004, \aj, 128, 955

\bibitem[{{Elbaz} {et~al.}(2007){Elbaz}, {Daddi}, {Le Borgne},
  {et~al.}}]{elbaz07}
{Elbaz}, D., {Daddi}, E., {Le Borgne}, D., {et~al.} 2007, \aap, 468, 33

\bibitem[{{Emonts} {et~al.}(2005){Emonts}, {Morganti}, {Tadhunter},
  {et~al.}}]{emonts05}
{Emonts}, B.~H.~C., {Morganti}, R., {Tadhunter}, C.~N., {et~al.} 2005, \mnras,
  362, 931

\bibitem[{{Engelbracht} {et~al.}(2007){Engelbracht}, {Blaylock}, {Su},
  {et~al.}}]{engelbracht07}
{Engelbracht}, C.~W., {Blaylock}, M., {Su}, K.~Y.~L., {et~al.} 2007, \pasp,
  119, 994

\bibitem[{{Evans} {et~al.}(2005){Evans}, {Mazzarella}, {Surace},
  {et~al.}}]{evans05}
{Evans}, A.~S., {Mazzarella}, J.~M., {Surace}, J.~A., {et~al.} 2005, \apjs,
  159, 197

\bibitem[{{Fanaroff} \& {Riley}(1974)}]{fanaroff74}
{Fanaroff}, B.~L., \& {Riley}, J.~M. 1974, \mnras, 167, 31P

\bibitem[{{Fazio} {et~al.}(2004){Fazio}, {Hora}, {Allen}, {et~al.}}]{fazio04}
{Fazio}, G.~G., {Hora}, J.~L., {Allen}, L.~E., {et~al.} 2004, \apjs, 154, 10

\bibitem[{{Feruglio} {et~al.}(2010){Feruglio}, {Maiolino}, {Piconcelli},
  {et~al.}}]{feruglio10}
{Feruglio}, C., {Maiolino}, R., {Piconcelli}, E., {et~al.} 2010, \aap, 518,
  L155

\bibitem[{{Fisher} {et~al.}(2013){Fisher}, {Bolatto}, {Drory}, {Combes},
  {Blitz}, \& {Wong}}]{fisher13}
{Fisher}, D.~B., {Bolatto}, A., {Drory}, N., {Combes}, F., {Blitz}, L., \&
  {Wong}, T. 2013, \apj, 764, 174

\bibitem[{{Garc{\'{\i}}a-Burillo} {et~al.}(2014){Garc{\'{\i}}a-Burillo},
  {Combes}, {Usero}, {et~al.}}]{garcia14}
{Garc{\'{\i}}a-Burillo}, S., {Combes}, F., {Usero}, A., {et~al.} 2014, \aap,
  567, A125

\bibitem[{{Goddard Space Flight Center}(2004)}]{goddard04}
{Goddard Space Flight Center}. 2004, {\em GALEX} Observer's Guide, GSFC,
  Greenbelt, MD, {\url{http://galexgi.gsfc.nasa.gov/docs/galex/Documents/
  GALEXObserverGuide.pdf}}

\bibitem[{{Golombek} {et~al.}(1988){Golombek}, {Miley}, \&
  {Neugebauer}}]{golombek88}
{Golombek}, D., {Miley}, G.~K., \& {Neugebauer}, G. 1988, \aj, 95, 26

\bibitem[{{Gonz{\'a}lez-Mart{\'{\i}}n} \& {Vaughan}(2012)}]{gonzalez12}
{Gonz{\'a}lez-Mart{\'{\i}}n}, O., \& {Vaughan}, S. 2012, \aap, 544, A80

\bibitem[{{Goulding} {et~al.}(2011){Goulding}, {Alexander}, {Mullaney},
  {Gelbord}, {Hickox}, {Ward}, \& {Watson}}]{goulding11}
{Goulding}, A.~D., {Alexander}, D.~M., {Mullaney}, J.~R., {Gelbord}, J.~M.,
  {Hickox}, R.~C., {Ward}, M., \& {Watson}, M.~G. 2011, \mnras, 411, 1231

\bibitem[{{Griffin} {et~al.}(2010){Griffin}, {Abergel}, {Abreu},
  {et~al.}}]{griffin10}
{Griffin}, M.~J., {Abergel}, A., {Abreu}, A., {et~al.} 2010, \aap, 518, L3

\bibitem[{{Guainazzi} {et~al.}(2004){Guainazzi}, {Siemiginowska},
  {Rodriguez-Pascual}, \& {Stanghellini}}]{guainazzi04}
{Guainazzi}, M., {Siemiginowska}, A., {Rodriguez-Pascual}, P., \&
  {Stanghellini}, C. 2004, \aap, 421, 461

\bibitem[{{Guillard} {et~al.}(2015){Guillard}, {Boulanger}, {Lehnert},
  {et~al.}}]{guillard15}
{Guillard}, P., {Boulanger}, F., {Lehnert}, M.~D., {et~al.} 2015, \aap, 574,
  A32

\bibitem[{{Guillard} {et~al.}(2009){Guillard}, {Boulanger}, {Pineau Des
  For{\^e}ts}, \& {Appleton}}]{guillard09}
{Guillard}, P., {Boulanger}, F., {Pineau Des For{\^e}ts}, G., \& {Appleton},
  P.~N. 2009, \aap, 502, 515

\bibitem[{{Guillard} {et~al.}(2012){Guillard}, {Ogle}, {Emonts},
  {et~al.}}]{guillard12}
{Guillard}, P., {Ogle}, P.~M., {Emonts}, B.~H.~C., {et~al.} 2012, \apj, 747, 95

\bibitem[{{Hopkins} {et~al.}(2006){Hopkins}, {Hernquist}, {Cox}, {Di Matteo},
  {Robertson}, \& {Springel}}]{hopkins06}
{Hopkins}, P.~F., {Hernquist}, L., {Cox}, T.~J., {Di Matteo}, T., {Robertson},
  B., \& {Springel}, V. 2006, \apjs, 163, 1

\bibitem[{{Houck} {et~al.}(2004){Houck}, {Roellig}, {van Cleve},
  {et~al.}}]{houck04}
{Houck}, J.~R., {Roellig}, T.~L., {van Cleve}, J., {et~al.} 2004, \apjs, 154,
  18

\bibitem[{{Jansen} {et~al.}(2001){Jansen}, {Lumb}, {Altieri},
  {et~al.}}]{jansen01}
{Jansen}, F., {Lumb}, D., {Altieri}, B., {et~al.} 2001, \aap, 365, L1

\bibitem[{{Jarrett} {et~al.}(2003){Jarrett}, {Chester}, {Cutri}, {Schneider},
  \& {Huchra}}]{jarrett03}
{Jarrett}, T.~H., {Chester}, T., {Cutri}, R., {Schneider}, S.~E., \& {Huchra},
  J.~P. 2003, \aj, 125, 525

\bibitem[{{Kalberla} {et~al.}(2005){Kalberla}, {Burton}, {Hartmann},
  {et~al.}}]{kalberla05}
{Kalberla}, P.~M.~W., {Burton}, W.~B., {Hartmann}, D., {et~al.} 2005, \aap,
  440, 775

\bibitem[{{Karouzos} {et~al.}(2013){Karouzos}, {Trichas}, {Im}, {Malkan}, \&
  {the AKARI-NEP team}}]{karouzos13}
{Karouzos}, M., {Trichas}, M., {Im}, M., {Malkan}, M., \& {the AKARI-NEP team}.
  2013, ArXiv e-prints

\bibitem[{{Kellermann} {et~al.}(1969){Kellermann}, {Pauliny-Toth}, \&
  {Williams}}]{kellermann69}
{Kellermann}, K.~I., {Pauliny-Toth}, I.~I.~K., \& {Williams}, P.~J.~S. 1969,
  \apj, 157, 1

\bibitem[{{Kennicutt}(1998)}]{kennicutt98}
{Kennicutt}, Jr., R.~C. 1998, \apj, 498, 541

\bibitem[{{K\"{u}hr} {et~al.}(1981){K\"{u}hr}, {Witzel}, {Pauliny-Toth}, \&
  {Nauber}}]{kuhr81}
{K\"{u}hr}, H., {Witzel}, A., {Pauliny-Toth}, I.~I.~K., \& {Nauber}, U. 1981,
  \aaps, 45, 367

\bibitem[{{Labiano} {et~al.}(2013){Labiano}, {Garc{\'{\i}}a-Burillo}, {Combes},
  {et~al.}}]{labiano13}
{Labiano}, A., {Garc{\'{\i}}a-Burillo}, S., {Combes}, F., {et~al.} 2013, \aap,
  549, A58

\bibitem[{{Labiano} {et~al.}(2014){Labiano}, {Garc{\'{\i}}a-Burillo}, {Combes},
  {et~al.}}]{labiano14}
---. 2014, \aap, 564, A128

\bibitem[{{Lacy} {et~al.}(2004){Lacy}, {Storrie-Lombardi}, {Sajina},
  {Appleton}, {Armus}, {Chapman}, {Choi}, {Fadda}, {Fang}, {Frayer},
  {Heinrichsen}, {Helou}, {Im}, {Marleau}, {Masci}, {Shupe}, {Soifer},
  {Surace}, {Teplitz}, {Wilson}, \& {Yan}}]{lacy04}
{Lacy}, M., {et~al.} 2004, \apjs, 154, 166

\bibitem[{{Laing} \& {Peacock}(1980)}]{laing80}
{Laing}, R.~A., \& {Peacock}, J.~A. 1980, \mnras, 190, 903

\bibitem[{{Laing} {et~al.}(1983){Laing}, {Riley}, \& {Longair}}]{laing83}
{Laing}, R.~A., {Riley}, J.~M., \& {Longair}, M.~S. 1983, \mnras, 204, 151

\bibitem[{{Lanz} {et~al.}(2011){Lanz}, {Bliss}, {Kraft}, {et~al.}}]{lanz11}
{Lanz}, L., {Bliss}, A., {Kraft}, R.~P., {et~al.} 2011, \apj, 731, 52

\bibitem[{{Lanz} {et~al.}(2015){Lanz}, {Ogle}, {Evans}, {et~al.}}]{lanz15}
{Lanz}, L., {Ogle}, P.~M., {Evans}, D., {et~al.} 2015, \apj, 801, 17

\bibitem[{{Large} {et~al.}(1981){Large}, {Mills}, {Little}, {Crawford}, \&
  {Sutton}}]{large81}
{Large}, M.~I., {Mills}, B.~Y., {Little}, A.~G., {Crawford}, D.~F., \&
  {Sutton}, J.~M. 1981, \mnras, 194, 693

\bibitem[{{Leipski} {et~al.}(2009){Leipski}, {Antonucci}, {Ogle}, \&
  {Whysong}}]{leipski09}
{Leipski}, C., {Antonucci}, R., {Ogle}, P., \& {Whysong}, D. 2009, \apj, 701,
  891

\bibitem[{{Leroy} {et~al.}(2008){Leroy}, {Walter}, {Brinks},
  {et~al.}}]{leroy08}
{Leroy}, A.~K., {Walter}, F., {Brinks}, E., {et~al.} 2008, \aj, 136, 2782

\bibitem[{{Lintott} {et~al.}(2008){Lintott}, {Schawinski}, {Slosar},
  {et~al.}}]{lintott08}
{Lintott}, C.~J., {Schawinski}, K., {Slosar}, A., {et~al.} 2008, \mnras, 389,
  1179

\bibitem[{{Lutz} {et~al.}(2004){Lutz}, {Maiolino}, {Spoon}, \&
  {Moorwood}}]{lutz04}
{Lutz}, D., {Maiolino}, R., {Spoon}, H.~W.~W., \& {Moorwood}, A.~F.~M. 2004,
  \aap, 418, 465

\bibitem[{{Madden} {et~al.}(2006){Madden}, {Galliano}, {Jones}, \&
  {Sauvage}}]{madden06}
{Madden}, S.~C., {Galliano}, F., {Jones}, A.~P., \& {Sauvage}, M. 2006, \aap,
  446, 877

\bibitem[{{Mahony} {et~al.}(2013){Mahony}, {Morganti}, {Emonts}, {Oosterloo},
  \& {Tadhunter}}]{mahony13}
{Mahony}, E.~K., {Morganti}, R., {Emonts}, B.~H.~C., {Oosterloo}, T.~A., \&
  {Tadhunter}, C. 2013, \mnras, 435, L58

\bibitem[{{Martel} {et~al.}(1999){Martel}, {Baum}, {Sparks},
  {et~al.}}]{martel99}
{Martel}, A.~R., {Baum}, S.~A., {Sparks}, W.~B., {et~al.} 1999, \apjs, 122, 81

\bibitem[{{Martin} {et~al.}(2005){Martin}, {Fanson}, {Schiminovich},
  {et~al.}}]{martin05}
{Martin}, D.~C., {Fanson}, J., {Schiminovich}, D., {et~al.} 2005, \apjl, 619,
  L1

\bibitem[{{Mauch} {et~al.}(2003){Mauch}, {Murphy}, {Buttery},
  {et~al.}}]{mauch03}
{Mauch}, T., {Murphy}, T., {Buttery}, H.~J., {et~al.} 2003, \mnras, 342, 1117

\bibitem[{{Mel{\'e}ndez} {et~al.}(2014){Mel{\'e}ndez}, {Mushotzky}, {Shimizu},
  {Barger}, \& {Cowie}}]{melendez14}
{Mel{\'e}ndez}, M., {Mushotzky}, R.~F., {Shimizu}, T.~T., {Barger}, A.~J., \&
  {Cowie}, L.~L. 2014, \apj, 794, 152

\bibitem[{{Mittal} {et~al.}(2012){Mittal}, {Oonk}, {Ferland}, {Edge}, {O'Dea},
  {Baum}, {Whelan}, {Johnstone}, {Combes}, {Salom{\'e}}, {Fabian}, {Tremblay},
  {Donahue}, \& {Russell}}]{mittal12}
{Mittal}, R., {et~al.} 2012, \mnras, 426, 2957

\bibitem[{{Morganti} {et~al.}(2013){Morganti}, {Frieswijk}, {Oonk},
  {Oosterloo}, \& {Tadhunter}}]{morganti13}
{Morganti}, R., {Frieswijk}, W., {Oonk}, R.~J.~B., {Oosterloo}, T., \&
  {Tadhunter}, C. 2013, \aap, 552, L4

\bibitem[{{Morganti} {et~al.}(2015){Morganti}, {Oosterloo}, {Oonk},
  {Frieswijk}, \& {Tadhunter}}]{morganti15}
{Morganti}, R., {Oosterloo}, T.~A., {Oonk}, J.~B.~R., {Frieswijk}, W., \&
  {Tadhunter}, C.~N. 2015, ArXiv e-prints

\bibitem[{{Moshir} {et~al.}(1990){Moshir}, {Kopan}, {Conrow},
  {et~al.}}]{moshir90}
{Moshir}, M., {Kopan}, G., {Conrow}, T., {et~al.} 1990, in IRAS Faint Source
  Catalogue, version 2.0 (1990)

\bibitem[{{Mullaney} {et~al.}(2011){Mullaney}, {Alexander}, {Goulding}, \&
  {Hickox}}]{mullaney11}
{Mullaney}, J.~R., {Alexander}, D.~M., {Goulding}, A.~D., \& {Hickox}, R.~C.
  2011, \mnras, 414, 1082

\bibitem[{{M\"{u}ller} {et~al.}(2011){M\"{u}ller}, {Okumura}, \&
  {Klaas}}]{pacscolor}
{M\"{u}ller}, T., {Okumura}, K., \& {Klaas}, U. 2011, PACS Photometer Passbands
  and Colour Correction Factors for Various Source SEDs, Tech. Rep.
  PICC-ME-TN-038, European Space Agency

\bibitem[{{Narayanan} {et~al.}(2012){Narayanan}, {Krumholz}, {Ostriker}, \&
  {Hernquist}}]{narayanan12}
{Narayanan}, D., {Krumholz}, M.~R., {Ostriker}, E.~C., \& {Hernquist}, L. 2012,
  \mnras, 421, 3127

\bibitem[{{Nesvadba} {et~al.}(2010){Nesvadba}, {Boulanger}, {Salom{\'e}},
  {et~al.}}]{nesvadba10}
{Nesvadba}, N.~P.~H., {Boulanger}, F., {Salom{\'e}}, P., {et~al.} 2010, \aap,
  521, A65

\bibitem[{{Nyland} {et~al.}(2013){Nyland}, {Alatalo}, {Wrobel},
  {et~al.}}]{nyland13}
{Nyland}, K., {Alatalo}, K., {Wrobel}, J.~M., {et~al.} 2013, \apj, 779, 173

\bibitem[{{O'Brien} {et~al.}(2010){O'Brien}, {Reeves}, \& {Braito}}]{obrien10}
{O'Brien}, P.~T., {Reeves}, J.~N., \& {Braito}, V. 2010, in Bulletin of the
  American Astronomical Society, Vol.~42, AAS/High Energy Astrophysics Division
  \#11, 664

\bibitem[{{Oca{\~n}a Flaquer} {et~al.}(2010){Oca{\~n}a Flaquer}, {Leon},
  {Combes}, \& {Lim}}]{ocana10}
{Oca{\~n}a Flaquer}, B., {Leon}, S., {Combes}, F., \& {Lim}, J. 2010, \aap,
  518, A9

\bibitem[{{Ogle} {et~al.}(2007){Ogle}, {Antonucci}, {Appleton}, \&
  {Whysong}}]{ogle07}
{Ogle}, P., {Antonucci}, R., {Appleton}, P.~N., \& {Whysong}, D. 2007, \apj,
  668, 699

\bibitem[{{Ogle} {et~al.}(2010){Ogle}, {Boulanger}, {Guillard},
  {et~al.}}]{ogle10}
{Ogle}, P., {Boulanger}, F., {Guillard}, P., {et~al.} 2010, \apj, 724, 1193

\bibitem[{{Ogle} {et~al.}(2012){Ogle}, {Davies}, {Appleton}, {et~al.}}]{ogle12}
{Ogle}, P., {Davies}, J.~E., {Appleton}, P.~N., {et~al.} 2012, \apj, 751, 13

\bibitem[{{Ogle} {et~al.}(2014){Ogle}, {Lanz}, \& {Appleton}}]{ogle14}
{Ogle}, P.~M., {Lanz}, L., \& {Appleton}, P.~N. 2014, \apjl, 788, L33

\bibitem[{{Okuda} {et~al.}(2013){Okuda}, {Iguchi}, \& {Kohno}}]{okuda13}
{Okuda}, T., {Iguchi}, S., \& {Kohno}, K. 2013, \apj, 768, 19

\bibitem[{{Okuda} {et~al.}(2005){Okuda}, {Kohno}, {Iguchi}, \&
  {Nakanishi}}]{okuda05}
{Okuda}, T., {Kohno}, K., {Iguchi}, S., \& {Nakanishi}, K. 2005, \apj, 620, 673

\bibitem[{{Ott}(2010)}]{ott10}
{Ott}, S. 2010, in Astronomical Society of the Pacific Conference Series, Vol.
  434, Astronomical Data Analysis Software and Systems XIX, ed. Y.~{Mizumoto},
  K.-I. {Morita}, \& M.~{Ohishi}, 139

\bibitem[{{Paladini} {et~al.}(2012){Paladini}, {Linz}, {Altieri}, \&
  {Ali}}]{pacsunc}
{Paladini}, R., {Linz}, H., {Altieri}, B., \& {Ali}, B. 2012, Assessment
  analysis of the extended emission calibration for the PACS red channel, Tech.
  Rep. PICC-NHSC-TR-034, NASA Herschel Science Center

\bibitem[{{Papadopoulos} {et~al.}(2008){Papadopoulos}, {Kovacs}, {Evans}, \&
  {Barthel}}]{papadopoulos08}
{Papadopoulos}, P.~P., {Kovacs}, A., {Evans}, A.~S., \& {Barthel}, P. 2008,
  \aap, 491, 483

\bibitem[{{Papadopoulos} {et~al.}(2010){Papadopoulos}, {van der Werf}, {Isaak},
  \& {Xilouris}}]{papadopoulos10}
{Papadopoulos}, P.~P., {van der Werf}, P., {Isaak}, K., \& {Xilouris}, E.~M.
  2010, \apj, 715, 775

\bibitem[{{Pearson} {et~al.}(2014){Pearson}, {Lim}, {North},
  {et~al.}}]{pearson14}
{Pearson}, C., {Lim}, T., {North}, C., {et~al.} 2014, Experimental Astronomy,
  37, 175

\bibitem[{{Pilbratt} {et~al.}(2010){Pilbratt}, {Riedinger}, {Passvogel},
  {et~al.}}]{pilbratt10}
{Pilbratt}, G.~L., {Riedinger}, J.~R., {Passvogel}, T., {et~al.} 2010, \aap,
  518, L1

\bibitem[{{Poglitsch} {et~al.}(2010){Poglitsch}, {Waelkens}, {Geis},
  {et~al.}}]{poglitsch10}
{Poglitsch}, A., {Waelkens}, C., {Geis}, N., {et~al.} 2010, \aap, 518, L2

\bibitem[{{Press} {et~al.}(1986){Press}, {Flannery}, \& {Teukolsky}}]{numrec}
{Press}, W.~H., {Flannery}, B.~P., \& {Teukolsky}, S.~A. 1986, {Numerical
  recipes. The art of scientific computing} (Cambridge: University Press, 1986)

\bibitem[{{Punsly}(2005)}]{punsly05}
{Punsly}, B. 2005, \apjl, 623, L9

\bibitem[{{Rieke} {et~al.}(2004){Rieke}, {Young}, {Engelbracht},
  {et~al.}}]{rieke04}
{Rieke}, G.~H., {Young}, E.~T., {Engelbracht}, C.~W., {et~al.} 2004, \apjs,
  154, 25

\bibitem[{{Roussel}(2013)}]{roussel13}
{Roussel}, H. 2013, \pasp, 125, 1126

\bibitem[{{Sajina} {et~al.}(2012){Sajina}, {Yan}, {Fadda}, {Dasyra}, \&
  {Huynh}}]{sajina12}
{Sajina}, A., {Yan}, L., {Fadda}, D., {Dasyra}, K., \& {Huynh}, M. 2012, \apj,
  757, 13

\bibitem[{{Salom{\'e}} \& {Combes}(2003)}]{salome03}
{Salom{\'e}}, P., \& {Combes}, F. 2003, \aap, 412, 657

\bibitem[{{Salom{\'e}} {et~al.}(2006){Salom{\'e}}, {Combes}, {Edge},
  {Crawford}, {Erlund}, {Fabian}, {Hatch}, {Johnstone}, {Sanders}, \&
  {Wilman}}]{salome06}
{Salom{\'e}}, P., {et~al.} 2006, \aap, 454, 437

\bibitem[{{Sandstrom} {et~al.}(2013){Sandstrom}, {Leroy}, {Walter},
  {et~al.}}]{sandstrom13}
{Sandstrom}, K.~M., {Leroy}, A.~K., {Walter}, F., {et~al.} 2013, \apj, 777, 5

\bibitem[{{Saripalli} \& {Mack}(2007)}]{saripalli07}
{Saripalli}, L., \& {Mack}, K.-H. 2007, \mnras, 376, 1385

\bibitem[{{Scharw{\"a}chter} {et~al.}(2013){Scharw{\"a}chter}, {McGregor},
  {Dopita}, \& {Beck}}]{scharwachter13}
{Scharw{\"a}chter}, J., {McGregor}, P.~J., {Dopita}, M.~A., \& {Beck}, T.~L.
  2013, \mnras, 429, 2315

\bibitem[{{Schawinski} {et~al.}(2014){Schawinski}, {Urry}, {Simmons},
  {et~al.}}]{schawinski14}
{Schawinski}, K., {Urry}, C.~M., {Simmons}, B.~D., {et~al.} 2014, \mnras, 440,
  889

\bibitem[{{Scoville} {et~al.}(2014){Scoville}, {Aussel}, {Sheth},
  {et~al.}}]{scoville14}
{Scoville}, N., {Aussel}, H., {Sheth}, K., {et~al.} 2014, \apj, 783, 84

\bibitem[{{Shi} {et~al.}(2011){Shi}, {Helou}, {Yan}, {et~al.}}]{shi11}
{Shi}, Y., {Helou}, G., {Yan}, L., {et~al.} 2011, \apj, 733, 87

\bibitem[{{Silk} \& {Rees}(1998)}]{silk98}
{Silk}, J., \& {Rees}, M.~J. 1998, \aap, 331, L1

\bibitem[{{Skrutskie} {et~al.}(2006){Skrutskie}, {Cutri}, {Stiening},
  {et~al.}}]{skrutskie06}
{Skrutskie}, M.~F., {Cutri}, R.~M., {Stiening}, R., {et~al.} 2006, \aj, 131,
  1163

\bibitem[{{Smith} \& {Heckman}(1989)}]{smith89}
{Smith}, E.~P., \& {Heckman}, T.~M. 1989, \apjs, 69, 365

\bibitem[{{Smith} {et~al.}(2007){Smith}, {Draine}, {Dale},
  {et~al.}}]{smith07pah}
{Smith}, J.~D.~T., {Draine}, B.~T., {Dale}, D.~A., {et~al.} 2007, \apj, 656,
  770

\bibitem[{{Smith} {et~al.}(2001){Smith}, {Brickhouse}, {Liedahl}, \&
  {Raymond}}]{smith01}
{Smith}, R.~K., {Brickhouse}, N.~S., {Liedahl}, D.~A., \& {Raymond}, J.~C.
  2001, \apjl, 556, L91

\bibitem[{{Smol{\v c}i{\'c}} \& {Riechers}(2011)}]{smolcic11}
{Smol{\v c}i{\'c}}, V., \& {Riechers}, D.~A. 2011, \apj, 730, 64

\bibitem[{{Spergel} {et~al.}(2007){Spergel}, {Bean}, {Dor{\'e}},
  {et~al.}}]{spergel07}
{Spergel}, D.~N., {Bean}, R., {Dor{\'e}}, O., {et~al.} 2007, \apjs, 170, 377

\bibitem[{{Stanghellini} {et~al.}(1998){Stanghellini}, {O'Dea}, {Dallacasa},
  {et~al.}}]{stanghellini98}
{Stanghellini}, C., {O'Dea}, C.~P., {Dallacasa}, D., {et~al.} 1998, \aaps, 131,
  303

\bibitem[{{Stern} {et~al.}(2012){Stern}, {Assef}, {Benford},
  {et~al.}}]{stern12}
{Stern}, D., {Assef}, R.~J., {Benford}, D.~J., {et~al.} 2012, \apj, 753, 30

\bibitem[{{Stoughton} {et~al.}(2002){Stoughton}, {Lupton}, {Bernardi},
  {et~al.}}]{stoughton02}
{Stoughton}, C., {Lupton}, R.~H., {Bernardi}, M., {et~al.} 2002, \aj, 123, 485

\bibitem[{{Sutherland} \& {Bicknell}(2007)}]{sutherland07}
{Sutherland}, R.~S., \& {Bicknell}, G.~V. 2007, \apjs, 173, 37

\bibitem[{{Tonry} {et~al.}(2001){Tonry}, {Dressler}, {Blakeslee},
  {et~al.}}]{tonry01}
{Tonry}, J.~L., {Dressler}, A., {Blakeslee}, J.~P., {et~al.} 2001, \apj, 546,
  681

\bibitem[{{Tremblay} {et~al.}(2010){Tremblay}, {O'Dea}, {Baum},
  {et~al.}}]{tremblay10}
{Tremblay}, G.~R., {O'Dea}, C.~P., {Baum}, S.~A., {et~al.} 2010, \apj, 715, 172

\bibitem[{{Valtchanov}(2014)}]{spirehandbook}
{Valtchanov}, I. 2014, The Spectral and Photometric Imaging Receiver Handbook,
  European Space Agency, Madrid, Spain,
  {\url{http://herschel.esac.esa.int/Docs/SPIRE/spire_handbook.pdf}}

\bibitem[{{Varela} {et~al.}(2009){Varela}, {D'Onofrio}, {Marmo},
  {et~al.}}]{varela09}
{Varela}, J., {D'Onofrio}, M., {Marmo}, C., {et~al.} 2009, \aap, 497, 667

\bibitem[{{Wagner} \& {Bicknell}(2011)}]{wagner11}
{Wagner}, A.~Y., \& {Bicknell}, G.~V. 2011, \apj, 728, 29

\bibitem[{{Werner} {et~al.}(2004){Werner}, {Roellig}, {Low},
  {et~al.}}]{werner04}
{Werner}, M.~W., {Roellig}, T.~L., {Low}, F.~J., {et~al.} 2004, \apjs, 154, 1

\bibitem[{{Wright}(2006)}]{wright06}
{Wright}, E.~L. 2006, \pasp, 118, 1711

\bibitem[{{Wright} {et~al.}(2010){Wright}, {Eisenhardt}, {Mainzer},
  {et~al.}}]{wright10}
{Wright}, E.~L., {Eisenhardt}, P.~R.~M., {Mainzer}, A.~K., {et~al.} 2010, \aj,
  140, 1868

\bibitem[{{Wuyts} {et~al.}(2011){Wuyts}, {F{\"o}rster Schreiber}, {van der
  Wel}, {et~al.}}]{wuyts11}
{Wuyts}, S., {F{\"o}rster Schreiber}, N.~M., {van der Wel}, A., {et~al.} 2011,
  \apj, 742, 96

\bibitem[{{Wyder} {et~al.}(2005){Wyder}, {Treyer}, {Milliard},
  {et~al.}}]{wyder05}
{Wyder}, T.~K., {Treyer}, M.~A., {Milliard}, B., {et~al.} 2005, \apjl, 619, L15

\bibitem[{{York} {et~al.}(2000){York}, {Adelman}, {Anderson},
  {et~al.}}]{york00}
{York}, D.~G., {Adelman}, J., {Anderson}, Jr., J.~E., {et~al.} 2000, \aj, 120,
  1579

\end{thebibliography}

\appendix
\section{A. Imaging and SED Details}

This appendix contains UV, optical or NIR, MIR, and FIR images of each of our galaxies (Figures \ref{img1}-\ref{img6}). We preferentially show SDSS images for the optical morphology, but use 2MASS in its absence. Similarly, we show \wise images only in the absence of IRAC imaging, and MIPS only in the absence of \herschel imaging. The optical or NIR image also contains the extraction aperture used as well as the exclusion regions placed on foreground objects or companion galaxies (see \S 2.2.6 for discussion on how these were determined). 

Details of the observational parameters for all the images on which we measured photometry are given in Table \ref{obs_desc}, including an observation ID, the (mean) observation date, and the exposure time. For PACS observations, we note which bands were observed, while for MIPS observations we note which bands we opted to use. This is described in more detail in \S2.2.5. The measured photometry, as well as the additional photometry culled from the literature or IRS enhanced products is given in Table \ref{phot}.

The SEDs constructed from these photometry are shown in Figures \ref{seds1}-\ref{seds8}. For all galaxies, we tried fits both with and without an AGN component. However, we only show the AGN component if it improves the fit. For those galaxies where the fit is improved, we show both the best fit with and without an AGN component. The fitting process is described in \S3, along with some caveats. We note specific concerns with regards to particular galaxies in the comments below. In addition to the fitted SEDs, we show the PDFs for six parameters from the \magphys fits, again showing the results for both the best and non-AGN fits when an AGN component improves the fit.

\subsection{A1. Comments on Individual Galaxies}

\begin{figure*}[b]
\centerline{\includegraphics[width=0.925\linewidth]{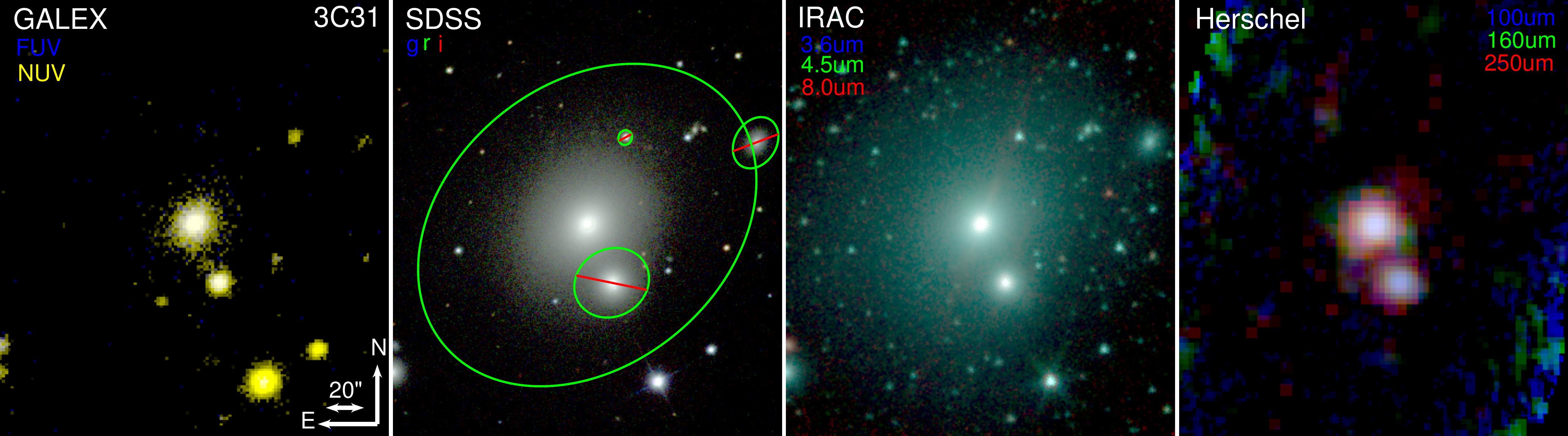}}
\centerline{\includegraphics[width=0.925\linewidth]{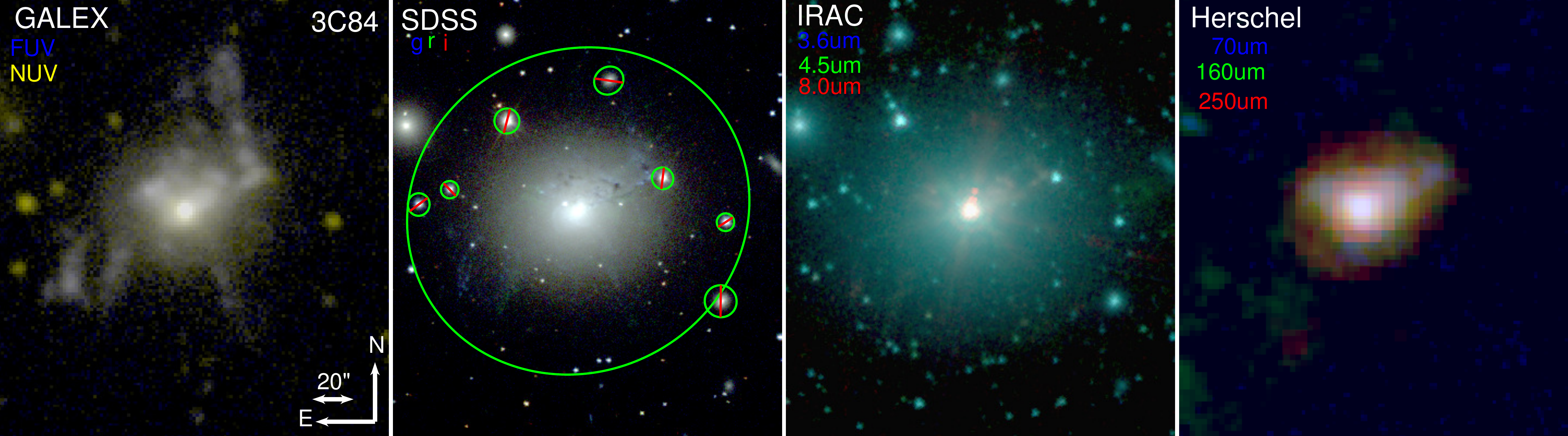}}
\centerline{\includegraphics[width=0.925\linewidth]{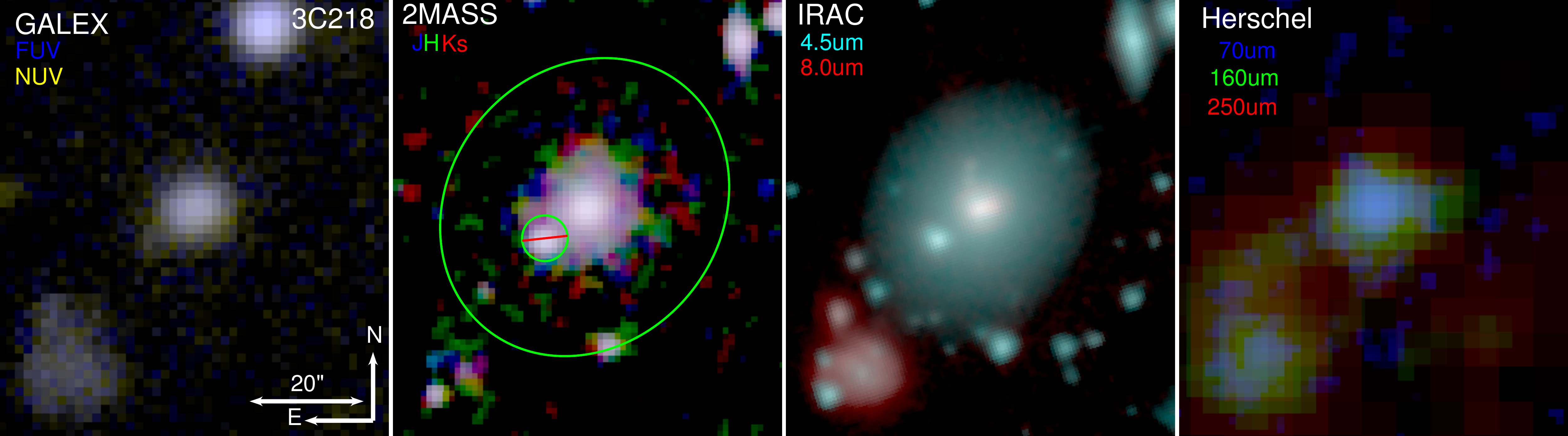}}
\centerline{\includegraphics[width=0.925\linewidth]{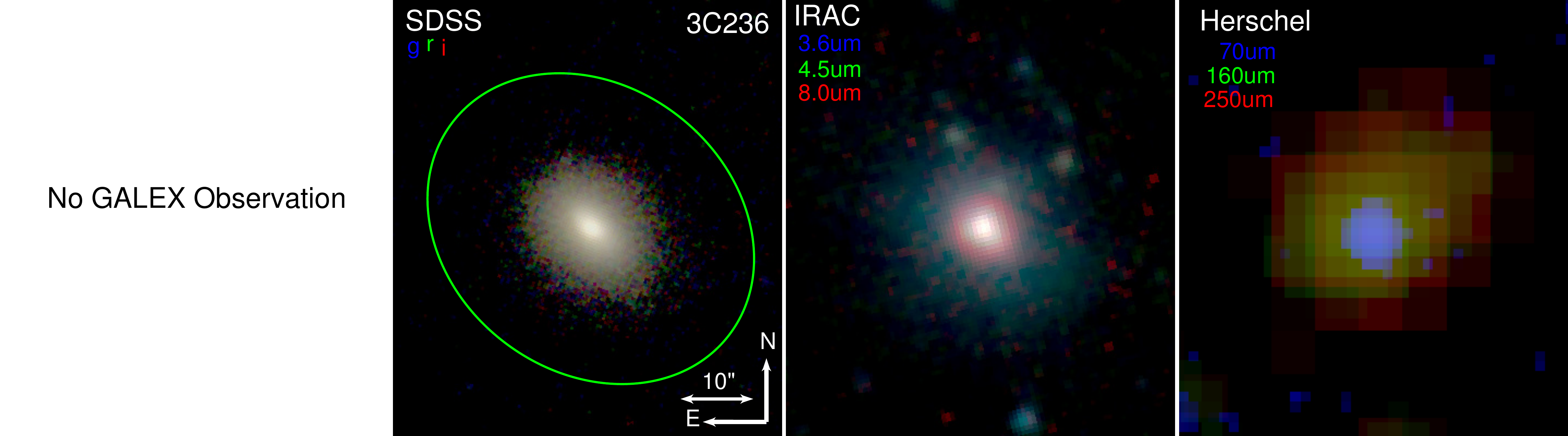}}
\caption{3C\,31, 3C\,84, 3C\,218, and 3C\,236  (from top to bottom) as observed by (from left to right) (1) \galex (NUV in yellow; FUV in blue), (2) SDSS or 2MASS ($g/J$ in blue, $r/H$ in green, and $i/Ks$ in red), (3) IRAC or \wise (3.6\um/3.4\um in blue, 4.5\um/4.6\um in green, and 8.0\um/12.0\um in red), and (4) \herschel (PACS70\um/100\um in blue, PACS160\um in green, and SPIRE 250\um in red) or MIPS (24\um in blue, 70\um in green, and 160\um in red). In the optical/NIR image, the extraction aperture and exclusion regions (those with a red line through them) are shown. Section 2.2.6 describes how these were determined. }
\label{img1}
\end{figure*}

\noindent
\textbf{3C\,31:} The top panel of Figure \ref{img1} shows that 3C\,31 (NGC\,383) has a close companion (NGC\,382). Due to the relative proximity of 3C\,31, these galaxies can be disentangled even at the resolution of SPIRE 500\um. Emission from its jet was previously detected by \citet{lanz11} at IRAC wavelengths but the contribution to the integrated SED is very small. A modest AGN contribution improves the SED fit (Figure \ref{seds1}), particularly in the FIR. Its IR spectrum has a normal PAH ratio but it has a silicate absorption feature  that we do not model \citep{ogle10}. \vspace{2mm}

\noindent
\textbf{3C\,84}: The brightest cluster galaxy (BCG) of the Perseus cluster, 3C\,84 (NGC\,1275) has significant filamentary structure in the UV (Figure \ref{img1}). Its MIR (e.g. IRAC 8\um) and FIR are dominated by the central region, although the PACS images may show some extended emission, particularly toward the northwest. Its SED (Fig. \ref{seds1}) shows significant MIR emission attributed to an AGN, which significantly improves the fit, as evidenced by the tighter PDFs, although we do not model the silicate emission at 10\um \citep{ogle10}. The IR luminosity of its host galaxy indicates that it is a LIRG. However, the SPIRE bands, especially at 500\um, show excess over the expected Rayleigh-Jeans dust emission, which is possibly due to synchrotron emission in this strong radio source. However, the SED analysis of \citet{leipski09} concluded that there was little synchrotron contribution to the MIR/FIR bump.  It is one of three galaxies whose molecular mass is estimated from a CO(2--1) observation. \vspace{2mm}

\noindent
\textbf{3C\,218}: 3C\,218 (Hydra A) is the BCG of the Abell\,780 cluster. It was not observed by SDSS, but we found {\em B} and {\em V} in the literature, and it was only observed in two IRAC bands. In addition to a small nearby companion visible in the 2MASS and IRAC images, but contributing little at UV and FIR wavelengths, there is another galaxy $\sim30''$ to the south--east, which is bright in the UV, MIR, and FIR. At PACS wavelengths, there exists a possible dusty bridge between these two galaxies. At the resolution of SPIRE 500\um, these sources begin to blend, so that photometry should be used with caution. Additionally, the longer-wavelength SPIRE bands show possible synchrotron contamination similar to 3C\,84. A MIR AGN component did little to improve the fit of this SED (Fig. \ref{seds1}), which is generally well fit by \magphys, which is consistent with its star formation-dominated IRS spectrum \citep{ogle10}. Its CO observations were performed with a single dish telescope (IRAM\,30m), so the extent of its molecular disk was not determined. We therefore estimate the extent of its star-forming/molecular disk for the extent of its 8\um (i.e. PAH) emission. \\\\\\ \vspace{2mm}

\begin{figure*}[b]
\centerline{\includegraphics[width=0.925\linewidth]{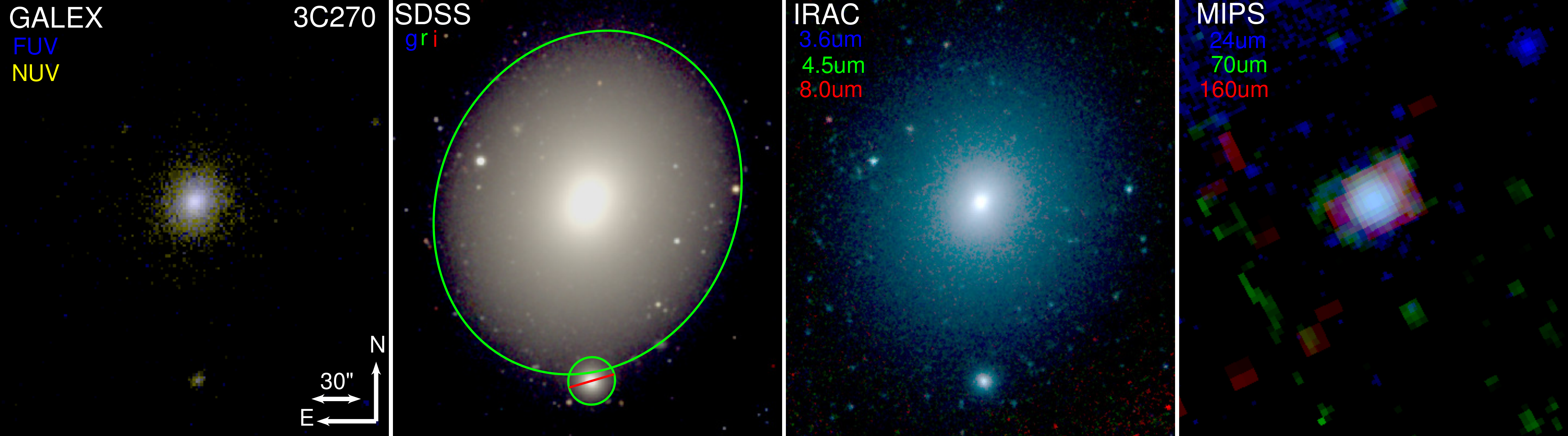}}
\centerline{\includegraphics[width=0.925\linewidth]{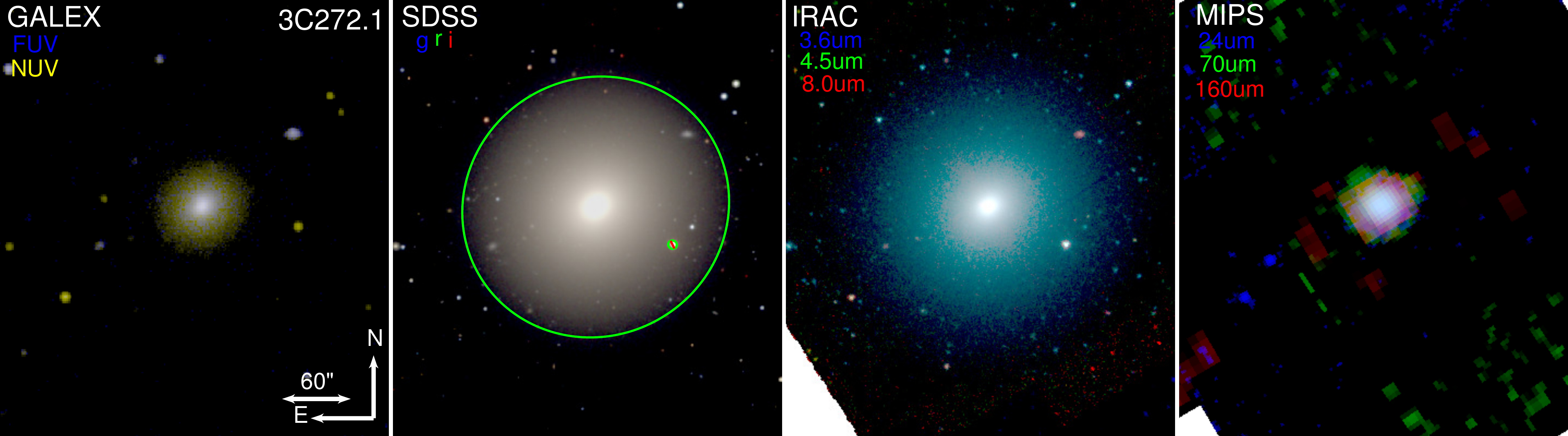}}
\centerline{\includegraphics[width=0.925\linewidth]{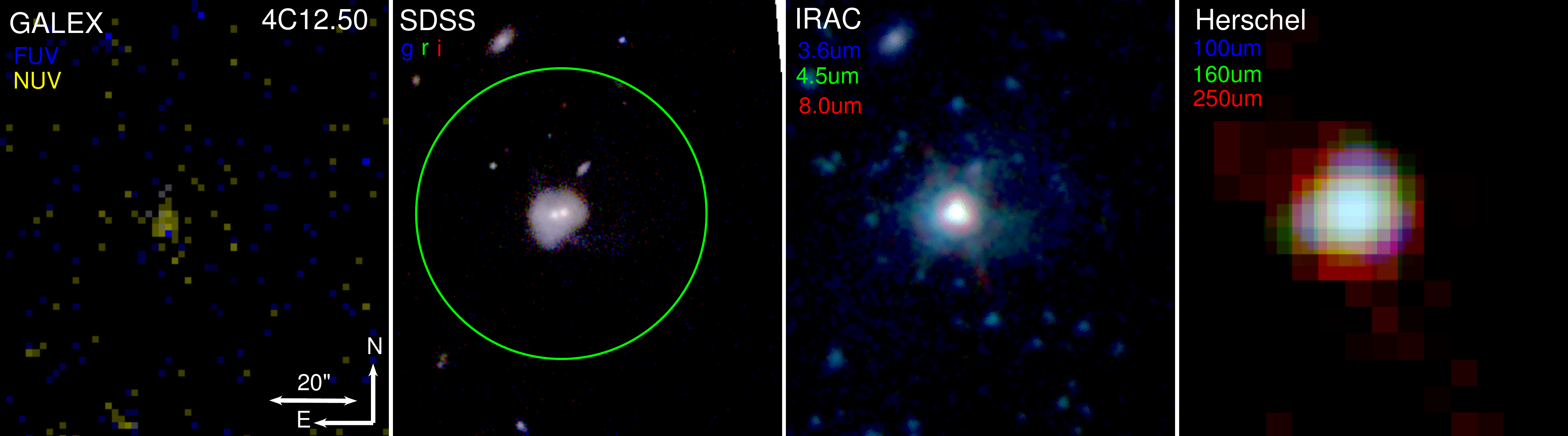}}
\centerline{\includegraphics[width=0.925\linewidth]{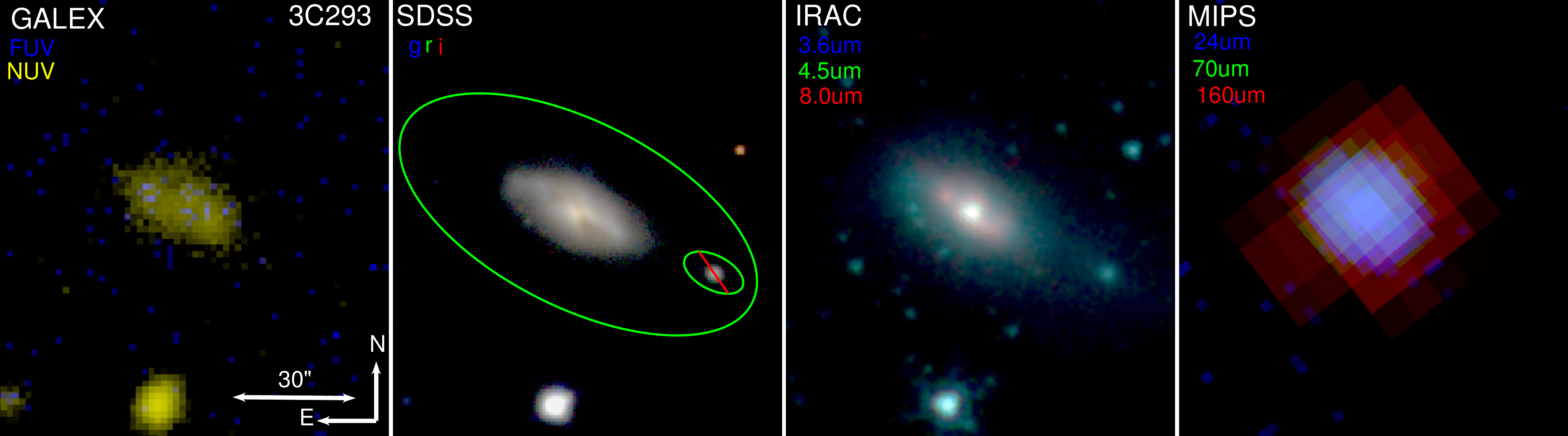}}
\caption{Panchromatic images of 3C\,270, 3C\,272.1, 4C\,12.50, and 3C\,293. See Fig. \ref{img1} for further details. }
\label{img2}
\end{figure*}

\begin{figure*}[b]
\centerline{\includegraphics[width=0.925\linewidth]{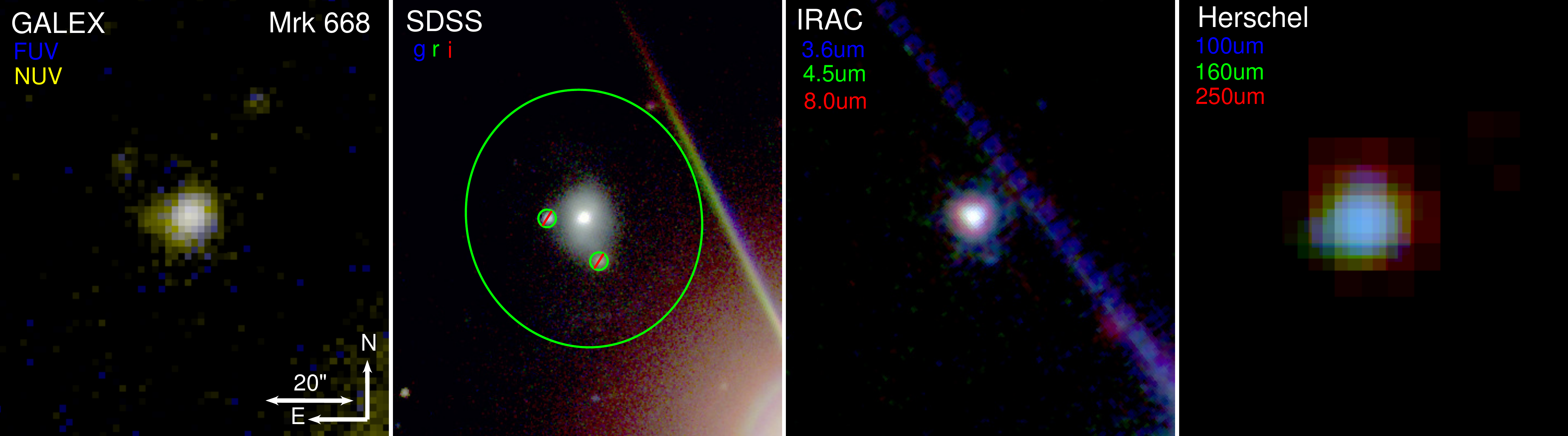}}
\centerline{\includegraphics[width=0.925\linewidth]{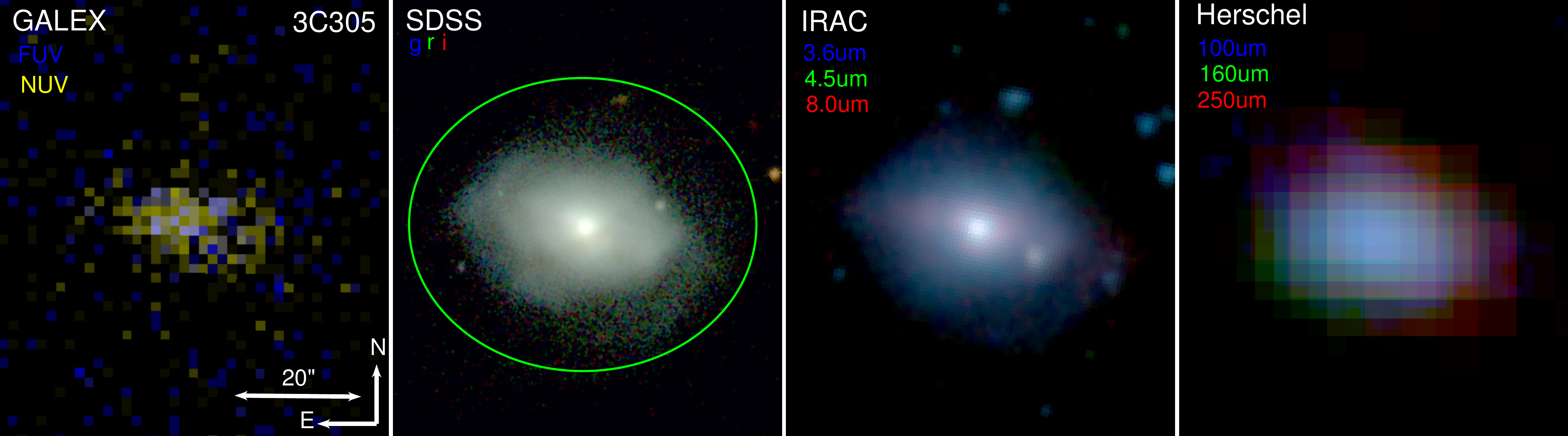}}
\centerline{\includegraphics[width=0.925\linewidth]{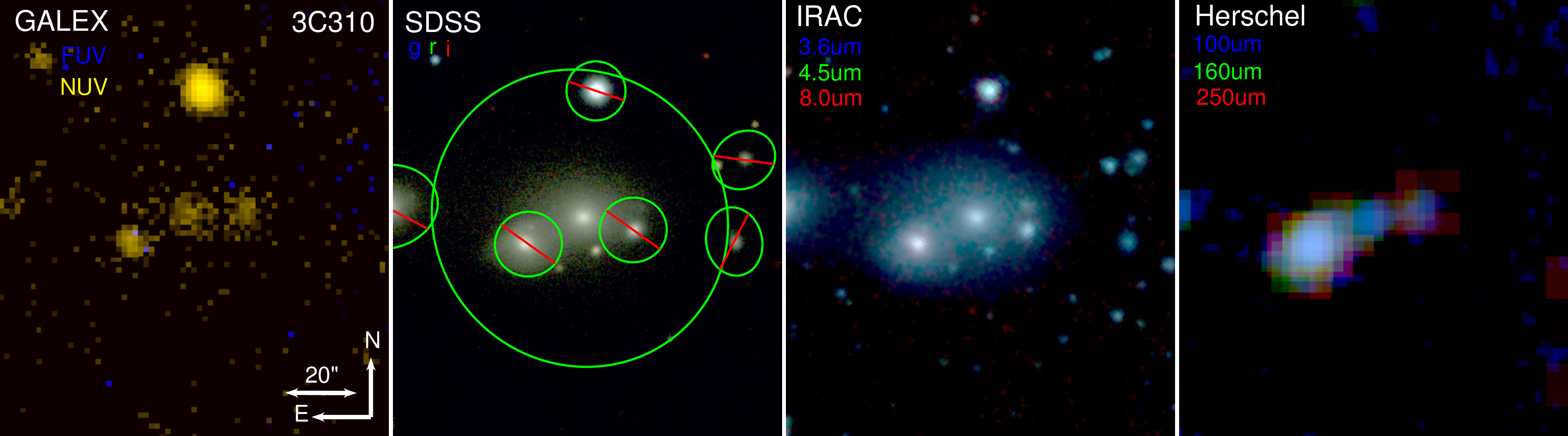}}
\centerline{\includegraphics[width=0.925\linewidth]{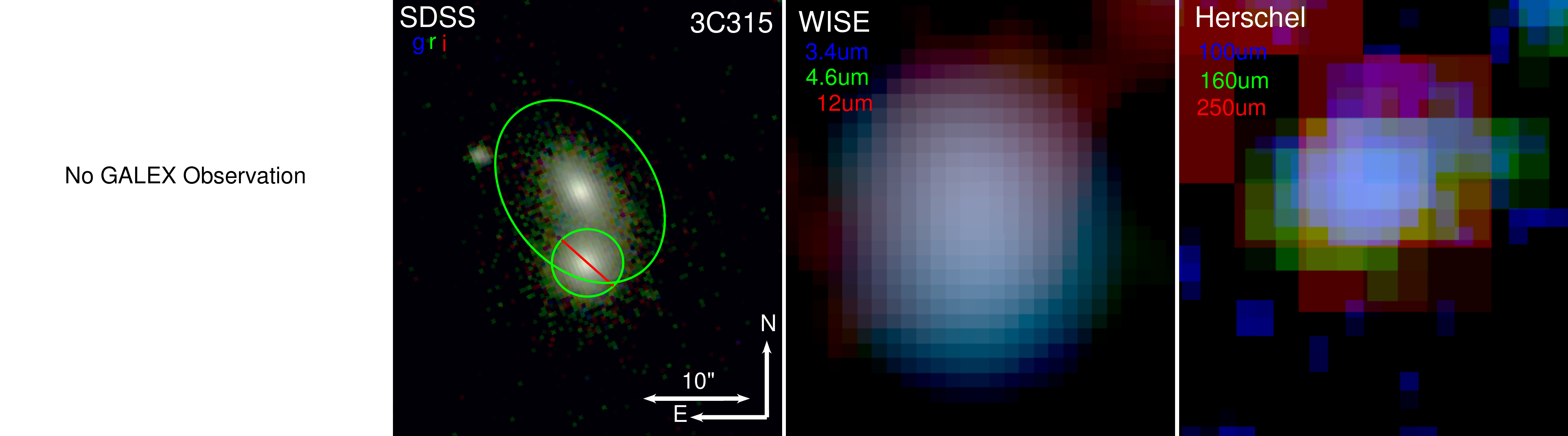}}
\caption{Panchromatic images of Mrk\,668,  3C\,305, 3C\,310, and 3C\,315. For 3C\,310, the emission at FIR wavelengths is dominated by the eastern companion, which becomes difficult to disentangle at wavelengths longer than 100\um, so we could only determine upper limits for those bands. See Fig. \ref{img1} for further details. }
\label{img3}
\end{figure*}

\begin{figure*}[b]
\centerline{\includegraphics[width=0.925\linewidth]{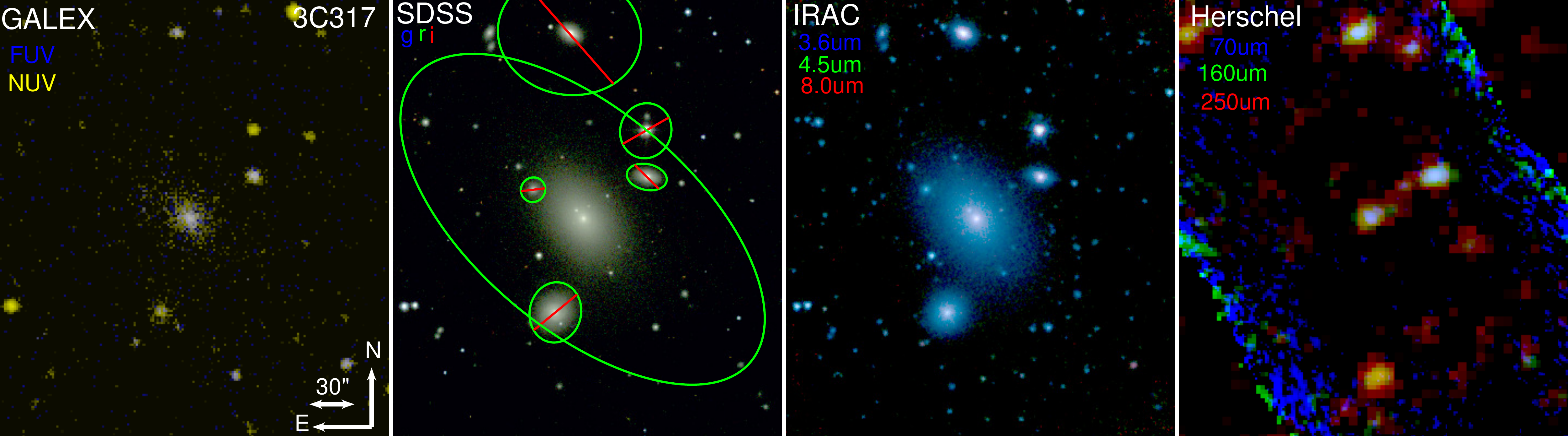}}
\centerline{\includegraphics[width=0.925\linewidth]{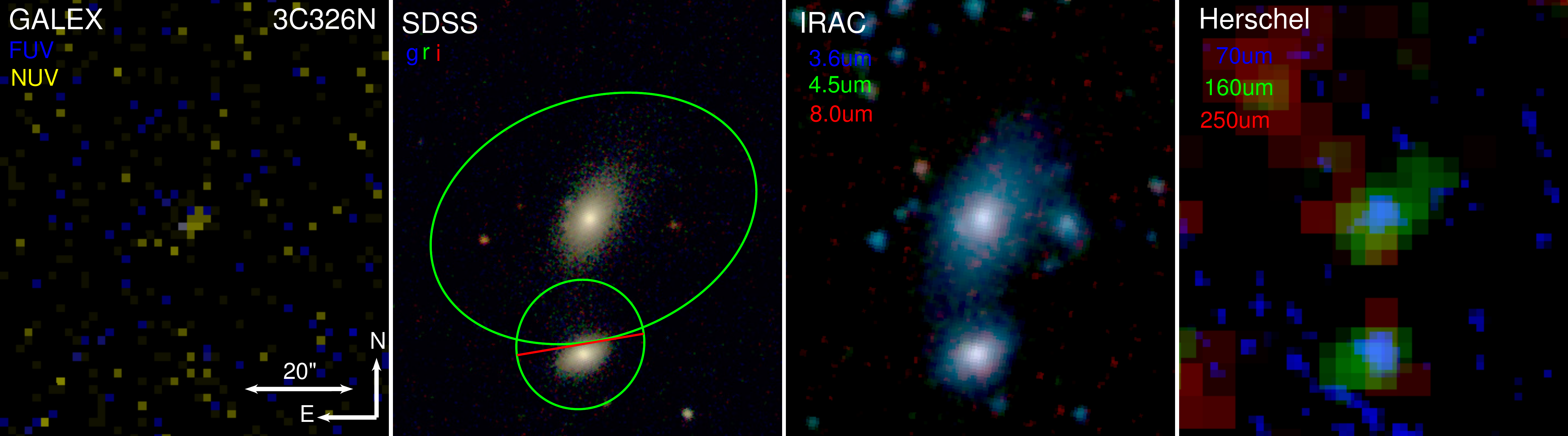}}
\centerline{\includegraphics[width=0.925\linewidth]{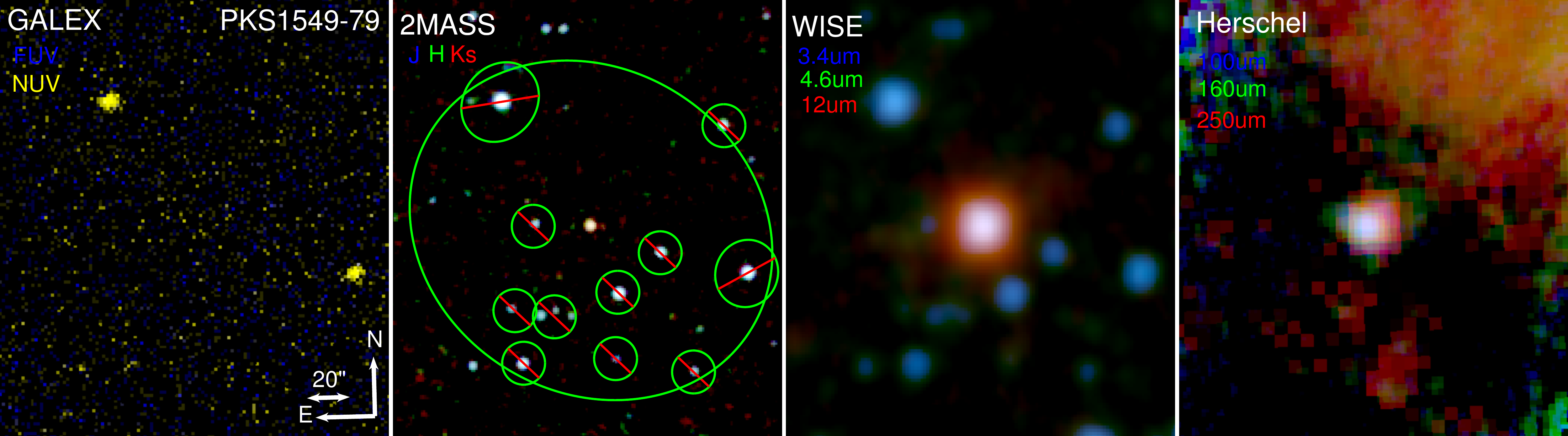}}
\centerline{\includegraphics[width=0.925\linewidth]{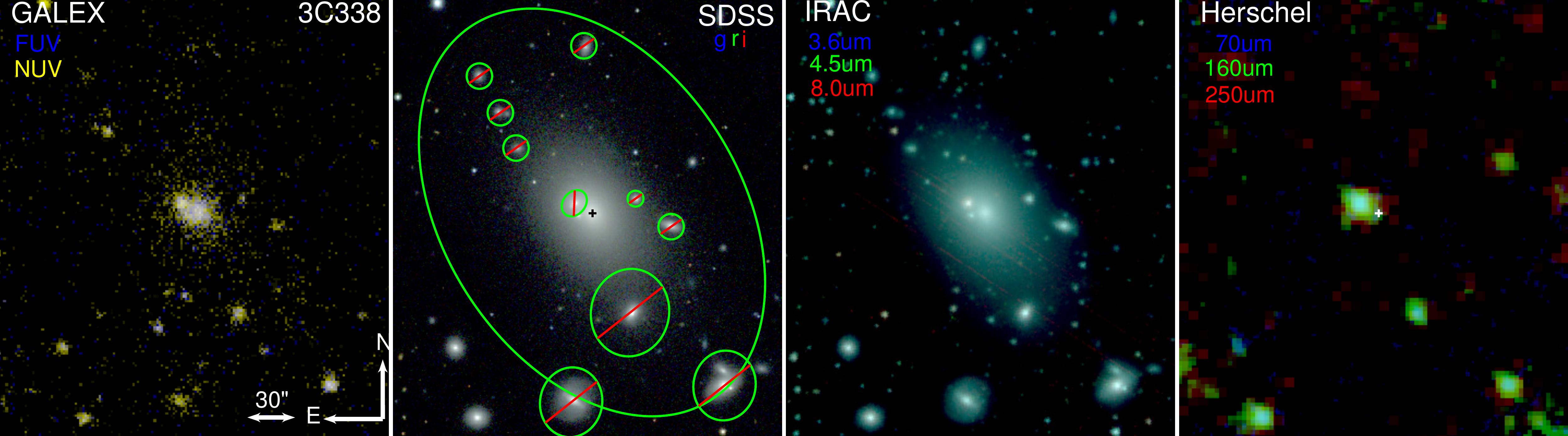}}
\caption{Panchromatic images of 3C\,317, 3C\,326N, PKS\,1549-79, and 3C\,338. For 3C\,317, the emission at FIR wavelengths is approximately equally dominated by 3C\,317 and the northwest companion; it becomes difficult to disentangle them at the longer SPIRE wavelengths.  The emission in the vicinity of 3C\,338 at \herschel wavelengths is dominated by the northeastern companion; the black cross on the SDSS image is in the same location as the white cross on the \herschel image. See Fig. \ref{img1} for further details. }
\label{img4}
\end{figure*}

\begin{figure*}[b]
\centerline{\includegraphics[width=0.925\linewidth]{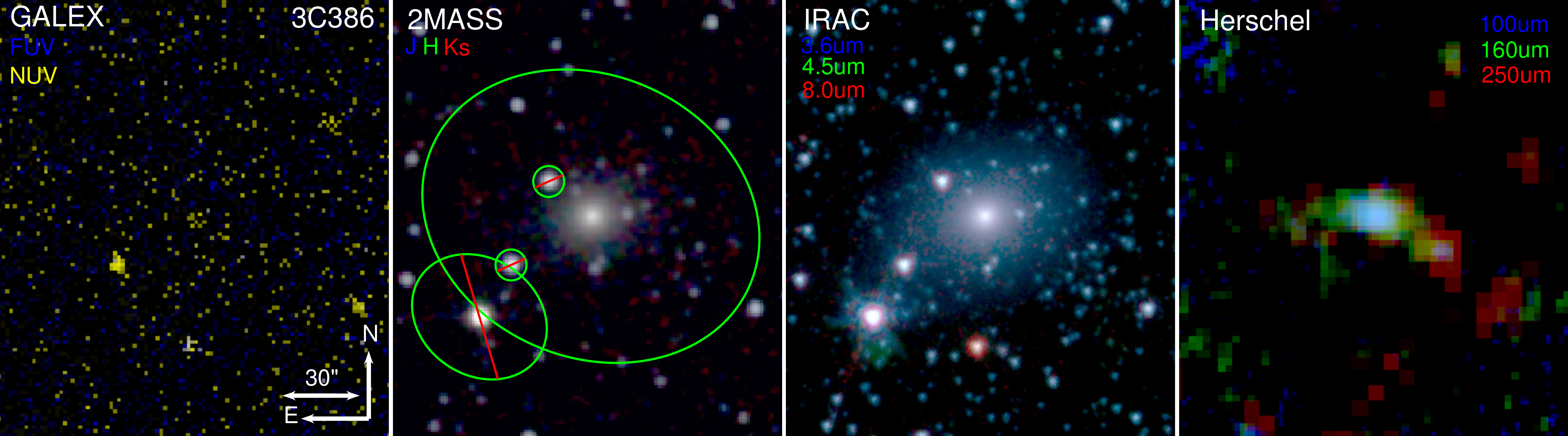}}
\centerline{\includegraphics[width=0.925\linewidth]{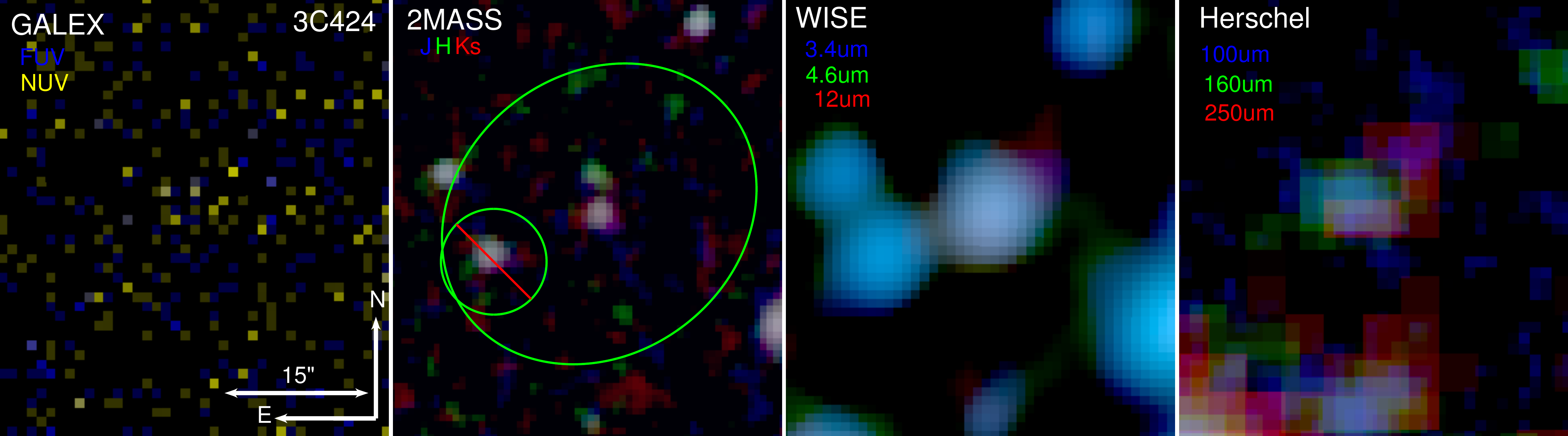}}
\centerline{\includegraphics[width=0.925\linewidth]{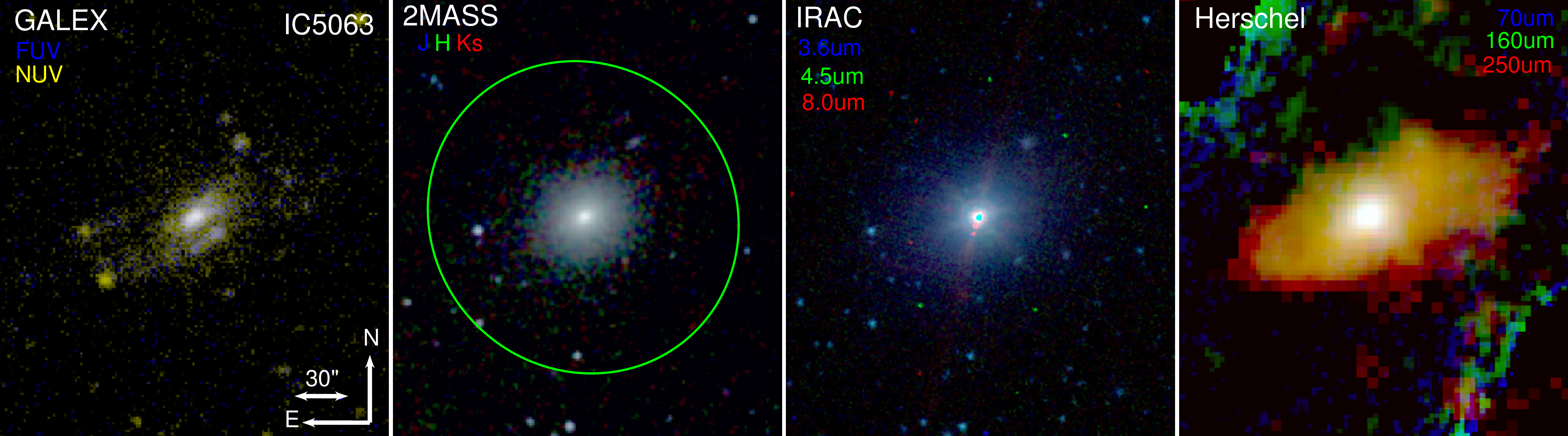}}
\caption{Panchromatic images of 3C\,386, 3C\,424, and IC\,5063. IC\,5063 is saturated at 8\um. See Fig. \ref{img1} for further details. }
\label{img5}
\end{figure*}

\begin{figure*}[b] 
\centerline{\includegraphics[width=0.925\linewidth]{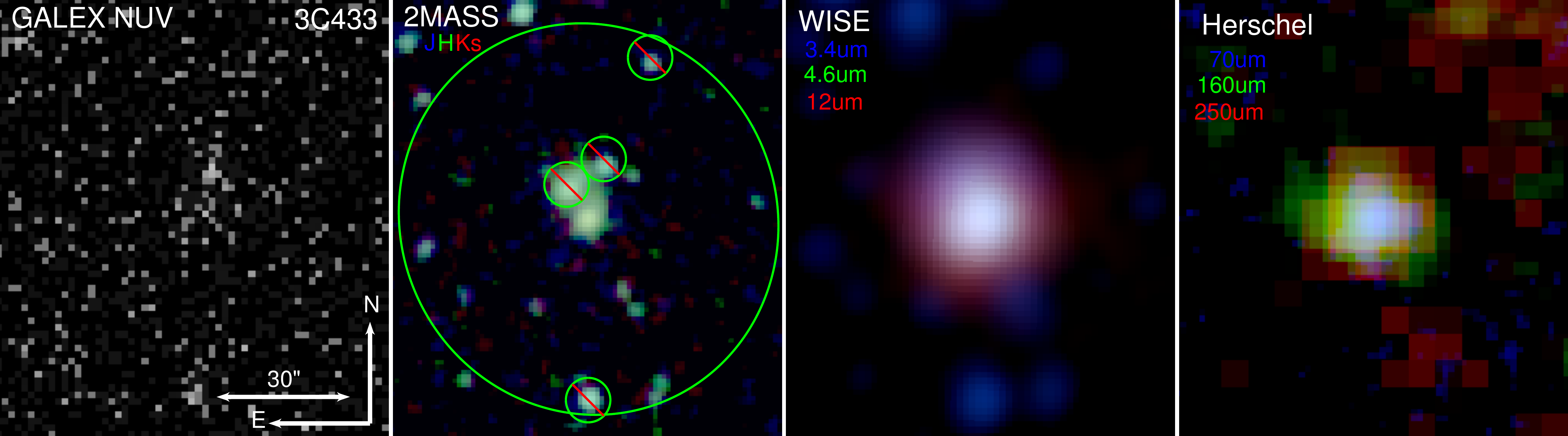}}
\centerline{\includegraphics[width=0.925\linewidth]{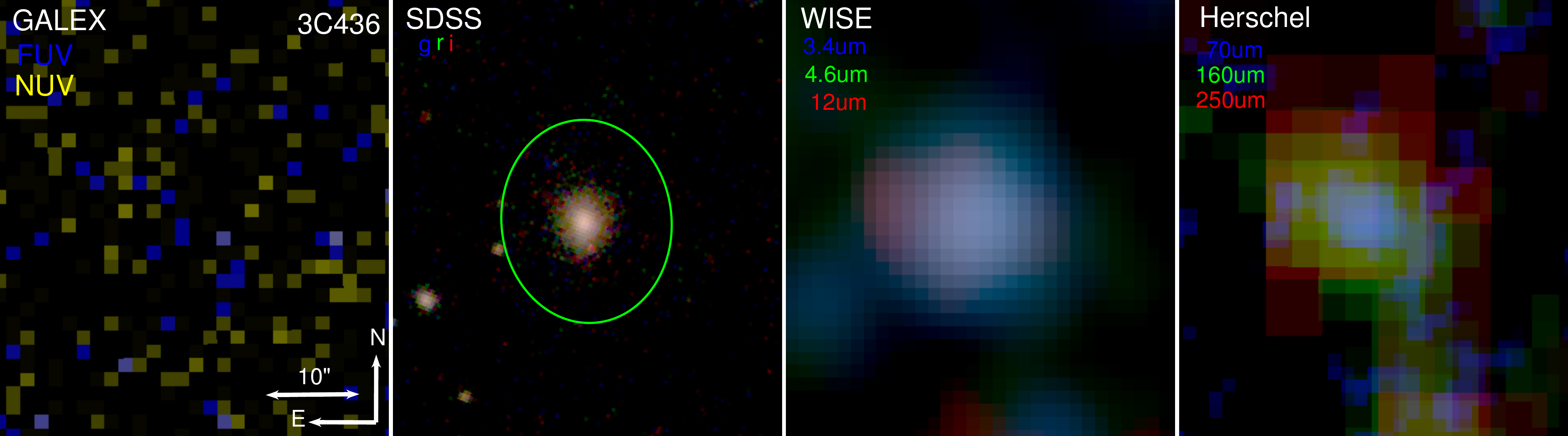}}
\centerline{\includegraphics[width=0.925\linewidth]{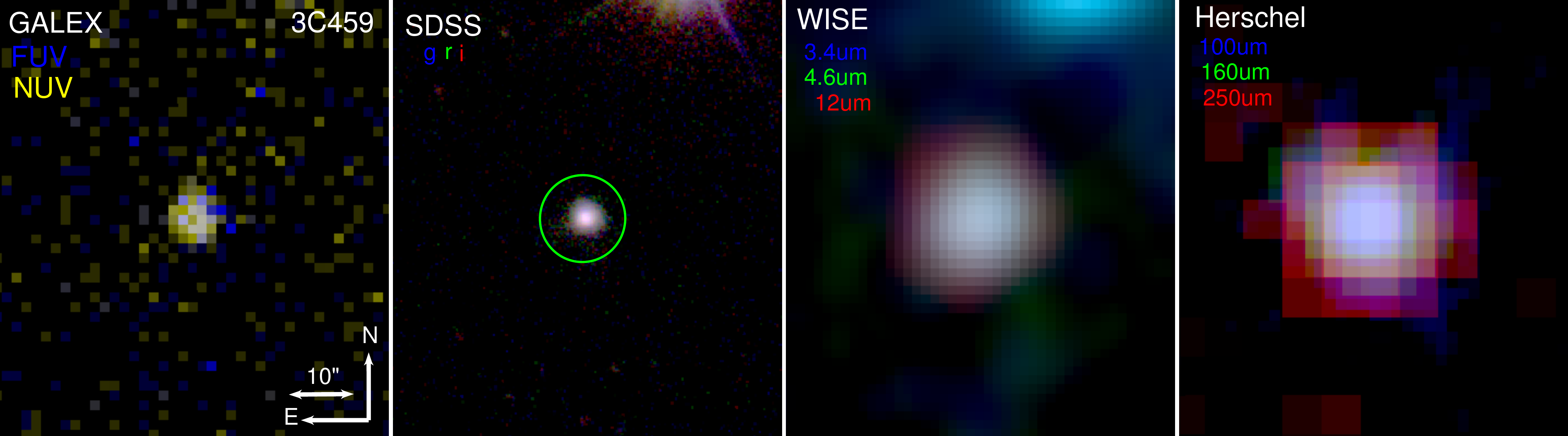}}
\caption{Panchromatic images of 3C\,433, 3C\,436, and 3C\,459. 3C\,433 was only observed in the NUV band with \galex. See Fig. \ref{img1} for further details. }
\label{img6}
\end{figure*}

\noindent
\textbf{3C\,236}: No GALEX observations exist of 3C\,236; however, we found UV photometry at very similar wavelengths in the literature. An isolated galaxy (Fig. \ref{img1}), its SED (Fig. \ref{seds2}) shows a  MIR excess requiring an AGN component. While this fit is an improvement overall, it is worse in the FUV and the shorter-wavelength PACS bands. There is also the possibility of synchrotron contribution at 500\um. \vspace{2mm}

\noindent
\textbf{3C\,270}: 3C\,270 (NGC\,4261; Fig. \ref{img2}) is a member of the Virgo cluster and is one of the three galaxies in our sample not observed by \herschel. As a result, its FIR SED (Fig. \ref{seds2}) is poorly sampled and is not very well fit even with the inclusion of an AGN. We are not convinced of the necessity of an AGN component in this SED fit, so we show both possibilities. Given the poor fit in FIR, we have concerns regarding the reliability of the derived parameters and use them with caution. Due to the proximity of this galaxy, the IRS slit only covers a small portion of this galaxy, so its \mh luminosity should be considered a lower limit. \\  \vspace{2mm}

\noindent
\textbf{3C\,272.1}: 3C\,272.1 (M84; Fig. \ref{img2}) is also a member of the Virgo cluster, and lacks \herschel imaging. Like 3C270, its FIR SED, although better defined, is poorly fit and the inclusion of an AGN, while shown in Figure \ref{seds2}, is not very convincing. The inclusion of this component improves the fit at 60-70\um, but worsens it at 100\um, and neither the FUV nor the MIPS 24\um is not well modeled by either fit. Therefore, we have concerns regarding the reliability of the derived parameters and use them with caution. Its proximity means that its IRS spectrum only comes from a small fraction of the galaxy, but it shows strong 11\um PAH emission \citep{ogle10}, explaining the 12\um bump in the SED that we do not match well. \\\\\\ \vspace{2mm}

\noindent
\textbf{4C\,12.50}: 4C\,12.50 (PKS\,1345+12) is a merging system with barely resolved centers in the SDSS images (Fig. \ref{img2}). At other wavelengths, it is typically unresolved. Its SED (Fig. \ref{seds3}) shows a strong IR bump. Indeed, it is one of three ULIRGs in our sample. Its fit is improved with the inclusion of a MIR AGN component; however the shape of the model in the FIR suggests this galaxy might be better fit with a combination of dust temperatures not currently implemented in \magphys (i.e. dust temperatures warmer than 60\,K). \\\\\\ \vspace{2mm}

\noindent
\textbf{3C\,293}: 3C\,293 (UGC\,8782) has a nearby companion to the southwest (Fig. \ref{img2}), which we exclude in measuring the photometry. Although 3C\,293 lacks \herschel imaging, its FIR SED (Fig. \ref{seds3}) is well constrained by IRAS and MIPS. Its fit is improved with a modest AGN. The inclusion of this component drives the \magphys fit  to slightly warmer cold dust temperatures and therefore a smaller derived dust mass than the non-AGN purely \magphys fit done in \citet{lanz15}. The derived SFR is likewise a little smaller, but agrees with the rate from \citet{lanz15} within the uncertainties. \vspace{2mm}

\noindent
\textbf{Mrk\,668}: Mrk\,668 (OQ\,208) has a nearby, bright foreground star (Fig. \ref{img3}), so care was taken to select a background region with similar levels of scattered optical light. Similarly, we added an additional exclusion region for the diffraction spike from that star in the IRAC images. Mrk\,668 has a strong MIR AGN contribution (Fig. \ref{seds3}). Even after the removal of this component, the remaining IR emission is sufficient to classify this galaxy as a LIRG. Its IR spectrum shows emission from the 10\um silicate feature \citep{guillard12}. \vspace{2mm}

\noindent
\textbf{3C\,305}: 3C\,305 (IC\,1065; Fig. \ref{img3}) has only been observed in CO with the IRAM\,30m, which measured flux but not extent. To estimate the extent of the star-forming and molecular disk, we use the size of the 8\um emission as a proxy for star-forming disk. Its SED (Fig. \ref{seds4}) is best fit including an AGN component. \vspace{2mm}

\noindent
\textbf{3C\,310}: Figure \ref{img3} shows that 3C\,310 (VV\,204b) lies at the center of a cluster with several nearby galaxies. We have excluded them as shown on the SDSS image. However, due to the proximity of these companions, we are likely excluding some of the source flux and possibly retaining some contamination from the companions. At UV--MIR wavelengths, these galaxies are typically resolved and 3C\,310 typically dominates the emission, so we will use the fluxes as measured. However, at FIR wavelengths, the situation becomes more complicated, because 3C\,310 no longer dominates the emission and the sources become increasingly blended with increasing wavelength. Therefore, while we do obtain a detection in the PACS\,160\um-SPIRE\,350\um bands, we do not find these fluxes to be trustworthy. Therefore, for these bands, we measured the total flux in the aperture without the exclusion regions and treat them as upper limits. As a result, the FIR SED (Fig. \ref{seds4}) of 3C\,310 is poorly constrained and we use its derived parameters with caution. \vspace{2mm}

\noindent
\textbf{3C\,315}: 3C\,315 was not observed with GALEX and has a nearby companion resolved by SDSS (Fig. \ref{img3}). At IR wavelengths, it only has \wise and \herschel imaging, so resolving the galaxies becomes very difficult. However,  the 12\um \wise image and the \herschel images strongly suggest that the MIR--FIR emission of this system is dominated by our host galaxy instead of its companion. Its SED (Fig. \ref{seds4}) does not require an AGN component, which is consistent with its IRS spectrum that show strong PAH emission from star forming activity \citep{ogle10}. \vspace{2mm}

\noindent
\textbf{3C\,317}: 3C\,317 (UGC\,9799) is the BCG of Abell 2052, and as such has a number of close companions (Fig. \ref{img4}). At FIR wavelengths, the emission of one of its smaller companions (to the north--west) becomes pronounced, so we took care to ensure that the exclusion region was sufficiently large to also include the \herschel emission. Its SED (Fig. \ref{seds5}) fit does not improve with the inclusion of an AGN component. \vspace{2mm}

\noindent
\textbf{3C\,326N}: 3C\,326N has a companion, which is resolved at all \herschel wavelengths (Fig. \ref{img4}). At PACS wavelengths, there is a hint of extended emission along the major axis of this galaxy. Its SED (Fig. \ref{seds5}) does not require an AGN to fit. Its \spitzer observations were discussed in detail by \citet{ogle07} who measured SFR from the 7.7\um PAH feature consistent with our SED-derived SFR.      \vspace{2mm}

\noindent
\textbf{PKS\,1549-79}: PKS\,1549-79 is a ULIRG, whose SED  (Fig. \ref{seds5}) is dominated by its IR emission. Its images (Fig. \ref{img4}) likewise show that it is much dimmer at UV--NIR wavelengths than in the MIR--FIR. Due to its declination, it is too far south to fall in the SDSS footprint and as a result, its UV--optical SED is poorly defined. Therefore, we treat its stellar mass with great caution (the PDF of $M_{*}$ is also quite broad). PKS\,1549-79 has a quite strong MIR AGN component.  In the \herschel bands, significant diffuse foreground emission can be seen in the top-right corner of the image, requiring the use of a point source aperture at those wavelengths to minimize contamination. \\\\ \vspace{2mm}

\begin{figure*}[b]
\centerline{\includegraphics[width=0.925\linewidth]{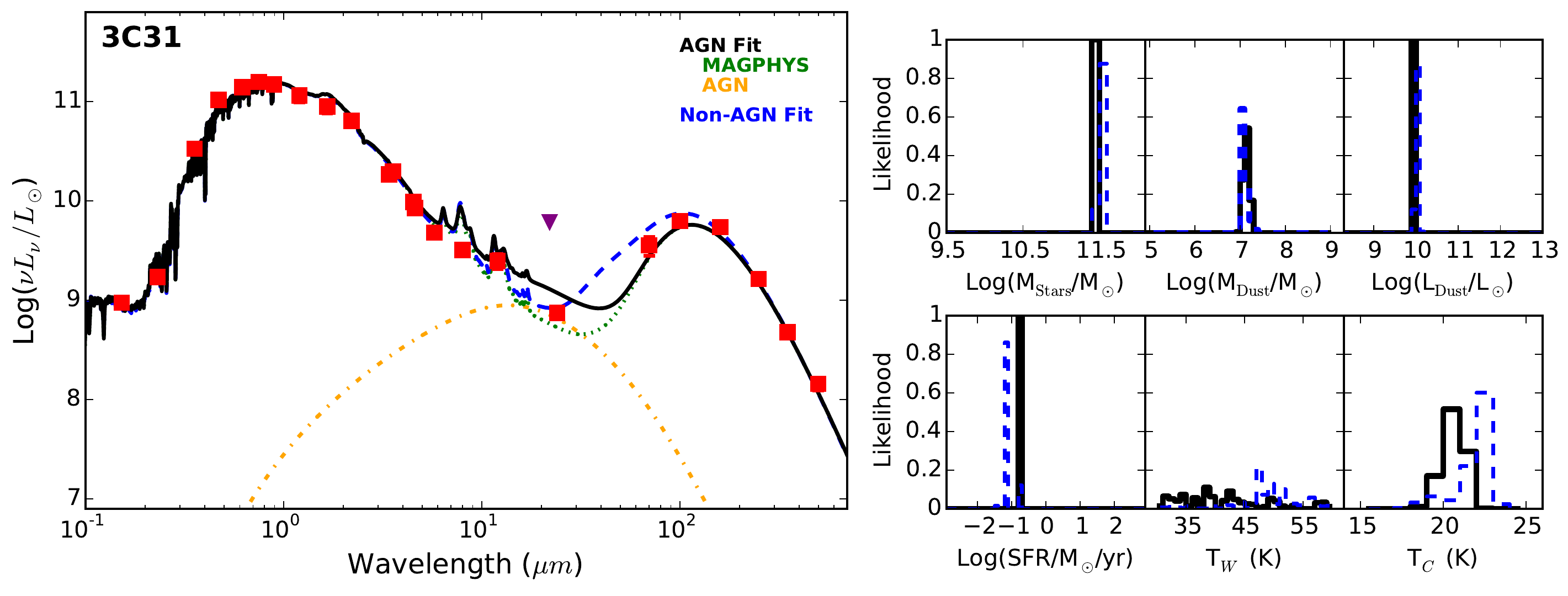}}
\centerline{\includegraphics[width=0.925\linewidth]{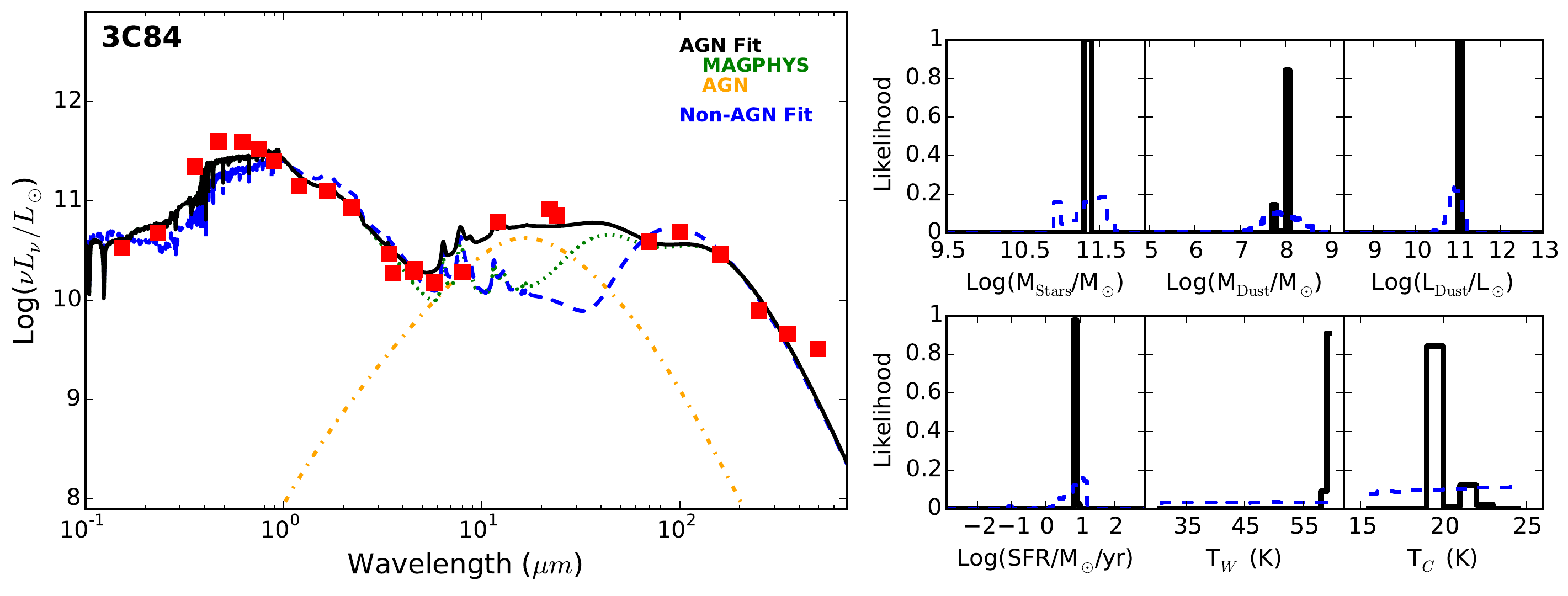}}
\centerline{\includegraphics[width=0.925\linewidth]{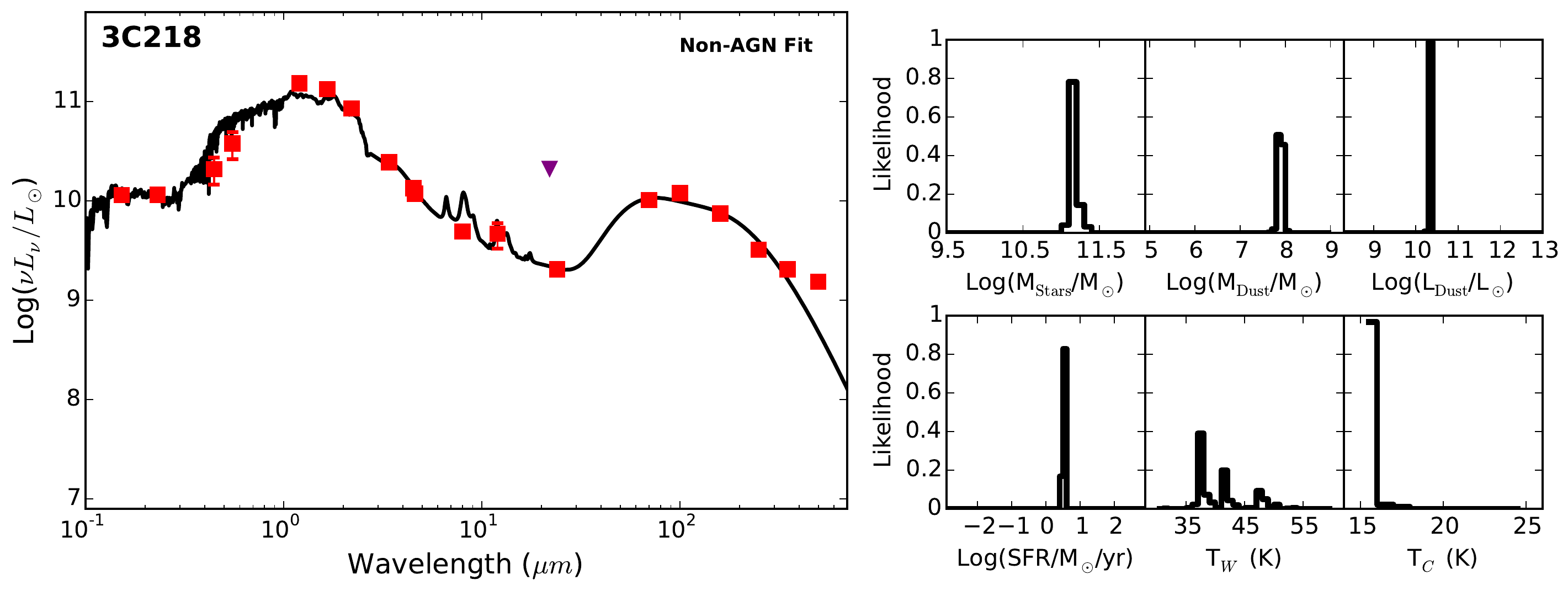}}
\caption{SEDs for 3C\,31 (top), 3C\,84 (middle), and 3C\,218 (bottom) 
with photometry shown as red squares (purple triangles are upper limits) and the best fit model plotted in black. When an AGN component is necessary to improve the fit in the MIR, we also show the AGN component (orange dash--dot line) and the host component (green dotted line), as well as the best fit without an AGN (blue dashed line).  To the right of the SED, we plot a subset of the PDFs of the fitted parameters for (from left to right): stellar mass,  dust mass, and dust luminosity (top) and SFR, warm dust temperature, and cold dust temperature (bottom). When an AGN is needed, we show the PDFs both for the best fit (black) and for the fit without an AGN (blue dashed).}
\label{seds1}
\end{figure*}

\begin{figure*}[b]
\centerline{\includegraphics[width=0.925\linewidth]{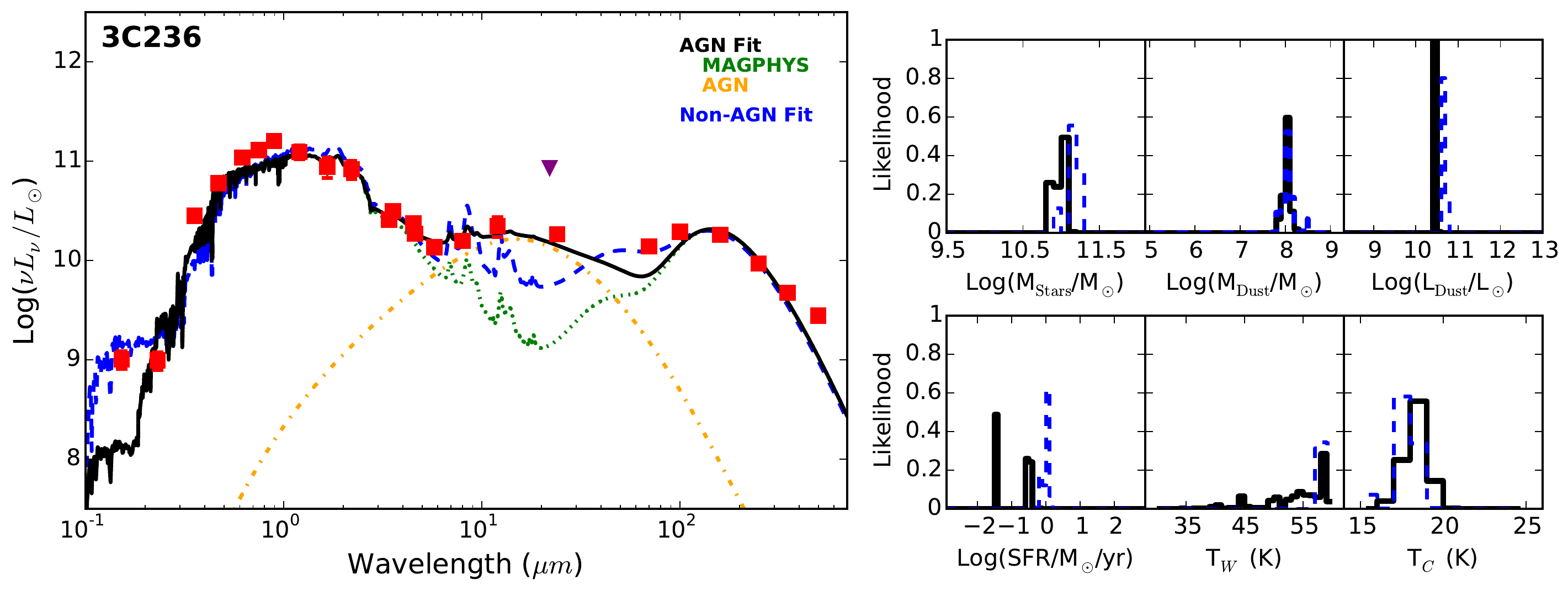}}
\centerline{\includegraphics[width=0.925\linewidth]{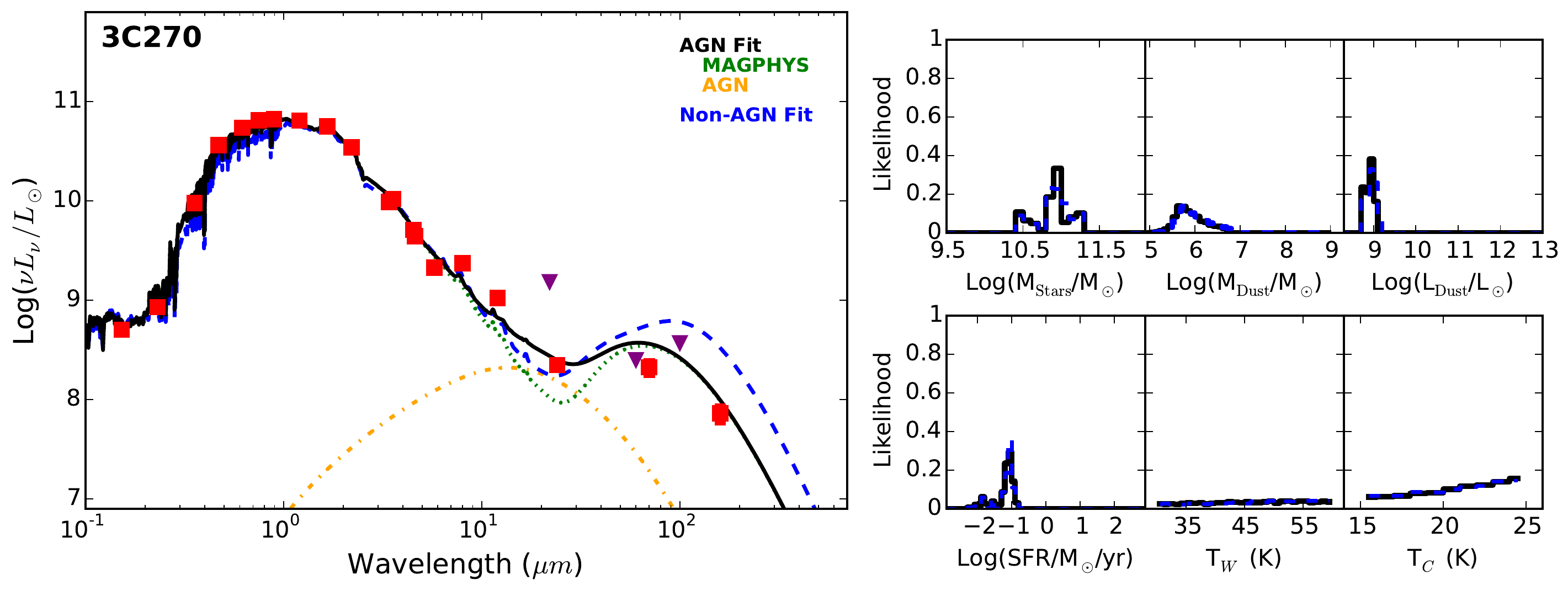}}
\centerline{\includegraphics[width=0.925\linewidth]{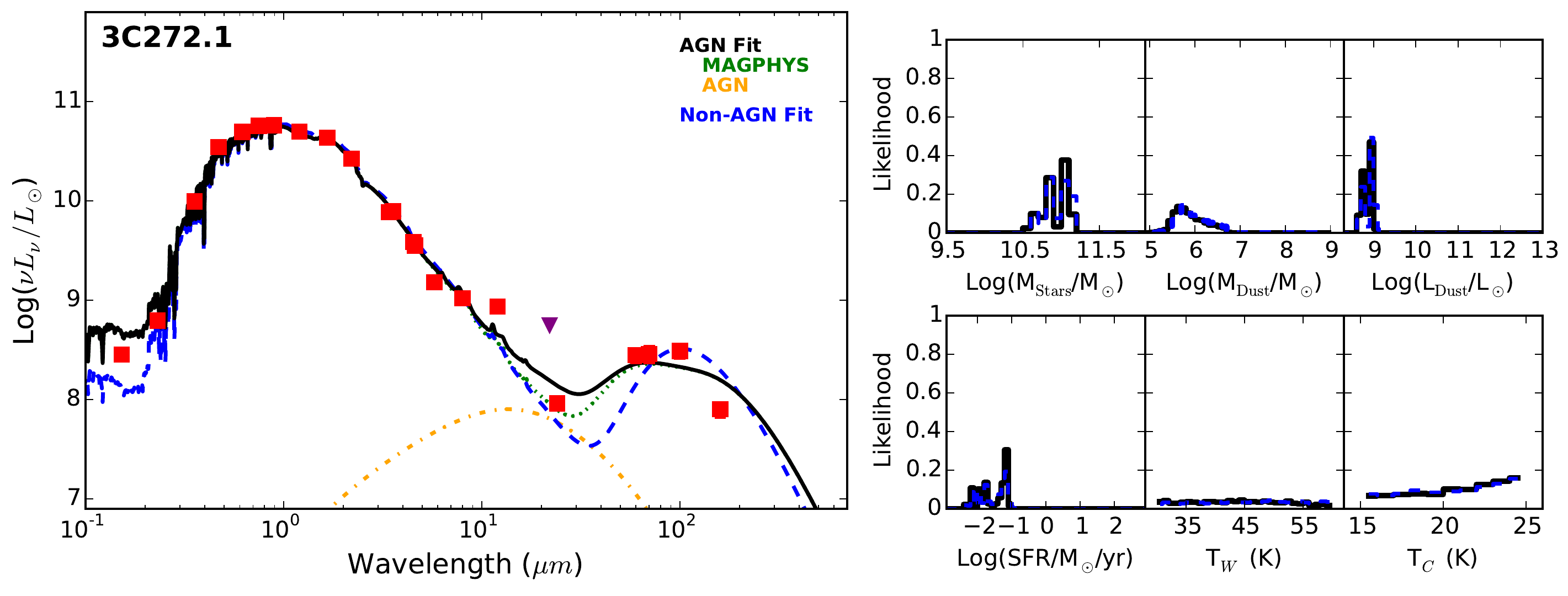}}
\caption{SEDs for 3C\,236 (top), 3C\,270 (middle), and 3C\,272.1 (bottom) 
with photometry shown as red squares (or purple triangles for upper limits) and the best fit model plotted in black. Further details are given in Fig. \ref{seds1} captions.}
\label{seds2}
\end{figure*}

\begin{figure*}[b]
\centerline{\includegraphics[width=0.925\linewidth]{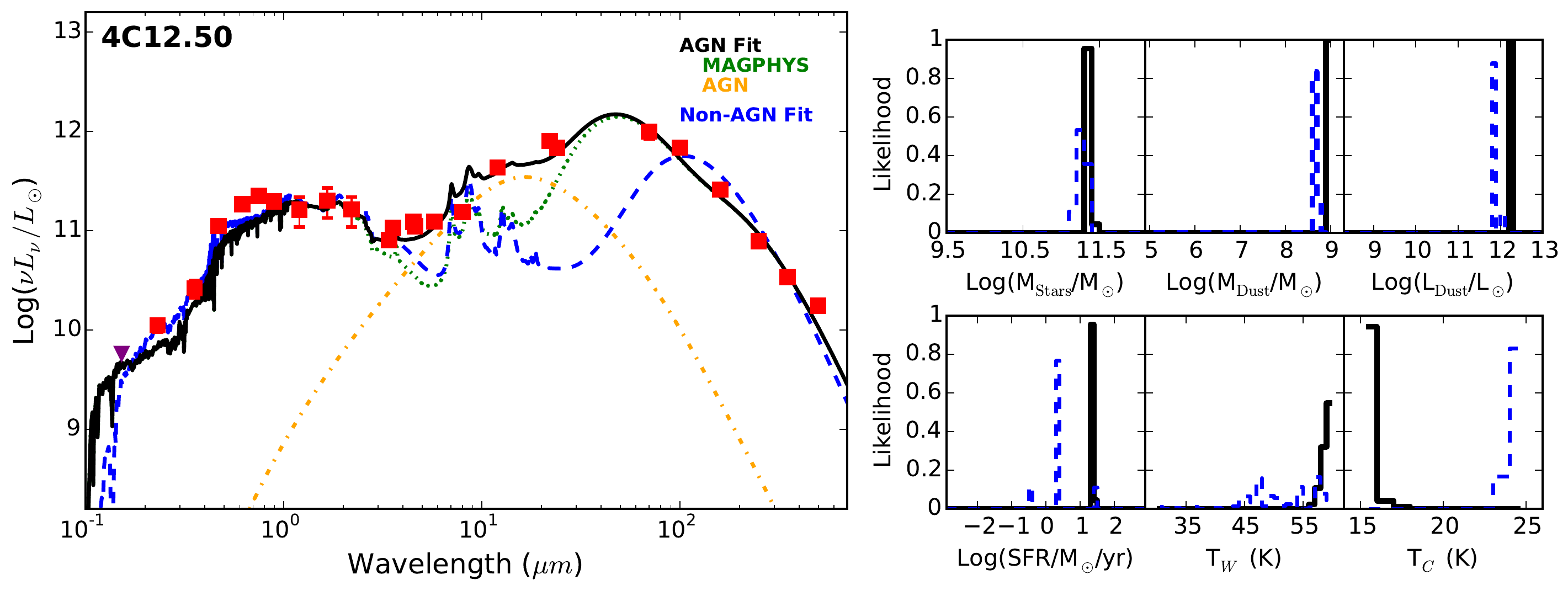}}
\centerline{\includegraphics[width=0.925\linewidth]{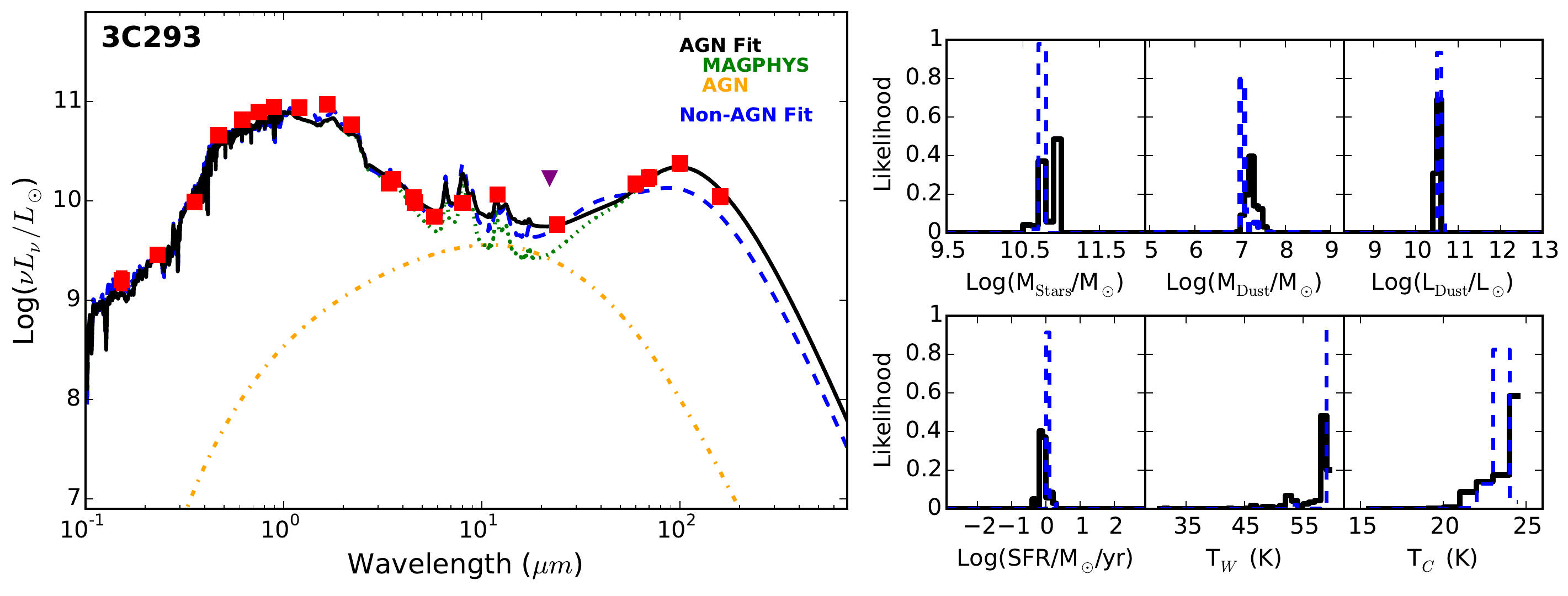}}
\centerline{\includegraphics[width=0.925\linewidth]{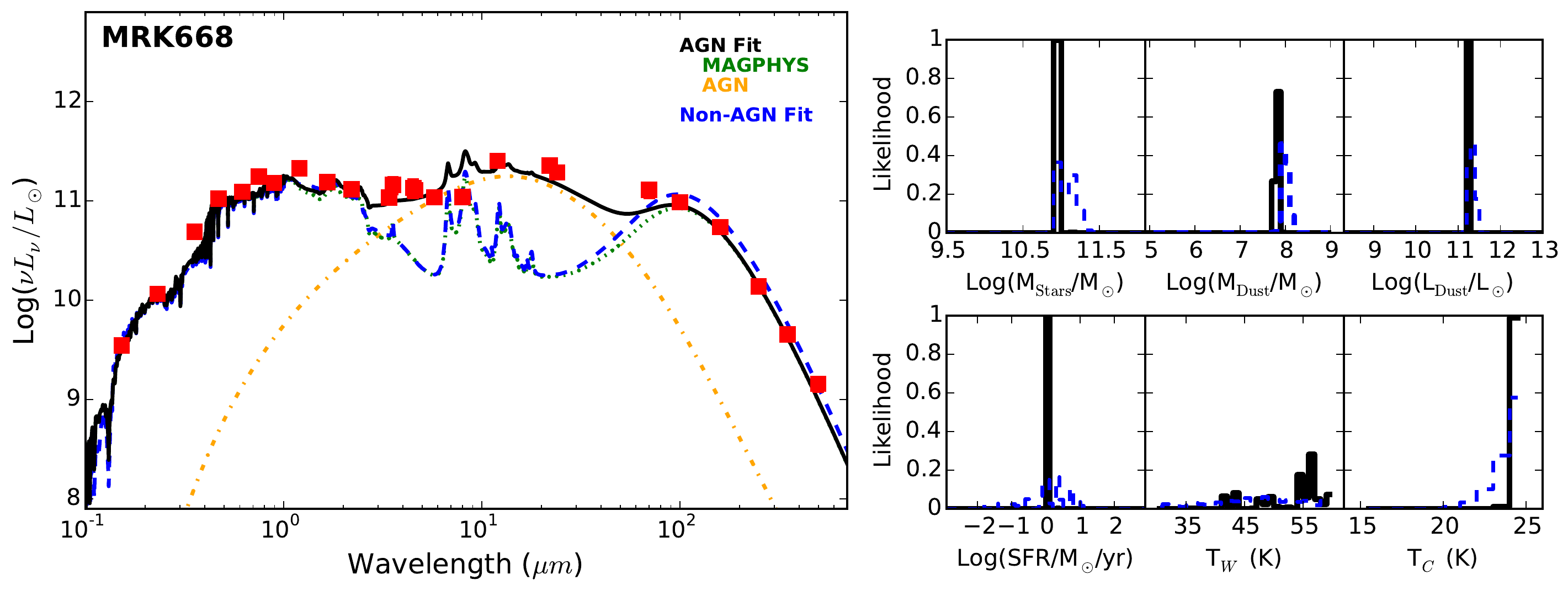}}
\caption{SEDs for 4C\,12.50 (top), 3C\,293 (middle), and Mrk\,668 (bottom) 
with photometry shown as red squares (or purple triangles for upper limits)  and the best fit model plotted in black. Further details are given in Fig. \ref{seds1} captions.}
\label{seds3}
\end{figure*}

\begin{figure*}[b]
\centerline{\includegraphics[width=0.925\linewidth]{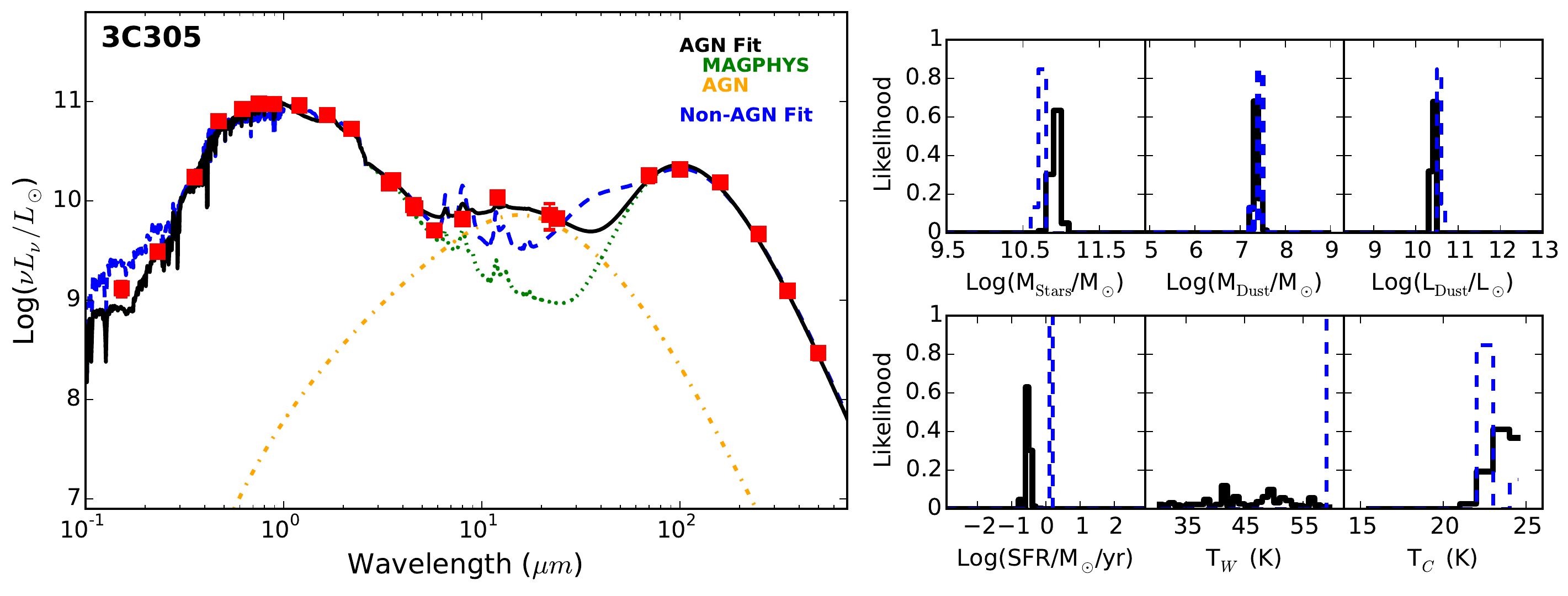}}
\centerline{\includegraphics[width=0.925\linewidth]{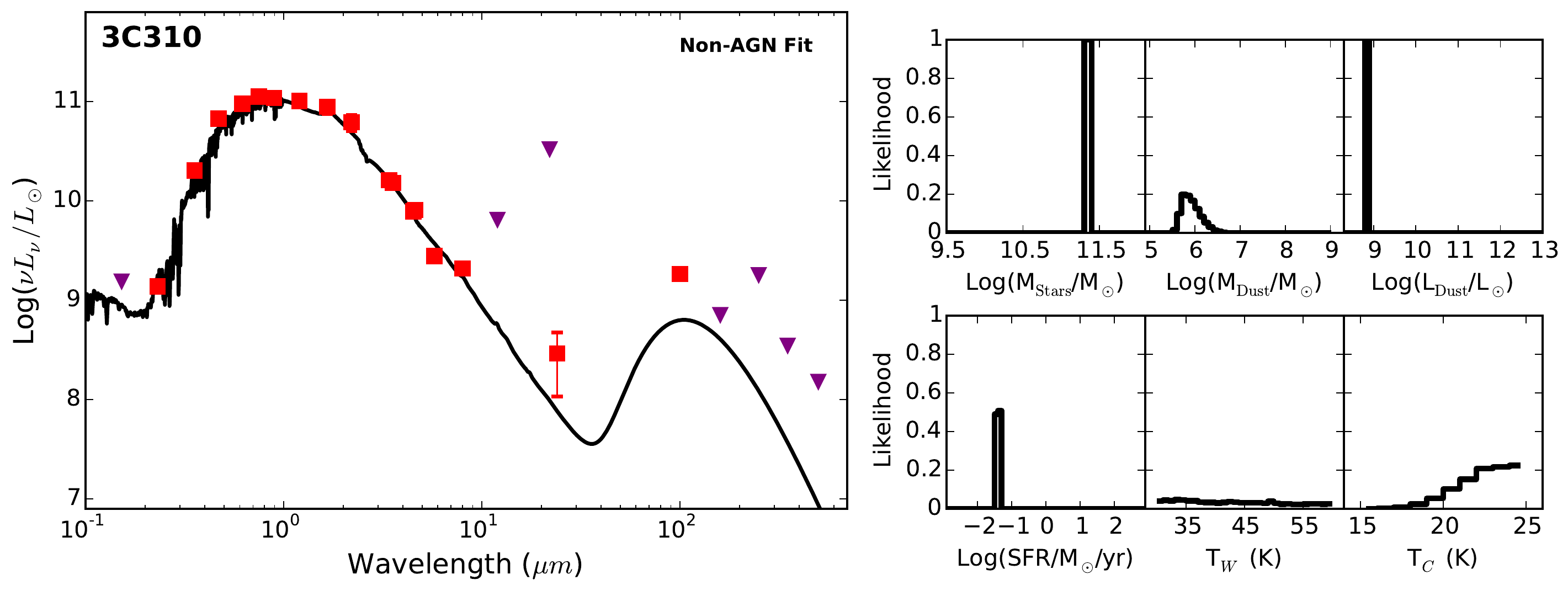}}
\centerline{\includegraphics[width=0.925\linewidth]{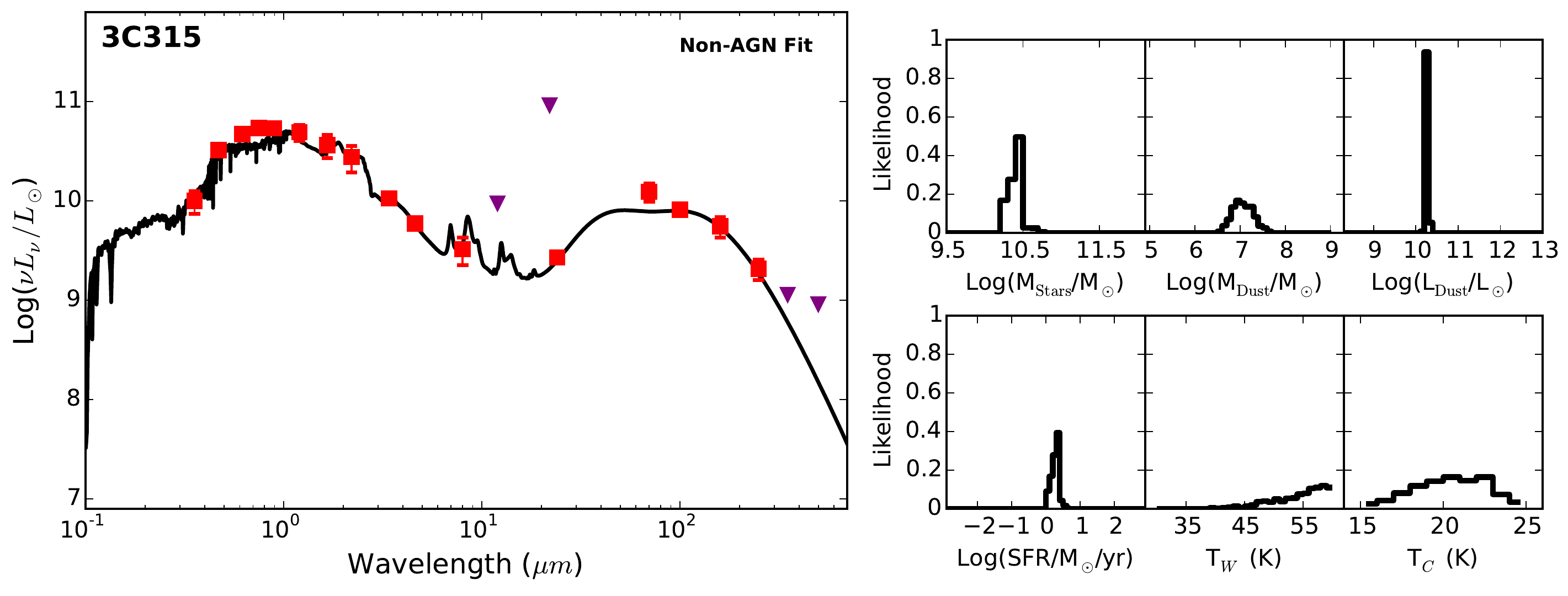}}
\caption{SEDs for 3C\,305 (top), 3C\,310 (middle), and 3C\,315 (bottom) 
with photometry shown as red squares (or purple triangles for upper limits)  and the best fit model plotted in black. Further details are given in Fig. \ref{seds1} captions.}
\label{seds4}
\end{figure*}

\begin{figure*}[b]
\centerline{\includegraphics[width=0.925\linewidth]{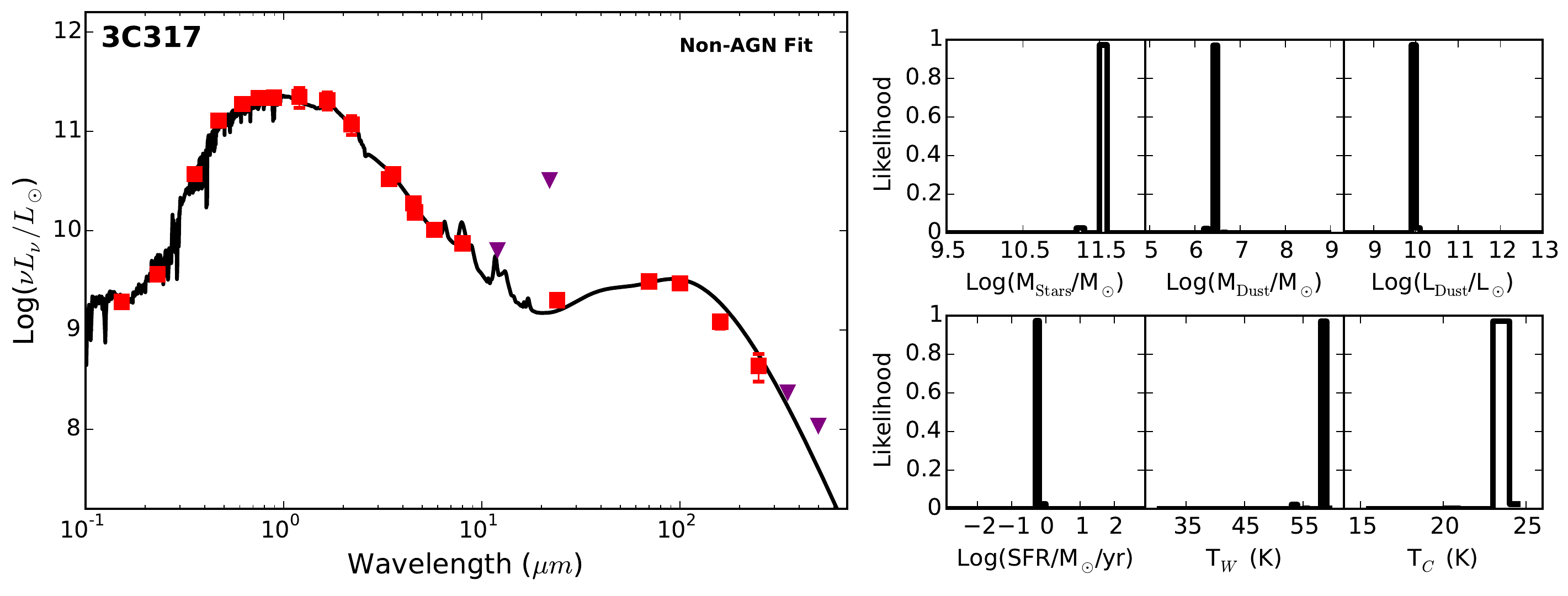}}
\centerline{\includegraphics[width=0.925\linewidth]{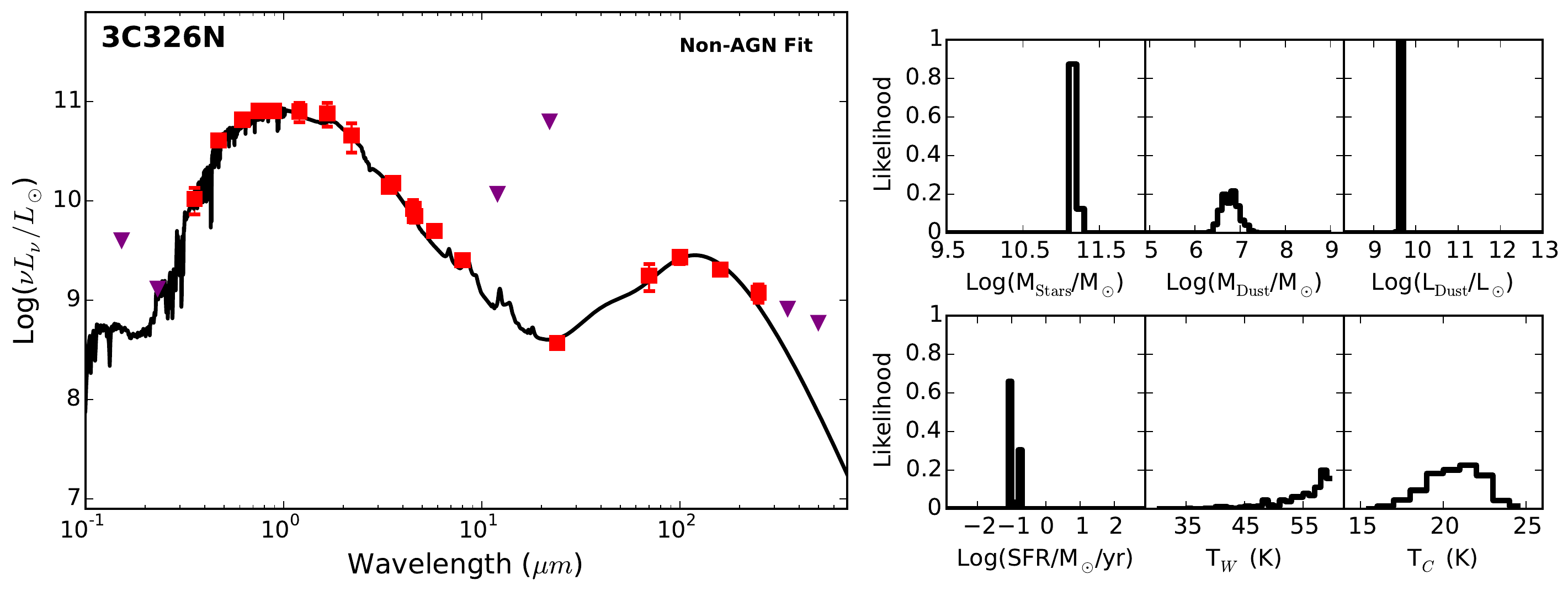}}
\centerline{\includegraphics[width=0.925\linewidth]{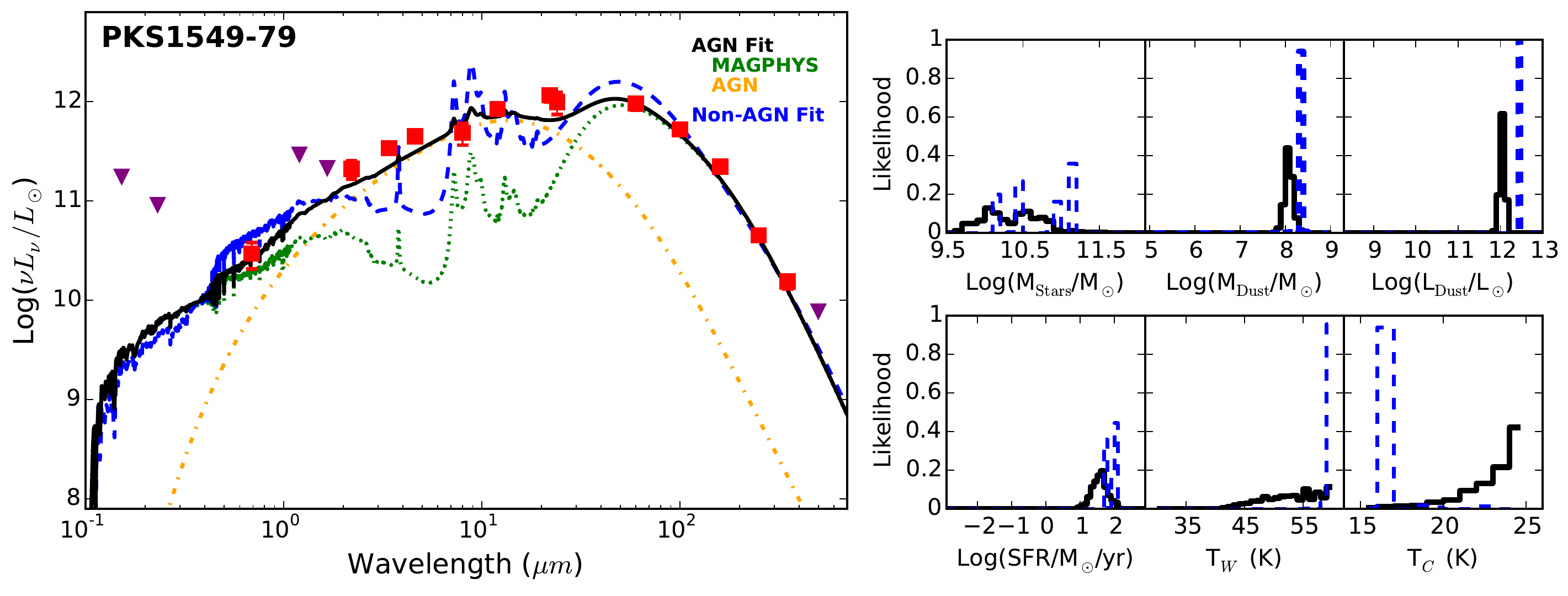}}
\caption{SEDs for 3C\,317 (top), 3C\,326N (middle), and PKS\,1549-79 (bottom) 
with photometry shown as red squares (or purple triangles for upper limits)  and the best fit model plotted in black. Further details are given in Fig. \ref{seds1} captions.}
\label{seds5}
\end{figure*}

\begin{figure*}[b]
\centerline{\includegraphics[width=0.925\linewidth]{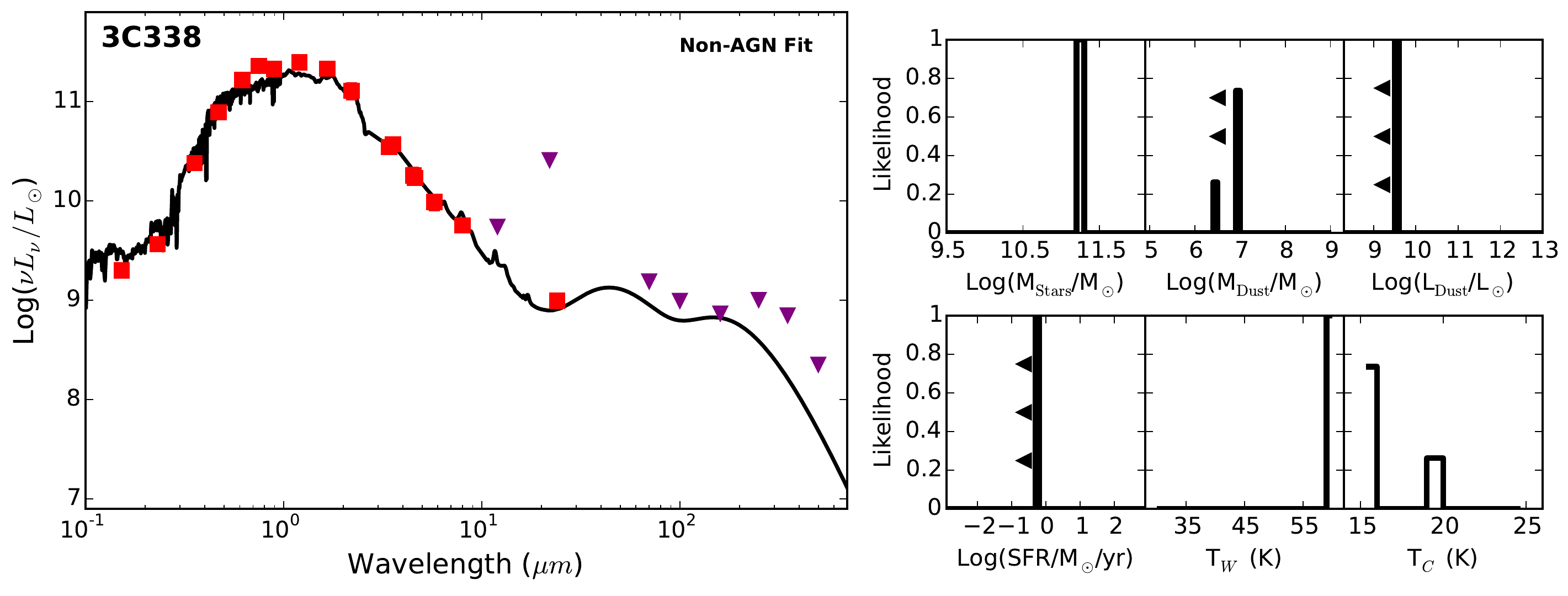}}
\centerline{\includegraphics[width=0.925\linewidth]{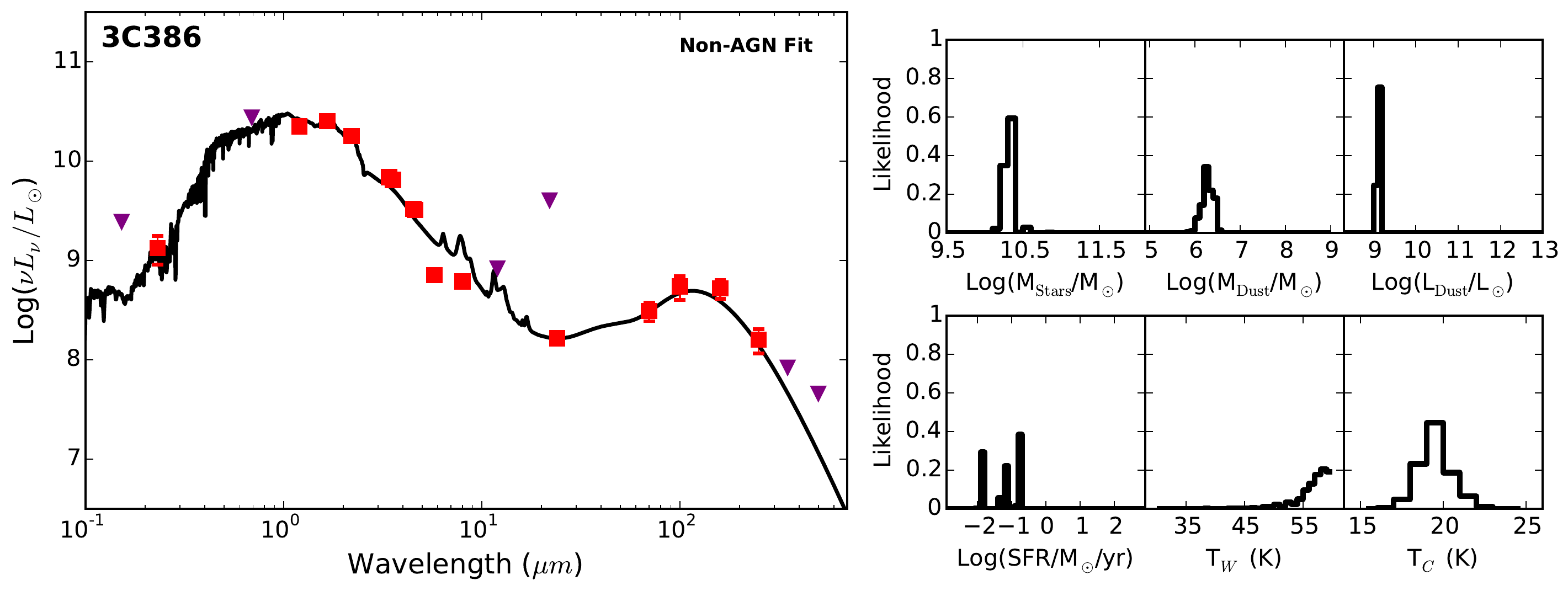}}
\centerline{\includegraphics[width=0.925\linewidth]{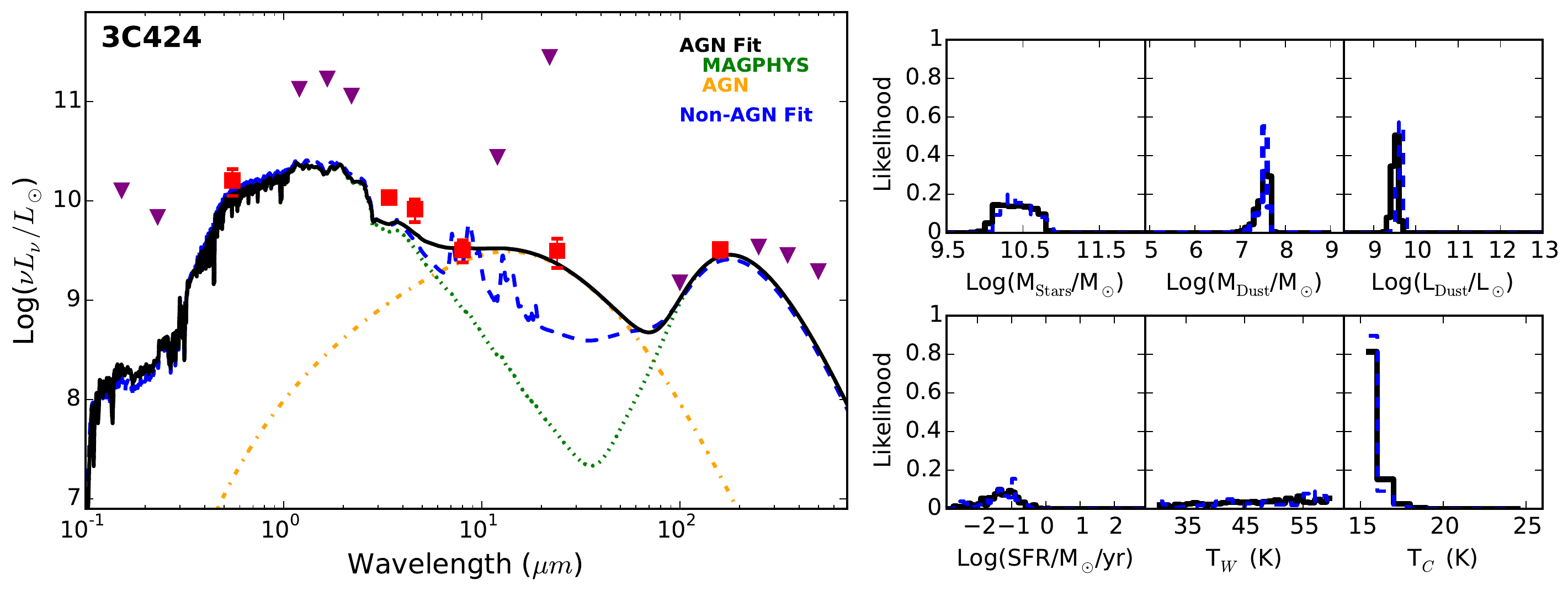}}
\caption{SEDs for 3C\,338 (top), 3C\,386 (middle), and 3C\,424 (bottom) 
with photometry shown as red squares (or purple triangles for upper limits)  and the best fit model plotted in black. Further details are given in Fig. \ref{seds1} captions.}
\label{seds6}
\end{figure*}

\begin{figure*}[b]
\centerline{\includegraphics[width=0.925\linewidth]{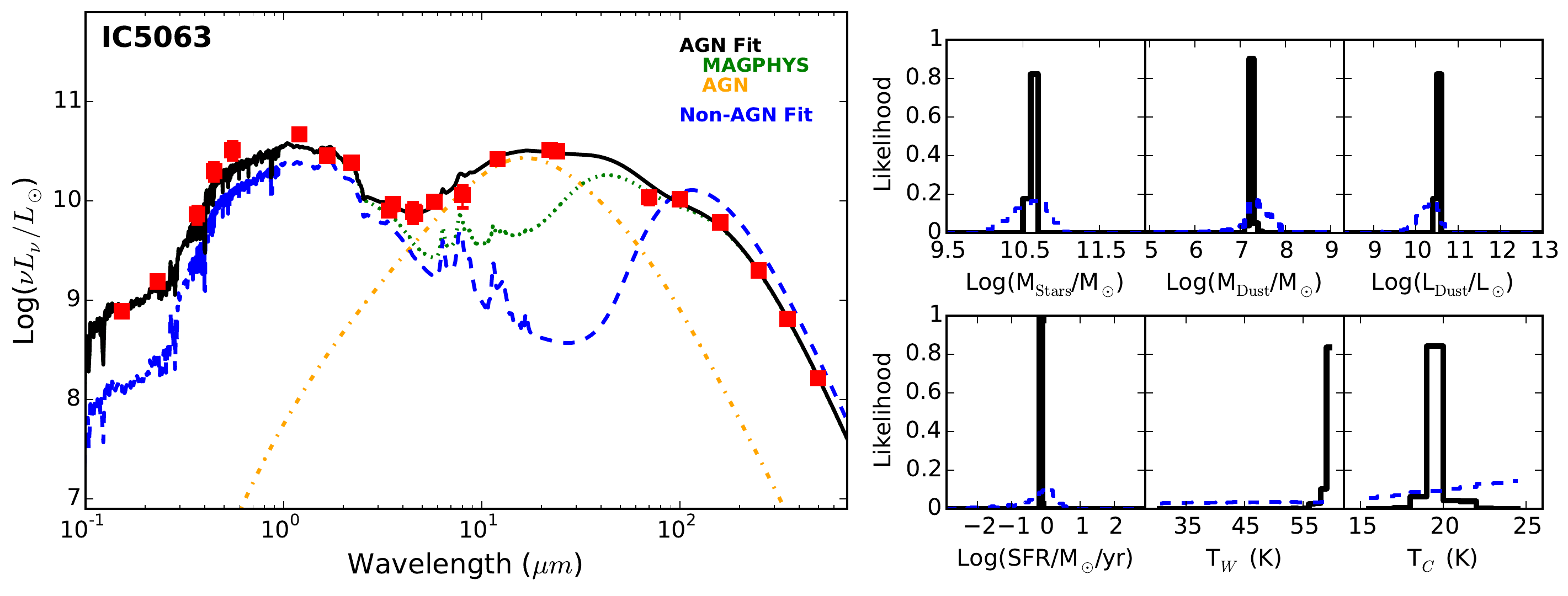}}
\centerline{\includegraphics[width=0.925\linewidth]{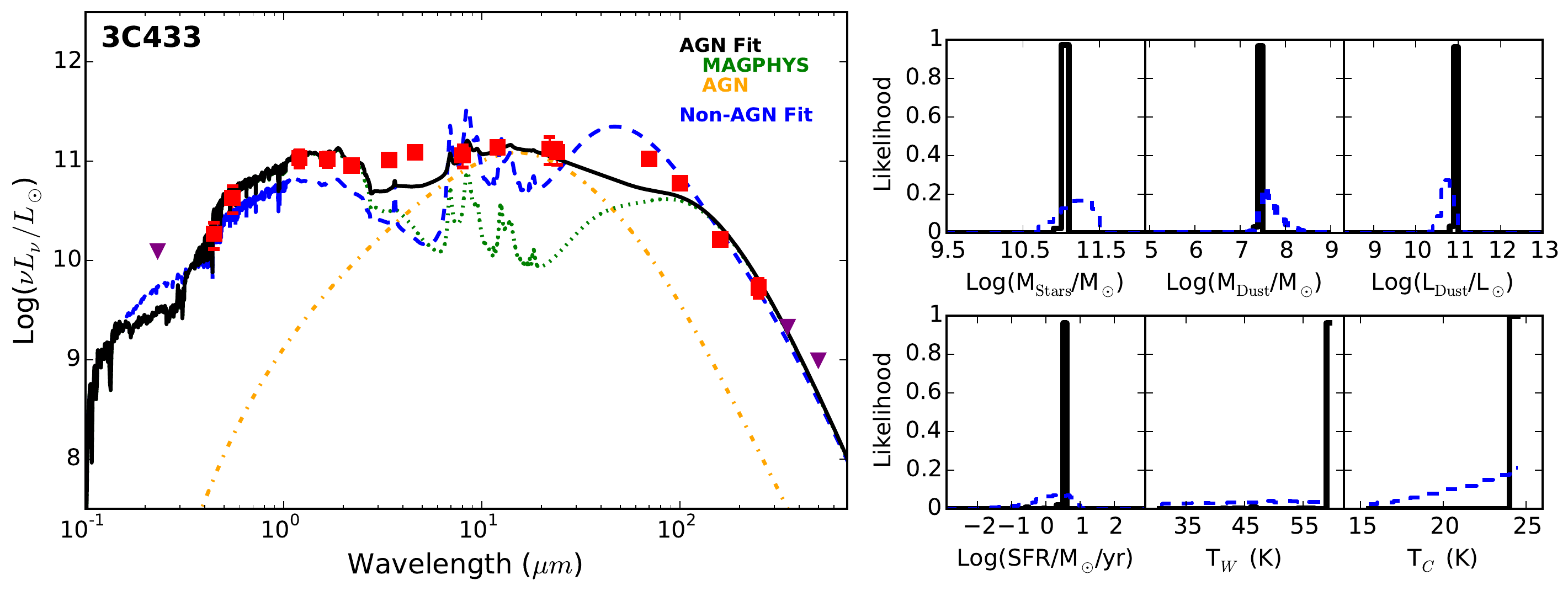}}
\caption{SEDs for IC\,5063 (top) and 3C\,433 (bottom) 
with photometry shown as red squares (or purple triangles for upper limits)  and the best fit model plotted in black. Further details are given in Fig. \ref{seds1} captions.}
\label{seds7}
\end{figure*}

\begin{figure*}[b]
\centerline{\includegraphics[width=0.925\linewidth]{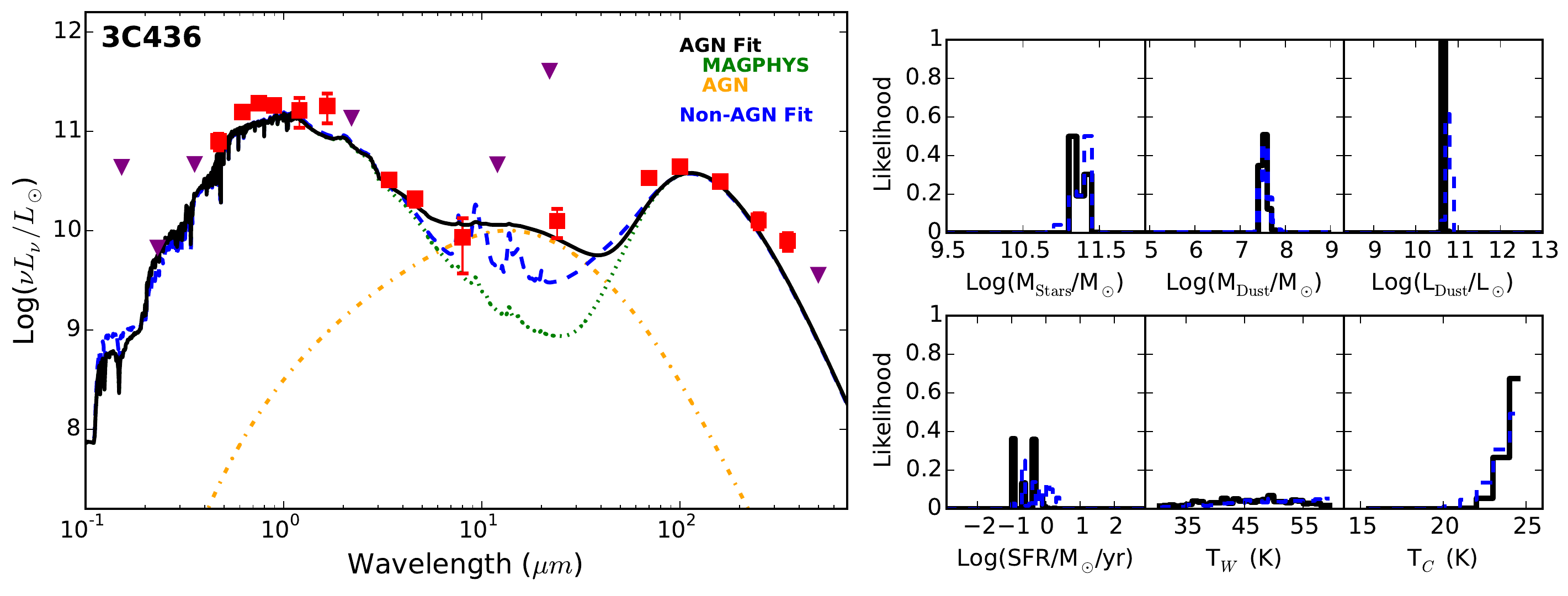}}
\centerline{\includegraphics[width=0.925\linewidth]{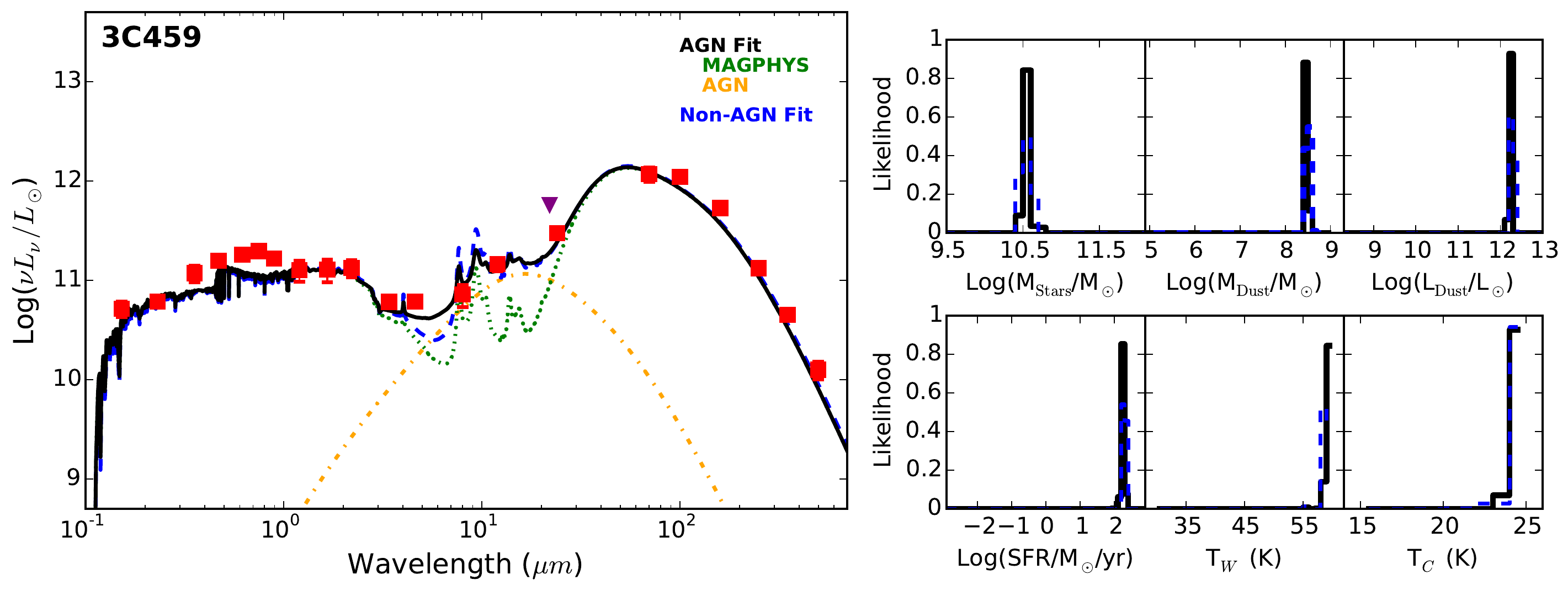}}
\caption{SEDs for 3C\,436 (top) and 3C\,459 (bottom) 
with photometry shown as red squares (or purple triangles for upper limits)  the best fit model plotted in black. Further details are given in Fig. \ref{seds1} captions.}
\label{seds8}
\end{figure*}

\noindent
\textbf{3C\,338}: 3C\,338 (NGC\,6166) is the BCG of Abell 2052 (Fig. \ref{img4}). As a result, it has numerous small companions, but still dominates the emission into the MIR. In the \herschel bands, however, the emission is dominated by the small galaxy just north--east of the center of 3C\,338, as can be seen by the location of the emission in that panel compared to the white cross which is in the same position as the black cross in the SDSS image. Given the proximity of that galaxy and the resolution of the \herschel instruments, we cannot disentangle any minor contribution from 3C\,338. Therefore, we only have upper limits, measured including the companions, on \herschel photometry. As a result, our FIR SED (Fig. \ref{seds6})  is completely undefined, and we treat the derived dust luminosity, dust mass, and SFR as upper limits. The little MIR information that we have does not support the inclusion of an AGN component.    \vspace{2mm}

\noindent
\textbf{3C\,386}: Figure \ref{img5} shows that 3C\,386 has indications of extended emission in the FIR, but it is not associated with its fat double radio structure.  Its CO observation does not have a measured extent. We therefore estimate the size of the molecular/star-forming region based on the extent of the 8\um emission, which, as can be seen in the IRAC panel of Figure \ref{img5} is centrally condensed compared to the stellar extent observed in the shorter IRAC bands. Its SED (Fig. \ref{seds6}) is sparsely sampled in the UV and optical. It is not very well fit at 5.6 and 8\um, which may be due to a difference between the PAH lines in the \magphys template and the reality in this system, in which both the 7.7 and 11.3\um PAH are weak \citep{ogle10}. \\ \vspace{2mm}

\noindent
\textbf{3C\,424}: 3C\,424, as can be seen in both its images (Fig. \ref{img5}) and its SED (Fig. \ref{seds6}) is poorly detected in only a few bands. Indeed of the six photometric points used in the fit, three come from the literature (or enhanced archive product). This is also visible in the large widths of its parameter PDFs, so the derived parameters have large uncertainties and should be treated with caution. The relative fluxes at 8 and 24\um are better fit with the inclusion of an AGN component. \citet{leipski09} concluded that star formation contributed little to IR spectrum, which \cite{ogle10} found to be flat. Indeed, the IR spectrum could be a continuation of the synchrotron emission from the radio into the IR.  \\\\\\  \vspace{2mm}

\noindent
\textbf{IC\,5063}: IC\,5063 is one of our closest galaxies and shows significant filamentary structure in the UV that may be tracing out a star-forming disk (Fig. \ref{img5}). The extended \herschel emission likewise suggests the presence of a dusty disk. Unfortunately, its IRAC 8\um image is saturated. Since its CO observations have not provided a measure of the extent of its molecular content, we estimate that size from the extent of the central region of strong UV emission. The SED of IC\,5063 (Fig. \ref{seds7}) requires a sizable MIR AGN contribution. \vspace{2mm}

\noindent
\textbf{3C\,433}: 3C\,433 was only observed by \galex in the NUV (Fig. \ref{img6}), but was not detected, so we only have limits on the UV emission. It has two nearby companions, resolvable only in our 2MASS images. However, both the \wise and \herschel emissions appear to be centered on the desired 2MASS source, therefore we assume that this galaxy dominates the IR emission. Its SED (Fig. \ref{seds7}) is better fit with a significant AGN component. Its IRS spectrum has strong silicate absorption and weak PAH emission, contrary to the SED fit \citep{ogle10}. \vspace{2mm} 

\noindent
\textbf{3C\,436}: Due to its distance, we only marginally resolve 3C\,436 in SDSS, where it appears to be a non-interacting ETG (Fig. \ref{img6}). Its MIR emission is poorly sampled, primarily with IRS-derived 8 and 24\um photometric points, which suggest the need for an AGN component in the SED fit (Fig. \ref{seds8}).  \vspace{2mm}

\noindent
\textbf{3C\,459}: 3C\,459 is our most distant source, and appears as a point source in all bands (Fig. \ref{img6}). Its SED (Fig. \ref{seds8}) is dominated by its IR emission, and indeed, it meets ULIRG criteria. The inclusion of a MIR AGN component improves the fit.

\clearpage
\LongTables
\begin{deluxetable*}{lllllrl}
\tabletypesize{\scriptsize}
\tablecaption{Observation Description\label{obs_desc}}
\tablewidth{0pt}
\tablehead{
\colhead{} & \colhead{Wavelength} & \colhead{Telescope/} & \colhead{} &\colhead{}& \colhead{Exposure} & \colhead{} \\
\colhead{Name} & \colhead{Region} & \colhead{Instrument} & \colhead{Obs. ID\tablenotemark{a}} 
	&\colhead{Date}& \colhead{(s; frames)\tablenotemark{b}} & \colhead{Notes}
}
\startdata
3C\,31		&	UV		& GALEX/(FUV;NUV)	&	GI2\_019002\_3C\,31	& 2005-11-05	&	4734.2; 4970.2			&	\\
			&	Optical	& SDSS				& 	008111-5-0175			&  2009-10-17	&	53.9					& Drift Mode \\
   			&	NIR		& 2MASS				&	981019n1010115		& 1998-10-19	& 273					&	\\
	   		&	MIR		& Spitzer/IRAC			& 3418/10918400			& 2005-01-16	& 24x30s 	 				&	\\
			& 	MIR		& WISE				& 0176p318\_ab41			& 2010-07-20	& 261$^{1,2}$/253$^{3,4}$  	&	\\
   			&	MIR/FIR	& Spitzer/MIPS			& 82/4691968				& 2004-12-26	& 28x2.6s; 28x10.5s			& 24, 70	\\
			&	FIR		& Herschel/PACS		& 1342224218-19 			& 2011-07-15 	& 445\,s$\times$2\,$^{2}$	 	&100, 160	\\ \vspace{1mm}
			&	FIR		& Herschel/SPIRE		& 1342236245 				& 2012-01-03 	& 307	 				&	\\  
\hline
3C\,84		&	UV		& GALEX/(FUV;NUV)	&	GI1\_098001\_A0426	& 2004-10-07	&	14990.2; 16249.3 		&	\\
			&	Optical	& SDSS				& 	003629-1-0067			&  2003-01-28	&	53.9					& Drift Mode \\
   	   		&	NIR		& 2MASS				&	 LGA(ngc1275) 		& ...\tablenotemark{c}	& ...\tablenotemark{c}		&	\\
	   		&	MIR		& Spitzer/IRAC			& 3228/10483456			& 2005-02-20	& 40x30s 	 				&	\\
			& 	MIR		& WISE				&  0494p408\_ab41			& 2010-02-11	& 132 					&	\\
			&	MIR/FIR	& Spitzer/MIPS			& 3351/11153920			& 2005-02-25 	& 42x2.6s					& 24	\\
			&	FIR		& Herschel/PACS\tablenotemark{d}	& 1342216022-23	& 2011-03-14 	& 153\,s$\times$2\,$^{1}$		& 70, 160	\\ 
			&	FIR		& Herschel/PACS\tablenotemark{d}	& 1342204217-18 	& 2010-09-09 	& 153\,s$\times$2\,$^{2}$		&100, 160	\\ \vspace{1mm}
			&	FIR		& Herschel/SPIRE		& 1342203614 				& 2010-08-24 	& 467	 				&	\\  
\hline
3C\,218		&	UV		& GALEX/(FUV;NUV)	&	GI3\_103007\_HydraA	& 2008-01-10 	&	2233.1				&	\\
 	 		&	NIR		& 2MASS				&	 990515s0180009 		& 1999-05-15	& 274					&	\\
			&	MIR		& Spitzer/IRAC			& 50795/26923008			& 2008-06-09	& 36x100s				&	\\
			& 	MIR		& WISE				& 1390m122\_ab41			& 2010-05-11	& 138$^{1,2}$/130$^{3,4}$	&	\\
			&	MIR/FIR	& Spitzer/MIPS			& 82	/4707584				& 2004-05-04	& 14x10s					& 24	\\
			&	FIR		& Herschel/PACS		& 1342207071-74 			& 2010-10-25 	& 571\,s$\times$4\,$^{3}$	 	&70, 100, 160	\\ \vspace{1mm}
			&	FIR		& Herschel/SPIRE		& 1342207041 				& 2010-10-24 	& 721	 				&	\\  
\hline
3C\,236		&	Optical	& SDSS				& 	004469-3-0269			&  2004-02-17	&	53.9					& Drift Mode \\
 			&	NIR		& 2MASS				&	 981212n1610056 		& 1998-12-12	& 273					&	\\
 			&	MIR		& Spitzer/IRAC			& 3418/10921216			& 2004-12-16	& 24x30s 	 				&	\\
			& 	MIR		& WISE				&  1516p348\_ab41			& 2010-05-06	& 142 					&	\\
			&	MIR/FIR	& Spitzer/MIPS			& 82	/4708096				& 2004-04-12	& 14x10s					& 24	\\
			&	FIR		& Herschel/PACS\tablenotemark{d}	& 1342270912-13	& 2013-04-26	& 266\,s$\times$2\,$^{3}$	 	&70, 100, 160	\\
			&	FIR		& Herschel/PACS\tablenotemark{d}	& 1342246697-98	& 2012-06-07	& 895\,s$\times$2\,$^{2}$		&100, 160	\\  \vspace{1mm}
			&	FIR		& Herschel/SPIRE		& 1342246613 				& 2012-06-03	& 997					&	\\  
\hline
3C\,270		&	UV		& GALEX/(FUV;NUV)	&	GI3\_079021\_NGC4261 	& 2008-03-04	& 	1655.0				&	\\
			&	Optical	& SDSS				& 	002126-5-0438			&  2001-02-20	&	53.9					& Drift Mode \\
   	 		&	NIR		& 2MASS				&	 LGA(ugc5360) 		& ...\tablenotemark{c}	& ...\tablenotemark{c}	&	\\
	 		&	MIR		& Spitzer/IRAC			& 69/4461056				& 2004-05-27	& 10x12s 	 				&	\\
			& 	MIR		& WISE				&  1853p060\_ab41			& 2010-06-17 	& 209$^{1}$/127$^{2}$/112$^{3}$/120$^{4}$&	\\  \vspace{1mm}
			&	MIR/FIR	& Spitzer/MIPS			& 82/4692736 				& 2005-06-22	& 28x10s; 128x10.5s; 68x10.5s	& 24, 70, 160	\\ 
\hline
3C\,272.1		&	UV		& GALEX/FUV			&	GI5\_057013\_NGC4388 	& 2009-05-07	& 	2538.0				&	\\
			&	UV		& GALEX/NUV			&	Virgo\_Epoque\_MOS01  	& 2006-03-20	&	15699.9				&	\\
			&	Optical	& SDSS				& 	003836-4-0249			&  2003-03-31	&	53.9					& Drift Mode \\
 	 		&	NIR		& 2MASS				&	 LGA(m84) 			& ...\tablenotemark{c}	& ...\tablenotemark{c}	&	\\
 	 		&	MIR		& Spitzer/IRAC			& 69/4463872				& 2004-05-27	& 10x12s 	 				&	\\
			& 	MIR		& WISE				&  862p136\_ab41			& 2010-06-15	& 144 					&	\\ \vspace{1mm}
			&	MIR/FIR	& Spitzer/MIPS			& 82/4692992 				& 2004-06-01	& 28x2.6s; 16x10.5s; 68x10.5s 	& 24, 70, 160	\\
\hline
4C\,12.50		&	UV		& GALEX/FUV			&	AIS\_220			 	& 2007-05-08	& 	132.1				&	\\
			&	UV		& GALEX/NUV			&	AIS\_220			 	& 2005-05-07	& 	224.1				&	\\
			&	Optical	& SDSS				& 	003836-5-0384			&  2003-03-31	&	53.9					& Drift Mode \\
 			&	NIR		& 2MASS				&	 980502n0390256 		& 1998-05-02	& 273					&	\\
 			&	MIR		& Spitzer/IRAC			& 32/3893760				& 2004-01-13	& 24x12s 	 				&	\\
			& 	MIR		& WISE				&  2070p121\_ab41			& 2010-07-06	& 256$^{1,2}$/248$^{3,4}$	&	\\
			&	MIR/FIR	& Spitzer/MIPS			& 30877/19167488 			& 2007-07-13	& 14x2.6s; 16x10.5s			& 24, 70	\\
			&	FIR		& Herschel/PACS		& 1342224349-50 			& 2011-07-17	& 445\,s$\times$2\,$^{2}$		&100, 160	\\ \vspace{1mm}
			&	FIR		& Herschel/SPIRE		& 1342234792				& 2011-12-17	& 307 	 				&	\\  
\hline	
3C\,293		&	UV		& GALEX/FUV			&	AIS\_238				& 2007-04-06	&	 119.0  				&	\\
			&	UV		& GALEX/NUV			&	MISGCN3\_02086\_0229	& 2011-06-01	&	 2632.3				&	\\
			&	Optical	& SDSS				& 	004623-6-0301			&  2004-05-12	&	53.9					& Drift Mode \\
			&	NIR		& 2MASS				&	 980310n1340068 		& 1998-03-10	& 273					&	\\
			&	MIR		& Spitzer/IRAC			& 3418/10922496			& 2005-06-11	& 24x30s 					&	\\
			& 	MIR		& WISE				&  2082p318\_ab41			& 2010-06-29	& 218 					&	\\  \vspace{1mm}
			&	MIR/FIR	& Spitzer/MIPS			& 82/4694016 				& 2005-06-28	& 56x2.6s; 28x10.5s; 68x10.5s	& 24, 70, 160	\\
\hline
MRK\,668		&	UV		& GALEX/FUV			&	GI1\_056017\_NGC5466	& 2007-05-01	& 	1838.1				&	\\
			&	UV		& GALEX/NUV			&	GI1\_056017\_NGC5466	& 2006-05-11	&	 3529.1				&	\\
			&	Optical	& SDSS				& 	004646-6-0117			&  2004-05-22	&	53.9					& Drift Mode \\
 			&	NIR		& 2MASS				&	 980505n0220209 		& 1998-05-05	& 273					&	\\
 			&	MIR		& Spitzer/IRAC			& 30443/17639168			& 2006-07-10	& 10x2s	 				&	\\
			& 	MIR		& WISE				&  2125p287\_ab41			& 2010-07-04	& 229 					&	\\
			&	MIR/FIR	& Spitzer/MIPS			& 30443/17640448			& 2006-07-14	& 14x2.6s; 16x3.2s			& 24, 70	\\
			&	FIR		& Herschel/PACS		& 1342223955-56 			& 2011-07-11	 & 445\,s$\times$2\,$^{2}$	 	&100, 160	\\ \vspace{1mm}
			&	FIR		& Herschel/SPIRE		& 1342234785				& 2011-12-17	 & 307 	 				&	\\  
\hline
3C\,305		&	UV		& GALEX/(FUV;NUV)	&	AIS\_23				& 2004-03-13	& 	196.0				&	\\	
			&	Optical	& SDSS				& 	001412-5-0275			&  2000-04-27	&	53.9					& Drift Mode \\
 	 		&	NIR		& 2MASS				&	 000221n0390150 		& 2000-02-21	& 273					&	\\
 	 		&	MIR		& Spitzer/IRAC			& 3418/10923008			& 2004-11-25	& 24x30s 	 				&	\\
			& 	MIR		& WISE				& 2212p636\_ab41			& 2010-06-01	& 391$^{1,2}$/370$^{3,4}$	&	\\
			&	MIR/FIR	& Spitzer/MIPS			& 82/4737280				& 2004-04-12	& 14x10s; 208x10.5s			& 24, 70	\\
			&	FIR		& Herschel/PACS		& 1342223959-60 	 	 	& 2011-07-11	 & 445\,s$\times$2\,$^{2}$	 	&100, 160	\\ \vspace{1mm}
			&	FIR		& Herschel/SPIRE		& 1342234915				& 2011-12-18	 & 307 	 				&	\\  
\hline
3C\,310		&	UV		& GALEX/FUV			&	AIS\_237				& 2007-04-10	& 	175.0				&	\\
			&	UV		& GALEX/NUV			&	MISGCSN3\_21467\_0238& 2011-05-16	& 	2373.0				&	\\
			&	Optical	& SDSS				& 	004588-4-0131			&  2004-04-22	&	53.9					& Drift Mode \\
 	 		&	NIR		& 2MASS				&	 990522n0590103		& 1999-05-22	& 273					&	\\
 	 		&	MIR		& Spitzer/IRAC			& 3418/10923264 			& 2005-07-16	& 24x30s 	 				&	\\
			& 	MIR		& WISE				& 2266p257\_ab41			& 2010-07-22	& 315$^{1,2}$/303$^{3,4}$	&	\\
			&	FIR		& Herschel/PACS		& 1342235116-17 	 	 	& 2011-12-24	 & 2020\,s$\times$2\,$^{2}$	&100, 160	\\ \vspace{1mm}
			&	FIR		& Herschel/SPIRE		& 1342234778				& 2011-12-17	 & 997 	 				&	\\  
\hline
3C\,315		&	Optical	& SDSS				& 	004576-2-0703			&  2004-04-16	&	53.9					& Drift Mode \\
			&	NIR		& 2MASS				&	 990527n0320103 		& 1999-05-27	& 273					&	\\
			& 	MIR		& WISE				& 2283p257\_ab41			& 2010-07-24	& 317$^{1,2}$/304$^{3,4}$	&	\\
			&	MIR/FIR	& Spitzer/MIPS			&82/4708864				& 2004-08-06	& 14x10s; 28x10.5s			& 24, 70	\\
			&	FIR		& Herschel/PACS		& 1342224636-37 			& 2011-07-21	 & 895\,s$\times$2\,$^{2}$	 	&100, 160	\\ \vspace{1mm}
			&	FIR		& Herschel/SPIRE		& 1342234777				& 2011-12-17	 & 997 	 				&	\\  
\hline
3C\,317		&	UV		& GALEX/(FUV;NUV)	&	GI3\_103015\_Abell2052	& 2007-06-04	& 	2857.1				&	\\
			&	Optical	& SDSS				& 	003903-3-0318			&  2003-04-27	&	53.9					& Drift Mode \\
 	 		&	NIR		& 2MASS				&	 000428s0700044 		& 2000-04-28	& 274					&	\\
 	 		&	MIR		& Spitzer/IRAC			& 30659/18654464			& 2006-08-10	& 18x12s 	 				&	\\
			& 	MIR		& WISE				& 2288p075\_ab41			& 2010-07-31	& 279$^{1,2}$/269$^{3,4}$	&	\\
			&	MIR/FIR	& Spitzer/MIPS			&30659/18641664 			& 2007-03-05	& 42x10s					& 24	\\
			&	FIR		& Herschel/PACS		& 1342237886-89			& 2012-01-05	 & 840\,s$\times$4\,$^{3}$ 	&70, 100, 160	\\ \vspace{1mm}
			&	FIR		& Herschel/SPIRE		& 1342238322				& 2012-01-28	 & 859 	 				&	\\  
\hline
3C\,326N		&	UV		& GALEX/FUV			&	AIS\_135				& 2007-04-13	&	 96.0					&	\\
			&	UV		& GALEX/NUV			&	AIS\_135				& 2005-06-17	& 	247.1				&	\\
			&	Optical	& SDSS				& 	004633-2-0076			&  2004-05-14	&	53.9					& Drift Mode \\
 	 		&	NIR		& 2MASS				&	 000422n0520173 		& 2000-04-22	& 273					&	\\
 	 		&	MIR		& Spitzer/IRAC			& 3418/10923776 			& 2005-03-27	& 24x30s 	 				&	\\
			& 	MIR		& WISE				& 2384p196\_ab41			& 2010-02-10	&  168$^{1,2}$/157$^{3,4}$	&	\\
			&	MIR/FIR	& Spitzer/MIPS			&3418/10930432			& 2005-08-28	& 80x2.6s					& 24	\\
			&	FIR		& Herschel/PACS\tablenotemark{d}	& 1342248732-33	& 2012-07-27 	&  2470\,s$\times$2\,$^{2}$	&100, 160	\\ 
			&	FIR		& Herschel/PACS\tablenotemark{d}	& 1342261315-18	& 2013-01-18 	 & 538\,s$\times$4\,$^{3}$	 	&70, 100, 160	\\ \vspace{1mm}
			&	FIR		& Herschel/SPIRE		& 1342238327				& 2012-01-28	 & 1135					&	\\  
\hline
PKS\,1549-79	&	UV		& GALEX/(FUV;NUV)	&	AIS\_470				& 2006-08-16	& 	109.0				&	\\
			&	NIR		& 2MASS				&	 000408s0730068		& 2000-04-08	& 274					&	\\
			& 	MIR		& WISE				& 2376m788\_ab41			& 2010-03-10  	& 235$^{1,2}$/226$^{3,4}$	&	\\
			&	FIR		& Herschel/PACS		& 1342225387-88 	 	 	& 2011-07-24	 & 445\,s$\times$2\,$^{2}$	 	&100, 160	\\ \vspace{1mm}
			&	FIR		& Herschel/SPIRE		& 1342239890				& 2012-03-01	 & 307 	 				&	\\  
\hline
3C\,338		&	UV		& GALEX/(FUV;NUV)	&	NGA\_NGC6166		& 2004-08-06	& 	1437.0				&	\\
			&	Optical	& SDSS				& 	003225-4-0238			&  2002-06-09	&	53.9					& Drift Mode \\
 	 		&	NIR		& 2MASS				&	 990603n0430162 		& 1999-06-03	& 273					&	\\
 	 		&	MIR		& Spitzer/IRAC			& 25/3860992 				& 2004-07-06	& 30x200s				&	\\
			& 	MIR		& WISE				& 2477p393\_ab41			& 2010-02-13 	& 204$^{1,2}$/196$^{3,4}$	&	\\
			&	MIR/FIR	& Spitzer/MIPS			& 20651/14957056			& 2005-08-29	&80x2.6s					& 24	\\
			&	FIR		& Herschel/PACS		& 1342207019-22			& 2010-10-23	 &  571\,s$\times$4\,$^{3}$	&70, 100, 160	\\ \vspace{1mm}
			&	FIR		& Herschel/SPIRE		& 1342207033				& 2010-10-24	 & 721 					&	\\  
\hline
3C\,386		&	UV		& GALEX/(FUV;NUV)	&	AIS\_121				& 2006-07-31	& 	183.0				&	\\
 	 		&	NIR		& 2MASS				&	 990608n0730232 		& 1999-06-08	& 273					&	\\
 	 		&	MIR		& Spitzer/IRAC			& 3418/10924800			& 2005-05-06	& 24x30s 	 				&	\\
			& 	MIR		& WISE				& 2798p166\_ab41			& 2010-04-01	& 162$^{1,2}$/150$^{3,4}$	&	\\
			&	MIR/FIR	& Spitzer/MIPS			& 3418/12418048 			& 2005-04-10	&80x2.6s; 52x10.5s 			& 24, 70	\\
			&	FIR		& Herschel/PACS		& 1342231672-73	 	 	& 2011-10-30	& 670\,s$\times$2\,$^{2}$	 	&100, 160	\\ \vspace{1mm}
			&	FIR		& Herschel/SPIRE		& 1342239789				& 2012-02-29	& 997  	 				&	\\  
\hline
3C\,424		&	UV		& GALEX/(FUV;NUV)	&	AIS\_242				& 2006-09-03	&	 176.0				&	\\
 	  	    	&	NIR		& 2MASS				&	 000806s0550233 		& 2000-08-06	& 274					&	\\
			& 	MIR		& WISE				& 3127p075\_ab41			& 2010-05-09	&151$^{1,2}$/138$^{3}$/139$^{4}$	&	\\
			&	FIR		& Herschel/PACS		& 1342233349-50 	 	 	& 2011-12-01	& 2470\,s$\times$2\,$^{2}$	&100, 160	\\ \vspace{1mm}
			&	FIR		& Herschel/SPIRE		& 1342244149				& 2012-04-12	& 997 	 				&	\\  
\hline
IC\,5063		&	UV		& GALEX/(FUV;NUV)	&	GI3\_087016\_IC\,5063 	& 2007-06-25	& 	2951.1				&	\\
			&	NIR		& 2MASS				&	 000621s0070127		& 2000-06-21	& 274					&	\\
			&	MIR		& Spitzer/IRAC			& 3269/12455680			& 2005-05-09	& 2x12s 					&	\\
			& 	MIR		& WISE				& 3129m576\_ab41			& 2010-04-18 	&163$^{1,2,3}$/140$^{4}$		&	\\
			&	MIR/FIR	& Spitzer/MIPS			& 86/4858624				& 2005-05-20	& 28x2.6s					& 24	\\
			&	FIR		& Herschel/PACS		& 1342216469-72  			& 2011-03-20	& 276\,s$\times$4\,$^{3}$	 	&70, 100, 160	\\ \vspace{1mm}
			&	FIR		& Herschel/SPIRE		& 1342206208				& 2010-10-11	& 445 	 				&	\\  
\hline
3C\,433		&	UV		& GALEX/NUV			&	AIS\_154				& 2011-10-09	& 	128.0				& No FUV obs.	\\
 	 	    	&	NIR		& 2MASS				&	 971029n0280220 		& 1997-10-29	& 273					&	\\
			& 	MIR		& WISE				& 3216p257\_ab41			& 2010-05-25 	& 167$^{1,2}$/156$^{3,4}$	&	\\
			&	FIR		& Herschel/PACS\tablenotemark{d}	& 1342232731-	32	& 2011-11-10	& 445\,s$\times$2\,$^{2}$		&100, 160	\\
			&	FIR		& Herschel/PACS\tablenotemark{d}	& 1342219391-92	& 2011-04-19	 &  266\,s$\times$2\,$^{3}$	&70, 100, 160	\\ \vspace{1mm}
			&	FIR		& Herschel/SPIRE		& 1342234675				& 2011-12-18	& 307 					&	\\  
\hline
3C\,436		&	UV		& GALEX/(FUV;NUV)	&	AIS\_40				& 2007-07-17	&	 64.0/328.				&	\\
			&	Optical	& SDSS				& 	008155-3-0058			&  2009-11-17	&	53.9					& Drift Mode \\
		    	&	NIR		& 2MASS				&	 991110n0460080		& 1999-11-10	& 273					&	\\
			& 	MIR		& WISE				& 3257p287\_ab41			& 2010-05-31	& 190$^{1,2}$/179$^{3,4}$	&	\\
			&	FIR		& Herschel/PACS\tablenotemark{d}	& 1342235316-17	& 2011-12-25 	& 445\,s$\times$2\,$^{2}$		&100, 160	\\ 
			&	FIR		& Herschel/PACS\tablenotemark{d}	& 1342257734-37	& 2011-04-19	 &  266\,s$\times$4\,$^{3}$	&70, 100, 160	\\ \vspace{1mm}
			&	FIR		& Herschel/SPIRE		& 1342234676				& 2011-12-18	& 997 	  				&	\\  
\hline
3C\,459		&	UV		& GALEX/(FUV;NUV)	&	AIS\_149				& 2006-10-02	&	 186.0				&	\\
			&	Optical	& SDSS				& 	007807-2-0076			&  2008-11-17	&	53.9					& Drift Mode \\
 			&	NIR		& 2MASS				&	 000825s0250092		& 2000-08-25	& 274					&	\\
			& 	MIR		& WISE				& 3493p045\_ab41			& 2010-06-13	& 129$^{1,2}$/122$^{3,4}$	&	\\
			&	MIR/FIR	& Spitzer/MIPS			& 20233/14432512 			& 2005-11-30	& 28x2.6s; 28x10.5s			& 24, 70	\\
			&	FIR		& Herschel/PACS		& 1342237979-80 			& 2012-01-06	& 445\,s$\times$2\,$^{2}$	 	&100, 160	\\ 
			&	FIR		& Herschel/SPIRE		& 1342234756				& 2011-12-19	& 307 	 				&	  
\enddata
\tablenotetext{a}{SDSS Obs. IDs are in the form of a 6 digit run number, followed by a one digit camera column, and ending in a four digit field number. 2MASS Obs. IDs are in the form of six digits dates (yymmdd) followed by scan directions (n/s) followed by a three digits scan number and ending in a four digit image number, except for 3C84, 3C270, and 3C272.1. \spitzer Obs. IDs are given as Project ID/AOR number.}
\tablenotetext{b}{If only one exposure time is given, it is the same for all bands. For 2MASS and \wise, we give the exposure in terms of the number of frames that were co-added to create the image. When different bands had different numbers of frames, the super-script indicates which bands the coverage indicates. Exposure times for PACS are given as (Time per Obs. ID) $\times$ (Number of Observations). PACS always observes at 160\um in conjunction with either 70\um or 100\um. $^{1}$ indicates all observations were done at 160\um and 70\um. $^{2}$ indicates all observations were performed at 160\um and 100\um. $^{3}$ indicates the observations are evenly split between the two modes (i.e. the 70 and 100\um bands were each observed for half of the total 160\um time).}
\tablenotetext{c}{The 2MASS mosaics of these galaxies come from the Large Galaxy Atlas, which combines multiple observations and does not clearly indicate the dates and number of frames that went into each mosaic.}
\tablenotetext{d}{3C\,84, 3C\,236, 3C\,326N, 3C\,433, and 3C\,436 were all observed twice by PACS, with different configurations of bands. For each galaxy, we combine all the available data at each wavelength.}
\end{deluxetable*}

\clearpage
\begin{landscape}
\begin{deluxetable*}{llcccccll}
\tabletypesize{\scriptsize}
\tablecaption{Photometry\label{phot}}
\tablewidth{0pt}
\tablehead{
\colhead{} & \colhead{$\lambda$ Region} & \multicolumn{5}{c}{Flux (mJy or Jy)\tablenotemark{a} }  \\
\cline{3-7}
\colhead{Galaxy} & \colhead{(Flux Unit)}	& \multicolumn{4}{c}{}	& \colhead{or Lit. Ref.\tablenotemark{b}}
}
\startdata
Name		&	UV/Opt. (mJy)  	    & GALEX FUV 	& GALEX NUV		& SDSS{\em u}		& Johnson{\em  U}		& Lit. Ref.	\\
			&	Opt. (mJy)	    	    & Johnson{\em B}	& SDSS{\em g}		& Johnson{\em V}	& SDSS{\em r}			& Lit. Ref.	\\
			&	Opt./NIR (mJy) 	    & Johnson{\em R}	& SDSS{\em i}  	& SDSS{\em z}		& 2MASS\,J			& Lit. Ref.	\\
			&	NIR (mJy)		    & 2MASS{\em H}	& 2MASS{\em Ks}	& WISE3.4\uma	& IRAC3.6\uma			& ...		\\
			& 	MIR1 (mJy)	    & IRAC4.5\uma	& WISE4.6\uma	& IRAC5.8\uma		& IRAC8.0\uma\tablenotemark{c} & ... \\
			&	MIR2 (mJy)	    & WISE12\uma	&WISE22\uma	& MIPS24\uma\tablenotemark{d}&IRAS60\uma\tablenotemark{e}& Lit. Ref.	\\
			&	FIR1 (Jy)		    & MIPS70\uma\tablenotemark{f}& PACS70\uma& IRAS100\uma\tablenotemark{g}	&	PACS100\uma	& Lit. Ref. \\ \vspace{3mm}
			& 	FIR2 (Jy)		    & MIPS160\uma\tablenotemark{g}	& PACS160\uma		& SPIRE250\uma	& SPIRE350\uma	& SPIRE500\uma	\\ 
\hline
\hline
3C\,31		&	UV/Opt. (mJy)  & $~~~~\,0.279~~~\pm~\,0.029~~~$	& $~~~~\,0.769~\,\pm~~~0.077~\,$	& $~~~23.1~~~\,~\pm~~~1.3~~\,~~$	& \,~...						& ...			\\
			&	Opt. (mJy)	    	& ~~~...						& $~~~94.5~~~~\,\pm~~~3.0~~~~\,$	& ~\,...						& $~\,167~~~~~~\,\pm~~~5~~~~~~\,$	& ...			\\
			&	Opt./NIR (mJy) & ~~~...						& $~\,228~~~~~~\,\pm~~~7~~~~~~\,$& $~\,258~~~~\,~~\pm~\,13~~~~\,~~$	& $~\,268~~~~~~\,\pm~\,39~~~~~~\,$	& ...			\\
			&	NIR (mJy)	      & $~\,287~~~~~~~~\pm44~~~~~~~~$	& $~\,273~~~~~~\,\pm~\,36~~~~~~\,$& $~\,122~~~~\,~~\pm~~~3~~~~\,~~$	& $~\,137~~~~~~\,\pm~~~5~~~~~~\,$	& ... 			\\
			& 	MIR1 (mJy)	& $~\,85.5~~~~~~\pm~\,3.5~~~~\,$	& $~~~75.5~~~~\,\pm~~~2.6~~~~\,$	& $~~~53.6~~~\,~\pm~~~2.0~~~\,~$	& $~~~49.9~~~~\,\pm~~~1.6~~~\,~$	& ...			\\
			&MIR2 (mJy)	& $~~~57.6~~~~~~~\pm~\,9.8~~~~~~~$	& $<260~~~~~~~~~~~~~~~~~~~~~~~$ 	& $~~~34.8~~~\,~\pm~~~1.4~~~\,~$ & ~\,...						& ...			\\
			&	FIR1 (Jy)	&  $~~~~\,0.484~~~\pm~\,0.103~~~$	& ~\,...						& ~\, ...							& $~~~~\,1.22~~~\pm~~~0.12~~~$	& ...			\\\vspace{1 mm}  
			&	FIR2 (Jy)	& ~~~... 							& $~\,~~~1.69~~~\pm~~~0.17~~~$	& $~~~~\,0.794~\,\pm~~~0.083~\,$		& $~~~~\,0.323~\,\pm~~~0.037~\,$	& $0.140~\,\pm0.020~\,$		\\   
\hline 
3C\,84		&	UV/Opt. (mJy)  & $~~~~\,9.31~~~~\,\pm~\,0.93~~~~\,$& $~~~20.1~~~~\,\pm~~~2.0~~~~\,$& $~\,142~~~~\,~~\pm~~~8~~~~\,~~$	& \,~... 						& ...			\\
			&	Opt. (mJy)	    	& ~~~...						& $~\,337~~~~~~\,\pm~\,11~~~~~~\,$& ~\,...						& $~\,438~~~~~~\,\pm~\,13~~~~~~\,$	& ...			\\
			&	Opt./NIR (mJy) & ~~~...						& $~\,452~~~~~~\,\pm~\,14~~~~~~\,$& $~\,408~~~~\,~~\pm~\,21~~~~\,~~$	& $~\,307~~~~~~\,\pm~\,31~~~~~~\,$	& ...			\\
			&	NIR (mJy)	      & $~\,379~~~~~~~~\pm34~~~~~~~~$	& $~\,341~~~~~~\,\pm~\,28~~~~~~\,$& $~\,181~~~~\,~~\pm~~~5~~~~\,~~$	&$~\,120.\,~~~~~\,\pm~~~5~~~~~~\,$	& ... 			\\
			& MIR1 (mJy)		& $157~~~~~~~~\pm~\,6~~~~~~\,$	& $~\,170.~~~~~~\pm~~~6~~~~~~\,$ & $~\,157~~~~~\,~\pm~~~6~~~~\,~~$	& $~\,278~~~~~~\,\pm~~~9~~~~~~\,$	& ...			\\
			&MIR2 (mJy)	& $1330~~~~~~~~~\pm60~~~~~~~~~$    & $3310~~~~~~\,\pm220~~~~~~\,$		& $3110~~~\,~~~\pm120~~~~\,~~$	 &  ~\,...						& ...			\\
			&	FIR1 (Jy)	&  ~~~...							& $~\,~~~4.92~~~\pm~~~0.49~~~$	&  ~\,...							& $~~~~\,8.84~~~\pm~~~0.88~~~$	& ...			\\ \vspace{1 mm} 
			&	FIR2 (Jy)	& ~~~... 							& $~\,~~~8.37~~~\pm~~~0.84~~~$	& $~~~~\,3.55~~~\pm~~~0.36~~~$		& $~~~~\,2.91~~~\pm~~~0.29~~~$	& $2.92~~~\pm0.29~~~$		\\ 
\hline
3C\,218		&	UV/Opt. (mJy)  & $~~~~\,0.296~~~\pm~\,0.030~~~$	& $~\,~~~0.451~\,\pm~~~0.045~\,$	& ~\,...						& \,~...						& ...			\\
			&	Opt. (mJy)	    	& $~~~~\,1.58~~~~\,\pm~\,0.47~~~~\,$& ~\,...						& $~~~~\,3.53~~~\pm~~~1.06~~~$	& \,~...						& {\em B/V}: (1) 		\\
			&	Opt./NIR (mJy) & ~~~...						& ~\,...						& ~\,...						& $~~~31.0~~~~\,\pm~~~2.4~~~~\,$	& ...			\\
			&	NIR (mJy)	      & $~~~37.5~~~~~~\pm~\,3.3~~~~~~$	& $~~~31.9~~~~\,\pm~~~3.2~~~~\,$	& $~~~14.1~~~\,~\pm~~~0.4~~~\,~$	& \,~~\,...\tablenotemark{h}		& ...			\\
			& MIR1 (mJy)		& $~\,10.3~~~~~~\pm~\,0.4~~~~\,$	& $~~~~\,9.23~~~\pm~~~0.41~~~$	&  ~\,... 						& $~~~~\,6.69~~~\pm~~~0.21~~~$	& ...			\\
			&MIR2 (mJy)	& $~~~~\,9.49~~~~\,\pm~\,2.72~~~~\,$	& $<78~~~~~~~~~~~~~~~~~~~~~$		& $~~~~\,8.32~~~\pm~~~0.33~~~$&  ~\,...	& ...	\\
			&	FIR1 (Jy)	&  ~~~...							& $~\,~~~0.121~\,\pm~~~0.013~\,$	&  ~\,...							& $~~~~\,0.203~\,\pm~~~0.021~\,$	& ...			\\ \vspace{1 mm} 
			&	FIR2 (Jy)	& ~~~... 		& $~\,~~~0.203~\,\pm~~~0.021~\,$	& $~~~~\,0.137~\,\pm~~~0.018~\,$ & $~~~\,0.122~\,\pm~~~0.014\,$	& $\,0.130~\,\pm0.015\tablenotemark{k}$ \\  
\hline
3C\,236		&	UV/Opt. (mJy)  & $~~~~\,0.00704\pm~\,0.00141$	& $~\,~~~0.0104\pm~~~0.0021$	& $~~~~\,0.457~\,\pm~~~0.065~\,$		& \,~...						& FUV/NUV: (2) \\
			&	Opt. (mJy)	    	& ~~~...						& $~\,~~~1.28~~~\pm~~~0.09~~~$	& ~\,...						& $~~~~\,3.05~~~\pm~~~0.15~~~$	& ...			\\
			&	Opt./NIR (mJy) & ~~~...						& $~\,~~~4.41~~~\pm~~~0.19~~~$	& $~~~~\,6.48~~~\pm~~~0.37~~~$	& $~~~~\,6.76~~~\pm~~~1.07~~~$	& ...			\\
			&	NIR (mJy)	    & $~~~~\,6.66~~~~\,\pm~\,1.55~~~~\,$	& $~\,~~~8.33~~~\pm~~~1.74~~~$	& $~~~~\,3.99~~~\pm~~~0.16~~~$	& $~~~~\,5.10~~~\pm~~~0.21~~~$	& ... 			\\
			& MIR1 (mJy) 	     & $~~~~\,4.87~~~~\,\pm~\,0.23~~~~\,$& $~~~~\,3.90~~~\pm~~~0.28~~~$	& $~~~~\,3.56~~~\pm~~~0.14~~~$	& $~~~~\,5.73~~~\pm~~~0.19~~~$	& ... 			\\
			&MIR2 (mJy)	& $~~~12.2~~~~~~~\pm~\,2.7~~~~~~~$	& $<85~~~~~~~~~~~~~~~~~~~~~$		& $~~~20.1~~~\,~\pm~~~0.8~~~\,~	$&  ~\,...	& ...	\\
			&	FIR1 (Jy)	&  ~~~...							& $~\,~~~0.0443\pm~~~0.0052$	&  ~\,...							& $~~~~\,0.0888\pm~~~0.0092$	& ...			\\ \vspace{1 mm} 
			&	FIR2 (Jy)	& ~~~... 		& $~\,~~~0.132~\,\pm~~~0.014~\,$	& $~~~\,0.106~\,\pm~~~0.011\,$& $~~~~\,0.0753\pm~~~0.0087$& $0.0633\pm0.0076$ \\
\hline
3C\,270		&	UV/Opt. (mJy)  & $~~~~\,0.797~~~\pm~\,0.080~~~$	& $~\,~~~2.05~~~\pm~~~0.21~~~$	& $~~~35.2~~~\,~\pm~~~1.8~~~\,~$	& \,~... 						& ...			\\
			&	Opt. (mJy)	    	& ~~~...						& $~\,178~~~~~~\,\pm~~~5~~~~~~\,$& ~\,...						& $~\,350~~~~~~\,\pm~\,11~~~~~~\,$	& ...			\\
			&	Opt./NIR (mJy) & ~~~...						& $~\,506~~~~~~\,\pm~\,15~~~~~~\,$& $~\,620~~~~\,~~\pm~\,31~~~\,~~~$	& $~\,803~~~~~~\,\pm~\,32~~~~~~\,$	& ...			\\
			&	NIR (mJy)	      & $~\,975~~~~~~~~\pm40~~~~~~~~$	& $~\,791~~~~~~\,\pm~\,33~~~~~~\,$& $~\,345~~~\,~~~\pm~\,10~~~\,~~~$	& $~\,387~~~~~~\,\pm~\,18~~~~~~\,$	& ...			\\
			& MIR1 (mJy)		& $240.\,~~~~~~~\pm18~~~~~~\,$	& $~\,212~~~~~~\,\pm~~~7~~~~~~\,$& $~\,128~~~~\,~~\pm~\,13~~~~\,~~$	& $~\,196~~~~~~\,\pm~~~9~~~~~~\,$	& ...			\\
			&MIR2 (mJy)	&  $~\,132~~~~~~~~~\pm14~~~~~~~~~$	& $<350~~~~~~~~~~~~~~~~~~~~~~~$ 	& $~~~55.7~~~\,~\pm~~~2.2~~~\,~	$&  $<155~~~~~~~~~~~~~~~~~~~~\,~~$& ~\,60\uma: (3)	\\
			&	FIR1 (Jy)	&  $~~~~\,0.154~~~\pm~\,0.028~~~$	& ~\,...						&  $<0.385~~~~~~~~\,~~~~~~~$		& ~\,...						& 100\uma: (3)	\\ \vspace{1 mm} 
			&	FIR2 (Jy)	& $~~~~\,0.121~~~\pm~\,0.026~~~$ 	& ~\,...						& ~\,...		& ~\,...		& ...		\\				    
\hline
3C\,272.1		&	UV/Opt. (mJy)  & $~~~~\,1.43~~~~\,\pm~\,0.14~~~~\,$& $~\,~~~4.76~~~\pm~~~0.48~~~$& $~\,116~~~~\,~~\pm~~~6~~~~\,~~$	& \,~... 						& ...			\\
			&	Opt. (mJy)	    	& ~~~...						& $~\,539~~~~~~\,\pm~\,16~~~~~~\,$& ~\,... 						& $1010~~~~\,~~\pm~\,30~~~~~~\,$	& ...			\\
   			&	Opt./NIR (mJy) & ~~~...						& $1410~~~~~~\,\pm~\,40~~~~~~\,$	& $1700~~~~\,~~\pm~\,90~~~~\,~~$	& $1980~~~~\,~~\pm~\,70~~~~~~\,$	& ...			\\
			&	NIR (mJy)	      & $2370~~~~~~~~\pm90~~~~~~~~$	& $1930~~~~~~\,\pm~\,70~~~~~~\,$	& $~\,866~~~~\,~~\pm~\,24~~~~\,~~$	& $~\,923~~~\,~~~\pm~\,36~~~~~\,~$	& ... 			\\
			& MIR1 (mJy)		& $576~~~~~~~~\pm31~~~~~~\,$	& $~\,541~~~~~~\,\pm~\,17~~~~~~\,$& $~\,289~~~\,~~~\pm~\,20.\,~~~\,~~$	& $~\,276~~~\,~~~\pm~\,12~~~~~\,~$	& ...			\\
			&MIR2 (mJy)	&  $~\,343~~~~~~~~~\pm21~~~~~~~~~$	& $<410~~~~~~~~~~~~~~~~~~~~~~~$ 	& $~~~72.4~~\,~~\pm~~~2.9~~~\,~	$&  $~\,556~~~~\,~~\pm~\,83~~~\,~~~$	& ~\,60\uma: (3)	\\
			&	FIR1 (Jy)	&  $~~~~\,0.666~~~\pm~\,0.115~~~$	& ~\,...						& $~~~~\,1.02~~~\pm~~~0.15~~~$	& ~\,...						& 100\uma: (3)	\\ \vspace{1 mm} 
			&	FIR2 (Jy)	& $~~~~\,0.422~~~\pm~\,0.067~~~$ 	& ~\,...						& ~\,...		& ~\,...		& ...		\\				    
\hline
4C\,12.50		&	UV/Opt. (mJy)  & $<0.026~~~~~~~~~~~~~~~~~~\, $& $~~~~\,0.0765\pm~~~0.0106$	& $~~~~\,0.277~\,\pm~~~0.055~\,$		& \,~...						& ...			\\
			&	Opt. (mJy)	    	& ~~~...						& $~~~~\,1.55~~~\pm~~~0.10~~~$	& ~\,...						& $~~~~\,3.39~~~\pm~~~0.16~~~$	& ...			\\
			&	Opt./NIR (mJy) & ~~~...						& $~\,~~~4.96~~~\pm~~~0.21~~~$	& $~~~~\,5.21~~~\pm~~~0.30~~~$	& $~~~~\,5.79~~~\pm~~~1.93~~~$	& ...			\\
			&	NIR (mJy)	    & $~~~~\,9.92~~~~\,\pm~\,3.30~~~~\,$	& $~~~10.7~~~~\,\pm~~~3.56~~~$	& $~~~~\,8.12~~~\pm~~~0.29~~~$	& $~~~11.3~~~\,~\pm~~~0.6~~~\,~$		& ... 			\\
			& MIR1 (mJy)		& $~\,16.4~~~~~~\pm~\,0.8~~~~\,$	& $~~~15.1~~~~\,\pm~~~0.6~~~~\,$	& $~~~21.1~~~\,~\pm~~~0.8~~~~\,$ 	& $~~~36.4~~~\,~\pm~~~1.2~~~\,~$	& ...			\\
			&MIR2 (mJy)	&  $~\,154~~~~~~~~~\pm~\,8~~~~~~~~~$ & $~\,522~~~~~~\,\pm~\,47~~~~~~\,$ & $~\,487~~~\,~~~\pm~\,20.\,~~~\,~~$& ~\,...	& ...	\\
			&	FIR1 (Jy)	&  $~~~~\,2.07~~\,~~\pm~\,0.32~~\,~~$	& ~\,...						&  ~\,...							& $~~~~\,2.04~~~\pm~~~0.20~~~$	& ...			\\ \vspace{1 mm} 
			&	FIR2 (Jy)	& ~~~... 					& $~\,~~~1.23~~~\pm~~~0.12~~~$	& $~~~~\,0.584~\,\pm~~~0.059~\,$		& $~~~~\,0.356~\,\pm~~~0.036~\,$		& $0.260~\,\pm0.027~\,$\\  
\hline
3C\,293		&	UV/Opt. (mJy)  & $~~~~\,0.063~~~\pm~\,0.013~~~$	& $~~~~\,0.171~\,\pm~~~0.017~\,$	& $~~~~\,0.900~\,\pm~~~0.093~\,$		& \,~... 						& ...			\\
			&	Opt. (mJy)	    	& ~~~...						& $~~~~\,5.58~~~\pm~~~0.23~~~$	& ~\,...	 					& $~~~10.5~~~\,~\pm~~~0.4~~~\,~$	& ...			\\
			&	Opt./NIR (mJy) & ~~~...						& $~~~15.2~~~~\,\pm~~~0.5~~~~\,$	& $~~~20.6~~~\,~\pm~~~1.1~~~\,~$	& $~~~27.0~~~\,~\pm~~~1.9~~~\,~$	& ...			\\
			&	NIR (mJy)	      & $~~~40.4~~~~~~\pm~\,3.2~~~~~~$	& $~~~33.5~~~~\,\pm~~~3.0~~~~\,$	& $~~~13.3~~~\,~\pm~~~0.4~~~\,~$	& $~~~15.2~~~\,~\pm~~~0.6~~~\,~$	& ...			\\
			& MIR1 (mJy)		& $~\,12.7~~~~~~\pm~\,0.7~~~~\,$	& $~~~11.5~~~~\,\pm~~~0.5~~~~\,$	& $~~~10.4~~~\,~\pm~~~0.4~~\,~~$	& $~~~19.9~~~\,~\pm~~~0.9~~~\,~$	& ... 			\\
			&MIR2 (mJy)	&  $~~~36.1~~~~~~~\pm~\,3.7~~~~~~~$	& $<~\,96~~~~~~~~~~~~~~~~~~~~~$ 	& $~~~35.9~~\,~~\pm~~~1.4~~~\,~	$&$~\,233~~~~~~\,\pm~\,35~~~~~~\,$	& ~\,60\uma: (3)	\\
			&	FIR1 (Jy)	&  $~~~~\,0.313~~~\pm~\,0.052~~~$	& ~\,...						&  $~~~~\,0.621~\,\pm~~~0.093~\,$		& ~\,...						& 100\uma: (3)	\\ \vspace{1 mm} 
			&	FIR2 (Jy)	&$~~~~\,0.462~~~\pm~\,0.071~~~$	 	& ~\,...						& ~\,...		& ...		& ...		\\				    
\hline
MRK\,668		&	UV/Opt. (mJy)  & $~~~~\,0.0441~\,\pm~\,0.0050~\,$	& $~~~~\,0.222~\,\pm~~~0.022~\,$	& $~~~~\,1.44~~~\pm~~~0.14~~~$	& \,~...	 					& ...			\\
			&	Opt. (mJy)	    	& ~~~...						& $~~~~\,4.09~~~\pm~~~0.25~~~$	& ~\,... 						& $~~~~\,6.33~~~\pm~~~0.33~~~$	& ...			\\
			&	Opt./NIR (mJy) & ~~~...						& $~~~10.9~~~~\,\pm~~~0.5~~~~\,$	& $~~~11.2~~~\,~\pm~~~0.6~~~\,~$		& $~~~21.2~~~\,~\pm~~~1.8~~~\,~$	& ...			\\
			&	NIR (mJy)	      & $~~~21.3~~~~~~\pm~\,1.5~~~~~~$	& $~~~23.8~~~~\,\pm~~~2.8~~~~\,$	& $~~~30.5~~~\,~\pm~~~0.9~~~\,~$	& $~~~42.7~~~\,~\pm~~~7.7~~~\,~$	& ... 			\\
			& MIR1 (mJy)		& $~\,50.5~~~~~~\pm10.1~~~~\,$	& $~~~48.9~~~~\,\pm~~~1.6~~~~\,$	& $~~~51.8~~~\,~\pm~~~6.3~~\,~~$	& $~~~72.7~~~\,~\pm~~~8.8~~~\,~$	& ...			\\
			&MIR2 (mJy)	&  $~\,251~~~~~~~~~\pm12~~~~~~~~~$ & $~\,413~~~~~~\,\pm~\,42~~~~~~\,$	& $~\,385~~~\,~~~\pm~\,15~~~\,~~~$ & ~\,...	& ...	\\
			&	FIR1 (Jy)	&  $~~~~\,0.751~~~\pm~\,0.121~~~$	& ~\,...						&  ~\,...							& $~~~~\,0.804~\,\pm~~~0.080~\,$	& ...			\\ \vspace{1 mm} 
			&	FIR2 (Jy)	& ~~~... 				& $~\,~~~0.725~\,\pm~~~0.073~\,$	& $~~~~\,0.286~\,\pm~~~0.029~\,$	& $~~~~\,0.131~\,\pm~~~0.015\,\tablenotemark{j}$	& $~0.0597\pm0.0092\tablenotemark{j}$\\	  
\hline
3C\,305		&	UV/Opt. (mJy)  & $~~~~\,0.0610~\,\pm~\,0.0104~\,$	& $~~~~\,0.217~\,\pm~~~0.023~\,$	& $~~~~\,1.88~~~\pm~~~0.14~~~$	& \,~...	 					& ...			\\
			&	Opt. (mJy)	    	& ~~~...						& $~~~~\,9.09~~~\pm~~~0.34~~~$	& ~\,...	 					& $~~~15.8~~~\,~\pm~~~0.5~~~\,~$	& ...			\\
			&	Opt./NIR (mJy) & ~~~...						& $~~~21.6~~~~\,\pm~~~0.7~~~~\,$	& $~~~25.7~~\,~~\pm~~~1.3~~~\,~$	& $~~~33.5~~~\,~\pm~~~2.0~~~\,~$	& ...			\\
			&	NIR (mJy)	      & $~~~37.0~~~~~~\pm~\,3.0~~~~~~$	& $~~~35.6~~~~\,\pm~~~2.7~~~~\,$	& $~~~15.6~~~\,~\pm~~~0.5~~\,~~$	& $~~~17.5~~~\,~\pm~~~0.6~~~\,~$	& ...			\\
			& MIR1 (mJy)		& $~\,12.5~~~~~~\pm~\,0.5~~~~\,$	& $~~~11.8~~~~\,\pm~~~0.5~~~~\,$	& $~~~~\,8.85~~~\pm~~~0.36~~~$	& $~~~16.0~~~\,~\pm~~~0.7~~~\,~$	& ...			\\
			&MIR2 (mJy)	&  $~~~39.4~~~~~~~\pm~\,2.9~~~~~~~$	& $~~~48.5~~~~\,\pm~\,14.4~~~~\,$	 & $~~~48.6~~~\,~\pm~~~1.9~~~\,~	$& ~\,...	& ...	\\
			&	FIR1 (Jy)	&  $~~~~\,0.388~~~\pm~\,0.058~~~$	& ~\,...						&  ~\,...							& $~~~~\,0.632~\,\pm~~~0.063~\,$	& ...			\\ \vspace{1 mm} 
			&	FIR2 (Jy)	& ~~~... 				& $~\,~~~0.747~\,\pm~~~0.075~\,$	& $~~~~\,0.354~\,\pm~~~0.036~\,$	& $~~~~\,0.133~\,\pm~~~0.014\,\tablenotemark{j}$	& $~0.0450\pm0.0066\tablenotemark{j}$\\  
\hline
3C\,310		&	UV/Opt. (mJy)  & $<0.041~~~~~~~~~~~~~~~~~~\, $& $~~~~\,0.0562\pm~~~0.0066$	& $~~~~\,1.27~~~\pm~~~0.13~~~$	& \,~... 	 					& ...			\\
			&	Opt. (mJy)	    	& ~~~...						& $~~~~\,5.59~~~\pm~~~0.26~~~$	& ~\,...						& $~~~10.4~~~\,~\pm~~~0.4~~~\,~$	& ...			\\
			&	Opt./NIR (mJy) & ~~~...						& $~~~14.9~~~~\,\pm~~~0.5~~~~\,$	& $~~~17.2~~~\,~\pm~~~0.9~~~\,~$	& $~~~21.6~~~\,~\pm~~~2.5~~~\,~$	& ...			\\
			&	NIR (mJy)	      & $~~~26.0~~~~~~\pm~\,3.9~~~~~~$	& $~~~24.1~~~~\,\pm~~~4.4~~~~\,$	& $~~~~\,9.71~~~\pm~~~0.34~~~$	& $~~~~\,9.57~~~\pm~~~0.49~~~$	& ... 			\\
			& MIR1 (mJy) 		& $~~~~\,6.26~~~~\,\pm~\,0.58~~~~\,$& $~~~~\,6.58~~~\pm~~~0.49~~~$& $~~~~\,2.84~~~\pm~~~0.12~~~$	& $~~~~\,2.96~~~\pm~~~0.12~~~$	& ...			\\
			&MIR2 (mJy)	&  $<14~~~~~~~~~~~~~~~~~~~~~~~~\,~$ & $<130~~~~~~~~~~~~~~~~~~~~~~~$	& $~~~~\,1.24~~~\pm~~~0.78~~~$& ~\,...	& IRS24\uma\tablenotemark{d}	\\
			&	FIR1 (Jy)	&  ~~~...							&  ~\,...					&  ~\,...							& $~~~~\,0.0326\pm~~~0.0041	$	& ...			\\ \vspace{1 mm} 
			&	FIR2 (Jy)	& ~~~... 	& $<0.020 \tablenotemark{i}~~~~~~~~~~~~~~$ & $<0.079~\tablenotemark{i}~~~~~\,~~~~~~~~$	& $<0.022\,\tablenotemark{i}~~~~~~~~~~~~~~$ & $<0.013~~~~~~~~~~~~~~~~~$\\
\hline
3C\,315		&	UV/Opt. (mJy)  & ~~~...						& ~\,...						& $~~~~\,0.137~\,\pm~~~0.035~\,$		& \,~...						& ...			\\
			&	Opt. (mJy)	    	& ~~~...						& $~~~~\,0.588~\,\pm~~~0.066~\,$	& ~\,...						& $~~~~\,1.12~~~\pm~~~0.09~~~$	& ...			\\
			&	Opt./NIR (mJy) & ~~~...						& $~~~~\,1.56~~~\pm~~~0.10~~~$	& $~~~~\,1.86~~~\pm~~~0.13~~~$	& $~~~~\,2.28~~~\pm~~~0.42~~~$	& ...			\\
			&	NIR (mJy)	    & $~~~~\,2.34~~~~\,\pm~\,0.61~~~~\,$	& $~\,~~~2.35~~~\pm~~~0.70~~~$	& $~~~~\,1.39~~~\pm~~~0.08~~~$	& \,~...						& ... 			\\
			& MIR1 (mJy)		& ~~~...						& $~~~~\,1.05~~~\pm~~~0.12~~~$	&  ~\,... 				& $~~~~\,1.01~~~\pm~~~0.31~~~$& IRS 8\uma\tablenotemark{c} \\
			&MIR2 (mJy)	&  $<4.4~~~~~~~~~~~~~~~~~~~~~~$	& $<78~~~~~~~~~~~~~~~~~~~~~$	& $~~~~\,2.50~~~\pm~~~0.10~~~$	& ~\,...	& ...	\\
			&	FIR1 (Jy)	&  $~~~~\,0.0335~\,\pm~\,0.0071~\,$	&  ~\,...						&  ~\,...							& $~~~~\,0.0314\pm~~~0.0032	$	& ...			\\ \vspace{1 mm} 
			&	FIR2 (Jy)	& ~~~... & $~\,~~~0.0343\pm~~~0.0080	$& $~~\,~~~0.0201\pm~~~0.0047\tablenotemark{j}$& $<0.015\,\tablenotemark{j}~~~~~~~~~~~~~~$& $<0.018\,\tablenotemark{j}~~~~~~~~~~~~~~~$\\
\hline
3C\,317		&	UV/Opt. (mJy)  & $~~~~\,0.131~~~\pm~\,0.015~~~$	& $~~~~\,0.378~\,\pm~~~0.039~\,$	& $~~~~\,5.92~~~\pm~~~0.38~~~$	& \,~...	 				& ...			\\
			&	Opt. (mJy)	    	& ~~~...						& $~~~27.1~~~~\,\pm~~~0.9~~~~\,$	& ~\,... 	 					& $~~~52.2~~~\,~\pm~~~1.7~~\,~~$	& ...			\\
			&	Opt./NIR (mJy) & ~~~...						& $~~~73.1~~~~\,\pm~~~2.3~~~~\,$	& $~~~87.7~~\,~~\pm~~~4.4~~~\,~$	& $~\,120.\,~~~\,~~\pm~\,27~~~\,~~~$	& ...			\\
			&	NIR (mJy)	      & $~\,154~~~~~~~~\pm30~~~~~~~~$	& $~\,116~~~~~~\,\pm~\,25~~~~~~\,$& $~~~50.5~~~\,~\pm~~~1.5~~~\,~$	& $~~~58.9~~\,~~\pm~~~5.5~~\,~~$	& ...			\\
			& MIR1 (mJy)		& $~\,38.2~~~~~~\pm~\,4.7~~~~\,$	& $~~~31.5~~~~\,\pm~~~1.5~~~~\,$	& $~~~26.4~~~\,~\pm~~~1.8~~~\,~$	& $~~~26.9~~\,~~\pm~~~1.0~~\,~~$	& ...			\\
			&MIR2 (mJy)	&  $<34~~~~~~~~~~~~~~~~~~~~~~~~\,~$ & $<320~~~~~~~~~~~~~~~~~~~~~~~$	& $~~~21.5~~~\,~\pm~~~0.9~~\,~~$& ~\,...	& ...	\\
			&	FIR1 (Jy)	&  ~~~...							& $~\,~~~0.0968\pm~~~0.0122$	&  ~\,...						& $~~~~\,0.133~\,\pm~~~0.018~\,$	& ...			\\ \vspace{1 mm} 
			&	FIR2 (Jy)	& ~~~... 	& $~\,~~~0.0872\pm~~~0.0127	$	& $~~~~\,0.0491\pm~~~0.0151$							& $<0.037~~~~~~~~~~~~~~~$		& $<0.024~~~~~~~~~~~~~~~~~$	\\
\hline
3C\,326N		&	UV/Opt. (mJy)  & $<0.036~~~~~~~~~~~~~~~~~~\, $& $<0.018~~~~~~~~~~~~~~~\, $	& $~~~~\,0.219~\,\pm~~~0.065~\,$		& \,~...	 					& ...			\\
			&	Opt. (mJy)	    	& ~~~...						& $~~~~\,1.12~~~\pm~~~0.09~~~$	& ~\,...						& $~~~~\,2.38~~~\pm~~~0.14~~~$	& ...			\\
			&	Opt./NIR (mJy) & ~~~...						& $~~~~\,3.56~~~\pm~~~0.17~~~$	& $~~~~\,4.24~~~\pm~~~0.26~~~$	& $~~~~\,5.63~~~\pm~~~1.24~~~$	& ...			\\
			&	NIR (mJy)	    & $~~~~\,7.45~~~~\,\pm~\,2.00~~~~\,$	& $~\,~~~5.90~~~\pm~~~1.92~~~$	& $~~~~\,2.79~~~\pm~~~0.14~~~$	& $~~~~\,3.16~~~\pm~~~0.20~~~$	& ... 			\\
			& MIR1 (mJy) 		& $~~~~\,2.21~~~~\,\pm~\,0.50~~~~\,$& $~~~~\,1.91~~~\pm~~~0.27~~~$& $~~~~\,1.69~~~\pm~~~0.09~~~$	& $~~~~\,1.19~~~\pm~~~0.07~~~$	& ...			\\
			&MIR2 (mJy)	&  $<8.3~~~~~~~~~~~~~~~~~~~~~~$	& $<82~~~~~~~~~~~~~~~~~~~~~$ 	& $~~~~\,0.524~\,\pm~~~0.021~\,	$&~\,...	& ... 	\\
			&	FIR1 (Jy)	&  ~~~...							& $~\,~~~0.0073\pm~~~0.0022$	&  ~\,...							& $~~~~\,0.0160\pm~~~0.0023	$	& ...			\\ \vspace{1 mm} 
			&	FIR2 (Jy)	& ~~~... 					& $~\,~~~0.0192\pm~~~0.0026	$	& $~~~~\,0.0175\pm~~~0.0036$		& $<0.017~~~~~~~~~~~~~~~$			& $<0.017~~~~~~~~~~~~~~~~~$	\\	  
\hline
PKS\,1549-79	&	UV/Opt. (mJy)  & $<0.43~~~~~~~~~~~~~~~~~~~\, $& $<0.34~~~~~~~~~~~~~~~~\, $	& ~\,...						& \,~... 						& ...			\\
			&	Opt. (mJy)	    	& ~~~...						& ~~~...						& ~\,...						& \,~...						& ...			\\
			&	Opt./NIR (mJy) & $~~~~\,0.360~~~\pm~\,0.108~~~$	& ~~~...						& ~\,...						& $<6.2~~~~~~~~\,~~~~~~~~~~ $		& {\em R}: (4)		\\
			&	NIR (mJy)	      & $<6.3~~~~~~~~~~~~~~~~~~~~$	& $~\,~~~8.19~~~\pm~~~1.70~~~$	& $~~~15.2~~~\,~\pm~~~0.5~~~\,~$	& \,~...						& ...			\\
			& MIR1 (mJy)		& ~~~...						& $~~~32.7~~~~\,\pm~~~1.3~~~~\,$	&  ~\,... 			& $~~~68.9~~\,~~\pm~\,17.2~~~\,~$	& IRS 8\uma\tablenotemark{c}	\\
			&MIR2 (mJy)&  $~\,174~~~~~~~~~\pm10.\,~~~~~~~~$& $~\,424~~~~~~\,\pm~\,59~~~~~~\,$ & $~\,422~~~~\,~~\pm106~~~\,~~~$&$1020~~~~~\,~\pm150~~~~\,~~$ & IRS24\uma\tablenotemark{d}; 60\uma: (5) \\
			&	FIR1 (Jy)	&  ~~~...							& ~\,...						&  ...							& $~~~~\,0.928~\,\pm~~~0.093~\,$	& ...			\\ \vspace{1 mm} 
			&	FIR2 (Jy)	& ~~~... 					& $~\,~~~0.628\,~\pm~~~0.063\,~$	& $~~~~\,0.200~\,\pm~~~0.023\,\tablenotemark{l}$ & $~~~~\,0.096~\,\pm~~~0.015\,\tablenotemark{l}$& $<0.068\,\tablenotemark{l}~~~~~~~~~~~~~~~$	\\
\hline
3C\,338		&	UV/Opt. (mJy)  & $~~~~\,0.179~~~\pm~\,0.020~~~$	& $~~~~\,0.498~\,\pm~~~0.051~\,$	& $~~~~\,5.04~~~\pm~~~0.36~~~$	& ~~~...						& ...			\\
			&	Opt. (mJy)	    	& ~~~...						& $~~~21.6~~~~\,\pm~~~0.8~~~~\,$	& ~\,... 	 					& $~~~59.8~~~\,~\pm~~~1.9~~~\,~$	& ...			\\
			&	Opt./NIR (mJy) & ~~~...					       	& $~\,101~~~~~~\,\pm~~~3~~~~~~\,$& $~\,112~~~\,~~~\pm~~~6~~~\,~~~$	& $~\,176~~~~~\,~\pm~\,17~~~~\,~~$	& ...			\\
			&	NIR (mJy)	     & $~\,210.\,~~~~~~~\pm28~~~~~~~~$	& $~\,167~~~~~~\,\pm~\,28~~~~~~\,$	& $~~~69.9~~~\,~\pm~~~2.1~~~\,~$& $~~~77.9~~~\,~\pm~~~3.0~~~\,~$	& ... 			\\
			& MIR1 (mJy)		& $~\,47.8~~~~~~\pm~\,8.4~~~~\,$	& $~~~46.0~~~~\,\pm~~~2.0~~~~\,$	& $~~~33.2~~~\,~\pm~~~5.2~~~\,~$	& $~~~26.7~~~\,~\pm~~~1.3~~~\,~$	& ... 			\\
			&MIR2 (mJy)	&  $<39~~~~~~~~~~~~~~~~~~~~~~~~\,~$ & $<330~~~~~~~~~~~~~~~~~~~~~~~$	& $~~~14.0~~~\,~\pm~~~0.6~~~\,~$& ~\,...	& ...	\\
			&	FIR1 (Jy)	&  ~~~...							& $<0.064~~~~~~~~~~~~~~~	$				&  ...							& $<0.058	~~~~~~~~~~~~~~~$ 	& ...			\\ \vspace{1 mm} 
			&	FIR2 (Jy)	& ~~~... 					& $<0.069~~~~~~~~~~~~~~~$ 	& $<0.15\,\tablenotemark{i}~~~~~~~~~~~~~~~$	& $<0.14\,\tablenotemark{i}~~~~~~\,~~~~~~~~~$	& $<0.066\,\tablenotemark{i}~~~~~~~~~~~~~~~$ \\  
\hline
3C\,386		&	UV/Opt. (mJy)  & $<0.73~~~~~~~~~~~~~~~~~~~\, $& $~\,~~~0.606~\,\pm~~~0.194~\,$	& ~\,...						& \,~... 						& ...			\\
			&	Opt. (mJy)	    	& ~~~...						& ~\,...						& ~\,...						& \,~...						& ...			\\
			&	Opt./NIR (mJy) & $<37~~~~~~~~~~~~~~~~~~~~~~~~~ $	&~\, ...					& ~\,...						& $~~~52.9~~~\,~\pm~~~5.1~~\,~~$	& {\em R}: (6)		\\
			&	NIR (mJy)	      & $~~~82.5~~~~~~\pm~\,6.9~~~~~~$	& $~~~77.6~~~~\,\pm~~~9.3~~~~\,$	& $~~~46.1~~~\,~\pm~~~1.4~~~\,~$	& $~~~45.4~~~\,~\pm~~~1.7~~\,~~$	& ... 			\\
			& MIR1 (mJy)		& $~\,29.4~~~~~~\pm~\,1.7~~~~\,$	& $~~~29.1~~~~\,\pm~~~1.2~~~~\,$	& $~~~~\,8.06~~~\pm~~~0.36~~~$	& $~~~~\,9.66~~~\pm~~~0.40~~~$	& ...			\\
			&MIR2 (mJy)	&  $<20~~~~~~~~~~~~~~~~~~~~~~~~\,~$ & $<170~~~~~~~~~~~~~~~~~~~~~~~$	& $~~~~\,7.78~~~\pm~~~0.84~~~$&~\,...						& ... 			 \\
			&	FIR1 (Jy)	&  $~~~~\,0.0428~\,\pm~\,0.0091~\,$	& ~\,...						&  ~\,...							& $~~~~\,0.108~\,\pm~~~0.029~\,$	& ...			\\ \vspace{1 mm} 
			&	FIR2 (Jy)	& ~~~... 					& $~\,~~~0.165~\,\pm~~~0.035~\,$	& $~~~~\,0.0787\pm~~~0.0218$		& $<0.057~~~~~~\,~~~~~~~~~$								& $<0.045~~~~~~~~~~~~~~~~~$	\\
\hline
3C\,424		&	UV/Opt. (mJy)  & $<0.052~~~~~~~~~~~~~~~~~~\, $& $<0.043~~~~~~~~~~~~~~~\, $	& ~\,...						& \,~...						& ...			\\
			&	Opt. (mJy)	    	& ~~~...						& ~\,...						& $~~~~\,0.238~\,\pm~~~0.710~\,$		& \,~...						& {\em V}: (7) 		\\
			&	Opt./NIR (mJy) & ~~~...						& ~\,...						& ~\,...						& $<4.3~~~~~~~~\,~~~~~~~~~~ $	& ...			\\
			&	NIR (mJy)	      & $<7.6~~~~~~~~~~~~~~~~~~~~$	& $<6.7~~~~~~~~~~~~~~~~~~\,$	& $~~~~\,0.992~\,\pm~~~0.113~\,$		& \,~...						& ...			\\
			& MIR1 (mJy)		& ~~~...						& $~~~~\,1.02~~~\pm~~~0.26~~~$	&  ~\,... 			& $~~~~\,0.692~\,\pm~~~0.173~\,$		& IRS 8\uma\tablenotemark{c}	 \\
			&MIR2 (mJy)	&  $<8.9~~~~~~~~~~~~~~~~~~~~~~$	& $<170~~~~~~~~~~~~~~~~~~~~~~~$	& $~~~~\,2.03~~~\pm~~~0.66~~~$& ~\,...	& IRS24\uma\tablenotemark{d}	\\
			&	FIR1 (Jy)	&  ~~~...							& ~\,...						&  ~\,...							& $<0.0041~~~~~~~~~~~\,~~$		& ...			\\ \vspace{1 mm} 
			&	FIR2 (Jy)	& ~~~... 			& $~\,~~~0.0140\pm~~~0.0022	$	& $<0.023~~~~~\,~~~~~~~~~~$			& $<0.027\,\tablenotemark{j}~~~~\,~~~~~~~~~~$		& $<0.026\,\tablenotemark{j}~~~~~~~~~~~~~~~$		\\
\hline
IC\,5063		&	UV/Opt. (mJy)  & $~~~~\,0.526~~~\pm~\,0.054~~~$	& $~\,~~~1.59~~~\pm~~~0.16~~~$	& ~\,...						& $~~~11.9~~~\,~\pm~~~2.4~~\,~~$		& {\em U}: (8)		\\
			&	Opt. (mJy)	    	& $~~~39.6~~~~~~\pm~\,7.9~~~~~~$& ~\,...						& $~~~79.6~~~\,~\pm~\,15.9~~~\,~$		& \,~...						& {\em B/V}: (8) 		\\
			&	Opt./NIR (mJy) & ~~~...						& ~\,...						& ~\,...						& $~\,250.\,~~\,~~~\pm~\,14~~~~\,~~$	& ...			\\
			&	NIR (mJy)	     & $~\,211~~~~~~~~\pm20.~~~~~~~\,$	& $~\,237~~~~~~\,\pm~\,18~~~~~~\,$& $~\,121~~~\,~~~\pm~~~3~~~\,~~~$	& $~\,148~~~\,~~~\pm~\,21~~~~\,~~$	& ... 			\\
			& MIR1 (mJy)		& $156~~~~~~~~\pm36~~~~~~\,$	& $~\,153~~~~~~\,\pm~~~5~~~~~~\,$	& $~\,253~~~\,~~~\pm~\,34~~~\,~~~$	& $~\,408~~~~\,~~\pm102~~~~\,~~$& IRS 8\uma\tablenotemark{c}	 \\
			&MIR2 (mJy)	&  $1410~~~~~~~~~\pm70~~~~~~~~~$	& $3210~~~~~~\,\pm200~~~~~~\,$	& $3380~~~~\,~~\pm140~~~~\,~~	$&~\,...						& ...	\\
IC\,5063		&	FIR1 (Jy)	&  ~~~...							& $~\,~~~3.39~~~\pm~~~0.51~~~$	&  ~\,...							& $~~~~\,4.65~~~\pm~~~0.60~~~$	& ...			\\ \vspace{1 mm} 
			&	FIR2 (Jy)	& ~~~... 					& $~\,~~~4.34~~~\pm~~~0.52~~~$	& $~~~~\,2.22~~~\pm~~~0.23~~~$		& $~~~~\,1.01~~~\pm~~~0.11~~~$					& $0.365~\,\pm0.042~\,$	\\
\hline
3C\,433		&	UV/Opt. (mJy)  & ~~~...							& $<0.13~~~~~~~~~~~~~~~~\, $	& ~\,...						& \,~... 						& ...			\\
			&	Opt. (mJy)	    	& $~~~~\,0.368~~~\pm~\,0.110~~~$	& ~\,...						& $~~~\,~1.05~~~\pm~~~0.32~~~$	& \,~...						& {\em B/V}: (7) 		\\
			&	Opt./NIR (mJy) & ~~~...						& ~\,...						& ~\,...						& $~~~~\,5.74~~~\pm~~~1.14~~~$	& ...			\\
			&	NIR (mJy)	   & $~~~~\,7.76~~~~\,\pm~\,1.28~~~~\,$	& $~\,~~~8.84~~~\pm~~~1.10~~~$	& $~~~15.5~~\,~~\pm~~~0.5~~\,~~$	& \,~...						& ... 			\\
			& MIR1 (mJy)		& ~~~...						& $~~~25.1~~~~\,\pm~~~0.9~~~~\,$	&  ~\,... 			& $~~~40.9~~~\,~\pm~\,10.2~~\,~~$		& IRS 8\uma\tablenotemark{c}	 \\
			&MIR2 (mJy)	&  $~~~73.8~~~~~~~\pm~\,5.7~~~~~~~$	& $~\,131~~~~~~\,\pm~\,40.~~~~~~$	& $~\,132~~~\,~~~\pm~\,33~~~\,~~~$& ~\,...	& IRS24\uma\tablenotemark{d}	\\
			&	FIR1 (Jy)	&  ~~~...							& $~\,~~~0.329~\,\pm~~~0.034~\,$	&  ~\,...							& $~~~~\,0.267~\,\pm~~~0.027~\,$	& ...			\\ \vspace{1 mm} 
			&	FIR2 (Jy)	& ~~~... 					& $~\,~~~0.116~\,\pm~~~0.013~\,$	& $~~~~\,0.0596\pm~~~0.0127$		& $<0.033~~~~~~\,~~~~~~~~~$						& $<0.022~~~~~~~~~~~~~~~~~$	\\
\hline
3C\,436		&	UV/Opt. (mJy)  & $<0.052~~~~~~~~~~~~~~~~~~\, $& $<0.012~~~~~~~~~~~~~~~\, $	& $<0.13~~~\,~~~~~~~~~~~~~~ $		& \,~...						& ...			\\
			&	Opt. (mJy)	    	& ~~~...						& $~\,~~~0.293~\,\pm~~~0.056~\,$	& ~\,...						& $~~~~\,0.762~\,\pm~~~0.082~\,$		& ...			\\
			&	Opt./NIR (mJy) & ~~~...						& $~~~~\,1.13~~~\pm~~~0.09~~~$	& $~~~~\,1.28~~~\pm~~~0.11~~~$	& $~~~~\,1.54~~~\pm~~~0.51~~~$	& ...			\\
			&	NIR (mJy)	   & $~~~~\,2.36~~~~\,\pm~\,0.78~~~~\,$	& $<2.4~~~~~~~~~~~~~~~~~~\, $	& $~~~~\,0.870~\,\pm~~~0.063~\,$		& \,~...						& ...			\\
			& MIR1 (mJy)		& ~~~...						& $~~~~\,0.760~\,\pm~~~0.125~\,$	&  ~\,... 			& $~~~~\,0.539~\,\pm~~~0.305~\,$		& IRS 8\uma\tablenotemark{c}	 \\
			&MIR2 (mJy)	&  $<4.4~~~~~~~~~~~~~~~~~~~~~~$	& $<70.~~~~~~~~~~~~~~~~~~~~~$	& $~~~~\,2.36~~~\pm~~~0.76~~~$& ~\,...	& IRS24\uma\tablenotemark{d} \\
			&	FIR1 (Jy)	&  ~~~...							& $~~~~~~0.0187\pm~~~0.0023\tablenotemark{j}$&  ~\,...				& $~~~~~~0.0348\pm~~~0.0036\tablenotemark{j} $& ...	\\ \vspace{1 mm} 
			&	FIR2 (Jy)	& ~~~... 	& $~~\,~~~0.0393\pm~~~0.0042\tablenotemark{j}$	& $~~~~~\,0.0249\pm~~~0.0044\tablenotemark{j}$	& ~$~~~~\,0.0217\pm~~~0.0043\tablenotemark{j}$	& $<0.014\,\tablenotemark{j}~~~~~~~~~~~~~~~$	\\
\hline
3C\,459		&	UV/Opt. (mJy)  & $~~~~\,0.0585~\,\pm~\,0.0104~\,$	& $~\,~~~0.105~\,\pm~~~0.013~\,$	& $~~\,~~0.310~\,\pm~~~0.060~\,$		& \,~...						& ...			\\
			&	Opt. (mJy)	    	& ~~~...						& $~\,~~~0.549~\,\pm~~~0.068~\,$	& ~\,...						& $~~~~\,0.828~\,\pm~~~0.077~\,$		& ...			\\
			&	Opt./NIR (mJy) & ~~~...						& $~~~~\,1.09~~~\pm~~~0.08~~~$	& $~~~~\,1.10~~~\pm~~~0.09~~~$	& $~~~~\,1.13~~~\pm~~~0.27~~~$	& ...			\\
			&	NIR (mJy)	   & $~~~~\,1.58~~~~\,\pm~\,0.40~~~~\,$	& $~\,~~~2.18~~~\pm~~~0.44~~~$	& $~~~~\,1.53~~~\pm~~~0.08~~~$	& \,~...						& ... 			\\
			& MIR1 (mJy)		& ~~~...						& $~~~~\,2.08~~~\pm~~~0.16~~~$	&  ~\,... 			& $~~~~\,4.27~~~\pm~~~1.07~~~$	& IRS 8\uma\tablenotemark{c}	 \\
			&MIR2 (mJy)	&  $~~~13.0~~~~~~\pm~\,2.0~~~~~~$ & $<93~~~~~~~~~~~~~~~~~~~~~$	& $~~~53.3~~~\,~\pm~~~2.1~~\,~~$& ~\,...						& ...			\\
			&	FIR1 (Jy)	&  $~~~~\,0.611~~~\pm~\,0.100~~~$	& ~\,... 						&  ~\,...							& $~~~~~~0.817~\,\pm~~~0.082~\,\tablenotemark{j}$ & ...	\\ \vspace{1 mm} 
			&	FIR2 (Jy)	& ~~~... 			& $~~\,~~~0.634~\,\pm~~~0.063~\,\tablenotemark{j}$	& $~~~~~\,0.246~\,\pm~~~0.025~\,\tablenotemark{j}$	& $~~~~~\,0.117~\,\pm~~~0.013~\,\tablenotemark{j}$	& $~0.0466\pm0.0093\tablenotemark{j}$	
\enddata
\tablenotetext{a}{Upper limits (3$\sigma$) are given when flux was not detected with at least 3$\sigma$ of confidence.}
\tablenotetext{b}{Literature photometry is only obtained in the absence of observations in similar bands. References for literature photometry: (1) \citet{varela09}, (2) \citet{tremblay10}, (3) \citet{golombek88}, (4) \citet{drake04}, (5) \citet{moshir90}, (6) \citet{martel99}, (7) \citet{smith89}, and (8) \citet{devaucouleurs91}.}
\tablenotetext{c}{When \spitzer did not observe a galaxy with the IRAC instrument, we use IRAC 8\uma photometry estimated from the IRS spectrum, as part of the enhanced products of the Spitzer Heritage Archive. We also use this 8\uma photometry for IC\,5063 whose IRAC image is saturated.}
\tablenotetext{d}{When \spitzer did not observe a galaxy with the MIPS instrument, we use MIPS 24\uma photometry estimated from the IRS spectrum as part of the enhanced products of the Spitzer Heritage Archive. We also use this 24\uma photometry for 3C\,310 whose flux was not detected with 3$\sigma$ confidence on the MIPS image.}
\tablenotetext{e}{Used in the complete absence of \herschel data (3C\,270, 3C\,272.1, 3C\,293) or in the absence of either PACS or MIPS 70\um observations (PKS\,1549-79).}
\tablenotetext{f}{Used in the absence of PACS 70\uma photometry.}
\tablenotetext{g}{Used in the complete absence of \herschel data (3C\,270, 3C\,272.1, 3C\,293).}
\tablenotetext{h}{3C\,218 was only observed by the IRAC instrument at 4.5\um and 8.0\um.}
\tablenotetext{i}{Although detected at $>3\sigma$, we cannot fully disentangle the emission from the host galaxy from that of the other galaxies in the nest, so we consider this measurement an upper limit.}
\tablenotetext{j}{These photometry were extracted with the point source aperture (12$''$ for PACS bands; 22$''$ for 250\um; 30$''$ for 350\um; 42$''$ for 500\um), which is larger than the aperture used at shorter wavelengths.}
\tablenotetext{k}{3C\,218 has a nearby (30'') source that likely contaminates the 500\um photometry, so this photometry should be used with caution.}
\tablenotetext{l}{These photometry were extracted with the point source aperture to limit contamination by diffuse foreground structure. }
\end{deluxetable*}
\clearpage
\end{landscape}

\section{B. Additional X-ray Observations}

At the time of our last paper \citep{lanz15}, Mrk\,668 did not have a non-proprietary \chandra observation. PKS\,1549-79 has still not yet been observed with \chandra. For completeness, we have reduced the newly released \chandra observation of Mrk\,668 in the same manner and use the \xmm observation of PKS\,1549-79 to estimate diffuse X-ray emission. Below, we summarize these observations and the reduction done.

\subsection{B1. Mrk\,668}

Mrk\,668 was observed with \chandra for 34.6\,ks on 2014 September 04 (ObsID 16071; P.I. A. Siemiginowska), and will be discussed in detail by Sobolewska et al. (2016, in preparation). We retrieved the observation from the Chandra archive and reduced it in the manner described in \citet{lanz15}. The X-ray emission is clearly dominated by the central source. However, the hardness ratio (HR = (H--S)/(H$+$S) where H is the net counts in the 2--8\,keV band and S is the net counts in the 0.5--2\,keV band), of that emission is softer at $-0.12$ than expected for an AGN, suggesting that small scale diffuse emission may also be present. If we exclude that central source, we do not have sufficient counts remaining to fit a spectrum.

We therefore sought to estimate the diffuse X-ray luminosity in two ways. First, we measured the net (background-subtracted) count rate in the aperture, excluding the central $1\farcs0$ (in the same manner as was done for the galaxies in \citealt{lanz15}), in the 0.5--8\, keV, 0.5--2\,keV, and 2--8\,keV bands. Based on the hardness ratio which provides a sense of the temperature, we estimate its flux assuming a thermal \citep[APEC; ][]{smith01} model with the temperature from another radio galaxy with a similar hardness ratio (3C\,433; log$T=7.05$, $kT = 0.967$\,keV), using the WebPIMMS tool. We assume solar metallicity and a fixed foreground absorption due to the Milky Way's ISM ($1.6\times10^{20}$\,cm$^{-2}$, \citealt{kalberla05}). We also use the 2--8\,keV counts within the central $1\farcs0$ aperture to estimate the AGN's 2--10\,keV luminosity, assuming a power-law with $\Gamma=1.7$.

Second, we extracted a spectrum from the entire aperture (including the central source). We fit a combination of  thermal models and an absorbed power-law, all subject to absorption due to the Milky Way's ISM. Our best model required two thermal components (0.33\,keV and 1.6\,keV) as well as an absorbed power-law ($\Gamma=1.7,~N_{\rm H} = 8.3\times10^{21}\,{\rm cm^{-2}}$. From this fit, we calculate the 0.5--8\,keV luminosity of the diffuse emission (thermal components) and the unabsorbed 2--10\,keV emission of the AGN (power-law component).

These two methods yield consistent values for both the diffuse (log($L_{0.5-8\,keV}/{\rm erg\,s^{-1})} = 41.4-41.7$) and the AGN  (log($L_{2-10\,keV}/{\rm erg\,s^{-1})} = 42.3-42.5$) emission. Our AGN luminosity is in good agreement with the measurement of \citet{guainazzi04}, and the ratio of $L$(H$_{2}$)/$L_{\rm X, diffuse}$ that we measure for Mrk\,668 is consistent with those of other radio MOHEGs (see Fig. 9 of \citealt{lanz15}). Therefore, we believe that our modeling has yielded reliable values of the diffuse X-ray emission.

\subsection{B2. PKS\,1549-79}

Since \xmm has a much poorer spatial resolution than \chandra, we cannot hope to achieve the same spatial separation between the AGN and any diffuse emission in the \xmm observation of PKS\,1549-79. Instead, we estimate the diffuse emission based on a spectral decomposition. We attribute the power-law component to the AGN and any thermal component to the diffuse emission. As discussed in the previous section, we found good agreement for the estimates of diffuse X-ray luminosity from spatial and spectral decompositions for Mrk\,668. 

PKS\,1549-79 was observed with \xmm for 86.72\,ks on 2008 September 22 (ObsID 0550970101). We retrieved and analyzed the data taken with the European Photon Imaging Camera \citep[EPIC; ][]{jansen01} on both the metal oxide semiconductor (MOS) CCDs and the pn CCDs. There is significant background flaring in the last $\sim30$\,ks, which we filtered out reducing the exposure time to 55.15\,ks (MOS1/MOS2) and 53.69\,ks (pn). We only retained events with energies between 0.4 and 10\,keV with patterns between 0 and 12. We extracted the spectrum in a 45$''$ aperture in each data set and combined them to create the EPIC spectrum. We selected this aperture size, despite being smaller than the aperture in Table \ref{sample} because it contains all of the X-ray emission and does not cross the pn chip gap (as the Table \ref{sample} aperture would). This aperture contains $\sim55000$ counts, dominated by the hard component, likely the AGN.

Spectral modeling was accomplished using the {\sc sherpa} packages of {\sc ciao}. We fit a combination of thermal \citep[APEC; ][]{smith01} and an absorbed power-law, all of which is subject to a fixed foreground absorption due to the Milky Way's ISM ($9.4\times10^{20}$\,cm$^{-2}$, \citealt{kalberla05}) as well as a fitted intrinsic absorption. Our best model (well fit with $\chi^2$/dof = 949/1181) requires two thermal components (0.27\,keV and 2.3\,keV), a power-law index of $\Gamma=2.0$, an intrinsic column of $N_{\rm H}=1.5\times10^{22}$\,cm$^{-2}$, and an additional column of $N_{\rm H}=3.5\times10^{20}$\,cm$^{-2}$ on the power-law component. 

As noted by \citet{obrien10}, the \xmm spectrum is dominated by its buried AGN. Indeed, we find that 96\% of the 0.5--8\,keV luminosity comes from the power-law component. From our fit, we measure the 2--10\,keV luminosity of the absorption-corrected (both intrinsic and foreground) power-law component as well as the 0.5--8\,keV luminosity of the foreground-absorption corrected (as was done for the galaxies in \citealt{lanz15}) thermal component (given in Table \ref{litprop}). We find that our AGN luminosity is in good agreement with that reported by \citet{gonzalez12} and that the diffuse X-ray luminosity would place PKS\,1549-79 in the region occupied by radio MOHEGs in our plot of H$_{2}$ luminosity versus diffuse X-ray luminosity (Figure 9 of \citealt{lanz15}). As a result, we believe that our spectral decomposition provides a reliable value of the diffuse X-ray luminosity.

\end{document}